%% file: baryonPaperCDS.tex
\RequirePackage{fix-cm}
\RequirePackage{graphicx}
\RequirePackage{lineno}
\RequirePackage{color}
\RequirePackage[normalem]{ulem}
\RequirePackage{multirow}
\RequirePackage{tabularx}
\RequirePackage{sidecap}
\RequirePackage{amsmath}
\RequirePackage{placeins}
\RequirePackage{fixltx2e}

\documentclass[ALICE,manyauthors]{cernphprep}

\RequirePackage{booktabs}
\definecolor{dgreen}{cmyk}{1.,0.,1.,0.2}      
\definecolor{orange}{cmyk}{0.,0.353,1.,0.}    
\newcommand{\orange}[1]{{\color{orange}{\sout{#1}}}}
\newcommand{\blue}[1]{{\color{blue}{#1}}}

\def \new {\blue}
\def \old {\orange}

\newcommand{\hijingb}{HIJING/B}

\newcommand{\pt}{$p_{\rm{T}}$}
\newcounter{vers}
\newcommand{\ITS}{\rm{ITS}}
\newcommand{\SPD}{\rm{SPD}}
\newcommand{\SDD}{\rm{SDD}}

\newcommand{\TPC}{\rm{TPC}}

\newcommand{\VZERO}{\rm{VZERO}}

\newcommand{\pbarp}{$\overline{\rm p}/{\rm p}$}
\newcommand{\LbarL}{$\overline{\rm \Lambda}/{\rm \Lambda}$}
\newcommand{\XbarX}{${{\overline{\rm \Xi}}^{+}}/{\rm \Xi^{-}}$}
\newcommand{\ObarO}{${{\overline{\rm \Omega}}^{+}}/{\rm \Omega^{-}}$}

\begin{document}%
%
%
\begin{titlepage}
\PHnumber{2013-080}                 
\PHdate{May 03, 2013}              
%
%
\title{Mid-rapidity anti-baryon to baryon ratios in pp collisions \\ at $\mathbf{\sqrt{s}}$ = 0.9, 2.76 and 7~TeV measured by ALICE}
\ShortTitle{Anti-baryon to baryon ratios in pp collisions at $\sqrt{s}$ = 0.9, 2.76 and 7~TeV}   
%
\Collaboration{ALICE Collaboration%
         \thanks{See Appendix~\ref{app:collab} for the list of collaboration
                      members}}
\ShortAuthor{ALICE Collaboration}      
\begin{abstract}
The ratios of yields of anti-baryons to baryons probes the mechanisms of baryon-number transport. 
Results for \pbarp, \LbarL, \XbarX~and \ObarO~in pp collisions at $\sqrt{s} = 0.9$, $2.76$ and 
$7$~TeV, measured with the ALICE detector at the LHC, are reported. Within the experimental uncertainties and ranges covered by our measurement, these ratios are independent of rapidity, transverse momentum and multiplicity for all measured energies. The results are compared to expectations from event generators, such as PYTHIA and \hijingb, that are used to model the particle production in pp collisions. The energy dependence of \pbarp, \LbarL, \XbarX~and \ObarO, reaching values compatible with unity for $\sqrt{s} = 7$~TeV, complement the earlier \pbarp~measurement of ALICE. These dependencies can be described by exchanges with the Regge-trajectory intercept of $\alpha_{\rm {J}} \approx 0.5$, which are suppressed with increasing rapidity interval ${\rm \Delta} y$. Any significant contribution of an exchange not suppressed at large ${\rm \Delta} y$ (reached at LHC energies) is disfavoured.
\end{abstract}

\end{titlepage}
\setcounter{page}{2}
%

\input{Introduction.tex}


\input{Experimental.tex}

\input{Analysis.tex}

\input{Corrections.tex}


\input{Systematics.tex}


\input{Results.tex}

\input{Summary.tex}



%
\newenvironment{acknowledgement}{\relax}{\relax}
\begin{acknowledgement}
\section*{Acknowledgements}
\input{acknowledgements_march2013.tex}    
\end{acknowledgement}
%
%

\input{references.tex}%
\newpage
%
%
\appendix
\section{The ALICE Collaboration}
\label{app:collab}
\input{authorlist-2013-03-29-cernpreprint.tex}  
\end{document}

%% file: Introduction.tex
\section{Introduction}
\label{Sec:Intro}

Particle production at high transverse momentum (\pt) is well described by processes 
involving hard scattering between partons within the framework of perturbative Quantum Chromodynamics 
(pQCD) \cite{Ref:QCD}. In the low-\pt~region, though, where soft processes dominate, QCD inspired 
phenomenological models are commonly used. Studying the production of these soft particles, should then\old{,} shed light on the basic mechanisms responsible for particle production in this regime. 

In particular, the baryon production still lacks a complete QCD description. We do not have a clear view of whether the baryon number of a hadron should be associated with its valence quarks (as naively expected via analogy with the electric charge) or with its gluonic field. The gauge-invariant state operator representing the baryon in QCD can be interpreted as a configuration where the three (valence) quarks are connected with three strings (gluons), meeting at one point, called the string junction \cite{Ref:RossiVeneziano,Ref:StringJunction}. In this representation, the baryon number is associated with the gluonic field of the baryon, namely with the string junction itself: baryon--anti-baryon pair production from vacuum occurs by string junction and anti-string junction pair production accompanied by a combination of sea quarks and anti-quarks. This should be the mechanism for anti-baryon production in baryon--baryon collisions. The baryons, however, may also contain one of the valence quarks, di-quarks or the string junction (or a combination of the three) of the incoming baryon(s). If any of these constituents undergo a significant diffusion over large rapidity intervals, the spectrum of baryons can differ from the spectrum of anti-baryons at mid-rapidity. These problems have been debated in various theoretical papers for some time \cite{Ref:RossiVeneziano,Ref:QGSM,Ref:StringJunction,Ref:Kopeliovich,Ref:Kharzeev,Ref:QGSMMerino,Ref:HijingB}.

In Regge field theory \cite{Ref:ReggeBook}, the probability of finding the string junction of the beam baryon at relatively large rapidity distance ${\rm \Delta} y$ is given by $\exp\left[\left(\alpha_{\rm{J}}-1\right)\rm {\Delta} y\right]$ \cite{Ref:RossiVeneziano}, where ${\rm \Delta} y = y_{\rm beam} - y$, and $y_{\rm beam}=\ln{(\sqrt{s}/m_{\rm B})}$, is the rapidity of the incoming baryon, $y$ is the rapidity of the string junction and ${\alpha}_{\rm J}$ is the intercept of string-junction trajectory. Since the string junction is a non-perturbative QCD object, it is not possible, at present, to determine theoretically its intercept $\alpha_{\rm{J}}$. Depending on the value of the string-junction intercept, one expects a difference in the spectra of anti-baryons and baryons at mid-rapidity. In particular, if $\alpha_{\rm J} \approx 1$, as proposed in \cite{Ref:Kopeliovich}, then even at very high ${\rm \Delta} y$ values, one would expect a rapidity independent distribution of the incoming baryon string junction. However, if $\alpha_{\rm {J}} \approx 0.5$ as considered in \cite{Ref:RossiVeneziano}, the string-junction transport will approach zero with increasing ${\rm \Delta} y$.

Another source of the difference between the spectra of particles and anti-particles are Reggeon exchanges with negative C-parity \cite{Ref:ReggeBook}. One of the well known Regge poles is the $\rm \omega$ reggeon with intercept $\alpha_{\rm {\omega}} \approx 0.5$. The $\rm \omega$-reggeon exchange is also considered to be the main source of the difference between particle and anti-particle interaction total cross sections for low energy interactions. Since $\alpha_{\rm {\omega}} < 1$, its contribution at mid-rapidity decreases with increasing collision energy. However, if there exists a Regge pole with negative signature and $\alpha \approx 1$, it may also be a source of a difference between particle and anti-particle yields in the central region. In this case, both the inclusive cross sections of particles and anti-particles and the interaction cross sections at asymptotically high collision energies may be different.

One can gather information about the contribution of various mechanisms of baryon production from the spectra of baryons and anti-baryons in proton--proton collisions. In particular, one of the most direct ways to find constraints on different baryon production mechanisms is to measure the ratio of spectra of anti-baryons and baryons $\overline{\rm{B}}/\rm{B}$ with various (valence) quark content e.g. p, $\rm \Lambda$, charged $\rm \Xi$ and $\rm \Omega$ and at different collision energies. For instance, by increasing the strangeness of the observable, one reduces the contribution of the process related to the stopping of different constituents of beam particle. This would have a consequence of $\overline{\rm{B}}/\rm{B}$ ratio being closer to unity for higher strangeness.

The first results from the ALICE collaboration for the \pbarp~ratio in pp collisions at $\sqrt{s} = 0.9$ and $7$~TeV, reporting the measured ratio of the yields of anti-protons to protons at mid-rapidity as compatible with unity at $\sqrt{s} = 7$ TeV, have set stringent limits on the mechanisms of baryon production at LHC energies \cite{Ref:Alicepbarp}. In this article we complement these studies in pp collisions, by reporting the production ratio of \pbarp~at $\sqrt{s} = 2.76$~TeV and of baryons containing strange quarks \LbarL, \XbarX~and \ObarO~at $\sqrt{s} = 0.9$, $2.76$ and $7$~TeV. The results are presented as a function of the particle's rapidity defined as $y=0.5 \ln[(E+p_{\rm z})/(E-p_{\rm z})]$ and transverse momentum defined as $p_{\rm T}=\sqrt{p_{\rm x}^2 + p_{\rm y}^2}$. We also present the rapidity and transverse momentum integrated ratios as a function of the multiplicity (the definition of multiplicity will be given in Section~\ref{Sec:Results}). ALICE results at mid-rapidity are compared with lower energy data and with LHCb data at forward rapidities.

%% file: Experimental.tex
\section{Experimental setup}
\label{Sec:ExpSetup}

ALICE \cite{Ref:ALICE}, the dedicated heavy-ion experiment at the LHC, was designed to cope with the high 
charged-particle densities measured in central Pb--Pb collisions \cite{Ref:AlicedNdeta}. ALICE also provides 
excellent performance for proton--proton interactions \cite{Ref:ALICEPPR}. The experiment consists of a large 
number of detector subsystems \cite{Ref:ALICE} inside a solenoidal magnet (0.5~T). These subsystems 
are optimised to provide high-momentum resolution as well as excellent particle identification (PID) over a 
broad range in momentum. 

Collisions take place at the centre of the ALICE detector, inside a beryllium vacuum beam pipe (3~cm in radius 
and 800~$\mu$m thick). The tracking system in the ALICE central barrel covers the full azimuthal range in the 
pseudorapidity window $|\eta| < 0.9$. For more details on the ALICE experimental setup, see \cite{Ref:ALICE}. 
The following detector subsystems were used in this analysis:

\begin{itemize}
\item The Inner Tracking System (\ITS) \cite{Ref:ALICEITS}, the innermost detector of ALICE, consisting of six 
layers of silicon detectors. The two layers closest to the beam pipe are made of Silicon Pixel Detectors (\SPD) 
and are used for the determination of the primary vertex as well as for track reconstruction. The next two 
layers are made of Silicon Drift Detectors (\SDD), followed by two layers of double-sided Silicon Strip Detectors 
(SSD). Both detectors contribute to the tracking while providing particle identification for low-\pt~particles. The 
\ITS~covers the range $|\eta| < 0.9$.

\item The Time Projection Chamber (\TPC) \cite{Ref:ALICETPC} is the main tracking detector of the central barrel,  
providing, together with the other central-barrel detectors, charged-particle momentum measurements with good 
two-track separation, particle identification, and vertex determination. The phase space covered by the \TPC~in 
pseudorapidity is $|\eta| < 0.9$ for tracks of full radial track length, whereas for reduced track length (and reduced 
momentum resolution), an acceptance up to about $|\eta| = 1.5$ is accessible. The \TPC~ covers the full azimuth, 
with the exception of the dead zones between its sectors (in about 10\% of the azimuthal angle the detector is non-sensitive). 

\item The \VZERO~detector \cite{Ref:ALICE}, used in the trigger system, consists of two arrays of 32 scintillators each, placed 
around the beam pipe on both sides of the interaction region: one (\rm{VZERO-A}) at $z = 3.3$~m, covering $2.8 < \eta < 5.1$, and the other (\rm{VZERO-C}) at $z = 0.9$~m, covering $-3.7 < \eta < -1.7$. The time resolution of this detector is better than 1 ns. Its response 
is recorded in a time window of $\pm 25$~ns around the nominal beam crossing time.
\end{itemize}

%% file: Analysis.tex
\section{Data analysis}
\label{Sec:Analysis}
\subsection{Event sample and selection}

Data recorded during the 2010 and 2011 LHC pp runs at $\sqrt{s} = 0.9$, 2.76 and $7$~TeV were used for this analysis. The trigger required a hit in one of the \VZERO~counters or 
in the \SPD~detector \cite{Ref:ALICEITS}, in coincidence with the signals from two beam pick-up counters, one 
on each side of the interaction region, indicating the presence of passing bunches.

The luminosity at the ALICE interaction point was restricted between 0.6 and $1.2 \times 10^{29}$ $\rm{cm^{-2}}$ $\rm{s^{-1}}$ for all the data used in this analysis. This ensures a collision pile-up rate of 4$\%$ or lower, in each bunch crossing. Beam-induced background was reduced to a negligible level ($< 0.1\%$) \cite{Ref:ALICENchargedpapers} with the help of the timing information from the \VZERO~counters. In addition, in order to minimise acceptance and efficiency biases for tracks at the edge of the TPC detection volume, events are selected by requiring that the distance between the position 
of the primary vertex and the geometrical centre of the apparatus along the beam axis ($z$ position) is less than 
10 cm. The final number of analysed events for each energy is summarised in Table~\ref{tab:NEvents}.

\begin{table}[htbp]
\caption{Number of pp collisions before and after event selection.}
\centering
\begin{tabular}{l c c c}
\toprule
 $\sqrt{s}$ & 0.9~TeV & 2.76~TeV& 7~TeV   \\ 
\midrule
All & 11~M & 58~M & 230~M   \\
Analyzed & 6~M & 40~M & 180~M   \\
\bottomrule
\end{tabular}
\label{tab:NEvents}
\end{table}

\subsection{Selection of protons}

\begin{figure}
  \includegraphics[width=\textwidth]{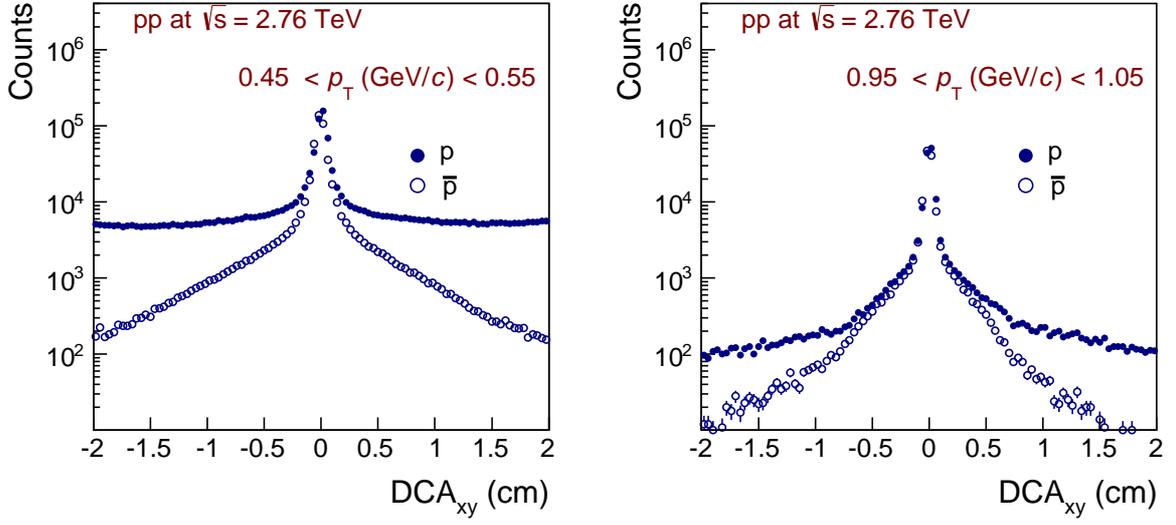}
  \caption{The $\rm{DCA_{xy}}$ distributions for pp at $\sqrt{s} = 2.76$~TeV for the lowest (left) and highest (right) \pt~bins. Protons (anti-protons) are shown with full (open) symbols.}
  \label{fig:DCA}
\end{figure}

Protons and anti-protons are reconstructed and identified by the \TPC, which measures the ionisation in the
\TPC~gas and the particle trajectory with up to 159 space points. Several selection criteria were imposed to 
ensure the quality of accepted tracks. The minimum number of associated \TPC~clusters (space points) 
per track was set to 80. In addition, the $\chi^2$ per \TPC~cluster of the momentum fit did not exceed the 
value of 2 per degree of freedom. A key element of the analysis was the reduction of the contamination of the track sample from background (i.e. particles originating from the interaction of a particle with the material) and secondary (i.e. protons and anti-protons originating from the weak decays of $\rm \Lambda$ and $\overline{\rm \Lambda}$, respectively) particles. To reduce the contamination from background, selected tracks were required to have at least two associated 
\ITS~clusters. Furthermore, a track must have at least one associated \ITS~cluster on either of the \SPD~layers. 
Finally, to further reduce the contamination from background and secondary tracks, a cut on the distance of closest 
approach (DCA) of the track to the primary vertex on the $xy$ plane was set to $0.2~\mathrm{cm}$ (of the order of the primary vertex resolution in $x$ and $y$ directions). The residual contamination is corrected by a data-driven method described in Section~\ref{Sec:Corrections}. Figure~\ref{fig:DCA} presents the DCA distributions for p and $\rm \overline{p}$ with full and open circles respectively, for the lowest ($0.45 < p_{\rm{T}} < 0.55$~GeV/$c$ --- top plot) and highest ($0.95 < p_{\rm{T}} < 1.05$~GeV/$c$ --- bottom plot) \pt~bins (intervals) used in this analysis. The distinct feature of the distribution of protons are long tails at large values of DCA that come predominantly from background protons. The effect is more pronounced for low \pt~values. On the other hand, the corresponding distribution of anti-protons is background free, with the main source of contamination being the weak decay of $\overline{\rm \Lambda}$. 

\begin{figure}
  \includegraphics[width=\textwidth]{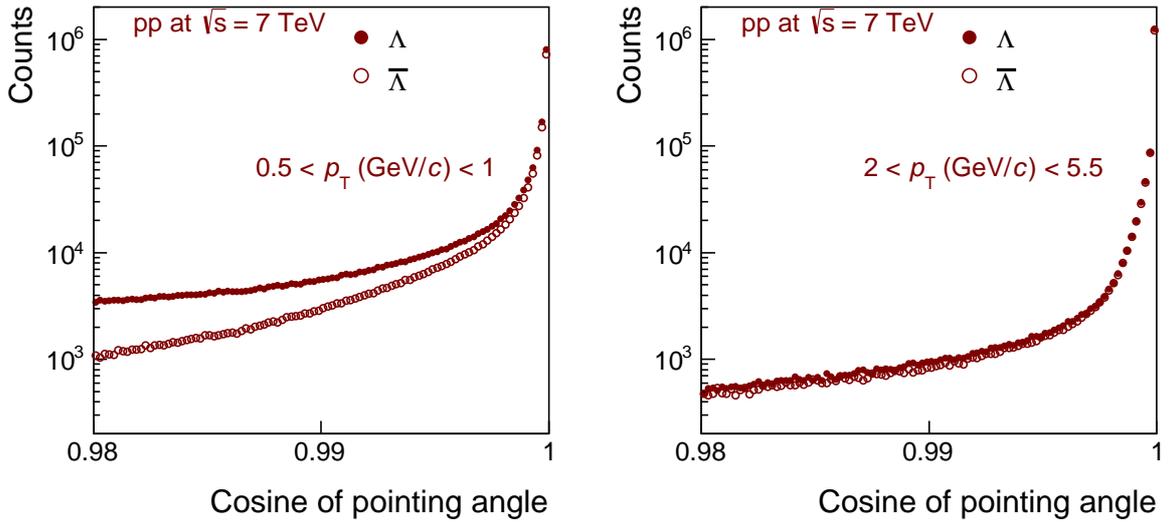}
  \caption{Cosine of pointing angle distributions for pp $\sqrt{s} = 7$~TeV in the lowest (left) and highest (right) $p_{\rm{T}}$ bins. $\rm \Lambda$ ($\overline{\rm \Lambda}$) are shown with full (open) symbols.}
  \label{fig:CPA}
\end{figure}

Particle identification was achieved by correlating the particle momenta as measured at the inner radius 
of the \TPC~and the specific ionisation (d$E$/d$x$) in the \TPC~gas \cite{Ref:ALICETPC}. The d$E$/d$x$ resolution of 
the \TPC~is about $5\%$, depending on the number of \TPC~clusters and the track inclination angle. For this analysis, 
(anti-)protons were selected by defining a band with a $3\sigma$ width with respect to the theoretical Bethe--Bloch 
parametrisation, similar to the procedure followed in \cite{Ref:Alicepbarp}. 

The phase space used for (anti-)protons was limited to $|y| < 0.5$ and $0.45 < p_{\rm{T}} < 1.05$~GeV/$c$. The lower limit of the \pt~value is driven by the systematic uncertainties that will be described later in this article, while the upper limit is chosen based on the increased contamination from the identification procedure due to the overlapping particle bands in the correlation between the d$E$/d$x$ and the momentum. The resulting contamination from other particle species in this \pt~region is negligible ($< 0.1\%$), compatible with the observation in \cite{Ref:Alicepbarp}.

\subsection{Topological reconstruction of $\rm \Lambda$, charged $\rm \Xi$ and $\rm \Omega$}

Baryons and their anti-particles containing strange quarks (i.e. $\rm{\Lambda}$, charged $\rm{\Xi}$ and
$\rm{\Omega}$), the hyperons, are reconstructed via their weak decay topologies in the charged decay channels as summarised 
in Table~\ref{tab:particleDecay}. The measurement of $\rm{\Lambda}$ and $\overline{\rm \Lambda}$ is based on 
the reconstruction of their decay vertexes which exhibits a characteristic V-shape, called V0, defined by the trajectories 
of the decay products. The corresponding measurement of $\rm \Xi$ and $\rm \Omega$ is performed based on the 
cascade topology of the decay, consisting of the aforementioned V-shape structure of the $\rm{\Lambda}$-decay 
and a charged bachelor particle (i.e. $\pi$ and $\rm K$ for the case of $\rm{\Xi}$ and $\rm{\Omega}$, respectively). The 
selection of $\rm \Lambda$, $\rm \Xi$ and $\rm \Omega$ is performed by applying criteria on both the quality 
of the candidates and on the decay products (i.e. the daughter candidates). These criteria, which are analysis 
and energy dependent, are described below and are also summarised in Table~\ref{tab:Cutspp}.

\begin{table*}[tb]
\caption{The valence quark content, mass, decay length and the main decay channel together with the branching 
ratio for baryons containing a strange quark \cite{Ref:PDG}.}
\centering
\begin{tabular}{ccccc}
\toprule
Particle & Mass (MeV/$c^2$) & Decay length (cm) & Decay channel & Branching ratio \\
\midrule
$\rm \Lambda$(uds) & \multirow{2}{*}{$1115.683 \pm 0.006$} & \multirow{2}{*}{7.89} & $\rm \Lambda \rightarrow \textrm{p} + \pi^{-}$ & \multirow{2}{*}{63.9$\%$} \\
$\overline{\rm \Lambda}$($\rm {\bar{u}\bar{d}\bar{s}}$) &  &  & $\overline{\rm \Lambda} \rightarrow \overline{\textrm{p}} + \pi^{+}$ &  \\
\midrule
$\rm \Xi^{-}$(dss) & \multirow{2}{*}{$1321.71 \pm 0.07$}& \multirow{2}{*}{4.91} & $\rm \Xi^{-} \rightarrow \rm \Lambda + \pi^{-}$ & \multirow{2}{*}{99.9$\%$ }\\
$\overline{\rm \Xi}^{+}$($\rm {\bar{d}\bar{s}\bar{s}}$) &  &  & $\overline{\rm \Xi}^{+} \rightarrow \overline{\rm \Lambda} + \pi^{+}$ &  \\
\midrule
$\rm \Omega^{-}$(sss) & \multirow{2}{*}{$ 1672.45 \pm 0.29$} & \multirow{2}{*}{2.46} & $\rm \Omega^{-} \rightarrow \rm \Lambda + \rm K^{-}$ & \multirow{2}{*}{67.8$\%$} \\
$\overline{\rm \Omega}^{+}$($\rm{\bar{s}\bar{s}\bar{s}}$) &  &  & $\overline{\rm \Omega}^{+} \rightarrow \overline{\rm \Lambda} + \rm K^{+}$ &  \\
\bottomrule
\end{tabular}
\label{tab:particleDecay}
\end{table*}

For all three hyperons, the V0 daughter candidates are required to have a minimum DCA 
to the primary vertex, enhancing the probability that they are not primary particles. In addition, a maximum DCA between the daughter candidates at the point of the V0 decay was required to ensure that 
they are products of the same decay. To reduce the contamination from secondary and background strange baryons, a 
minimum value of the cosine of the pointing angle is required. The pointing angle is defined as an angle between 
the momentum vector of the V0 candidate and the vector connecting the primary vertex and the production vertex 
of the V0. Figure~\ref{fig:CPA} presents the relevant distributions for $\rm \Lambda$ and $\overline{\rm \Lambda}$ 
candidates for pp collisions at $\sqrt{s} = 7$~TeV, for two \pt~regions. These distributions at the highest measured energy are also representative of those measured at lower collision energies. It is seen that for low values of \pt, there is a pronounced tail in the distribution of $\rm \Lambda$ originating from background particles. Finally, V0 candidates are required to have a transverse distance between the primary and the production vertex (V0 transverse decay radius) larger than a minimum value. All these cut parameters are listed in Table~\ref{tab:Cutspp}.

Additional selection criteria are applied for the multi-strange baryons (i.e. $\rm \Xi$ and $\rm \Omega$). 
In particular, the bachelor track is required to have a minimum DCA value to the primary vertex, increasing the 
probability that it is not a primary particle. A similar cut is applied to the DCA value of the V0 candidate relative to 
the primary vertex. Furthermore, a maximum value for the DCA between the V0 candidate and the bachelor track 
at the point of the cascade decay is also required. As in the case of the V0, to reduce the contamination from background 
particles, a minimum cut on the cosine of the pointing angle is applied. The cascade candidates are selected if the transverse distance between the primary and the decay vertex (cascade transverse decay radius) is larger than a minimum value. Also in this case, the cut parameters are listed in Table~\ref{tab:Cutspp}.

Particle identification of the daughter candidates helps to substantially decrease the background, especially in the 
low $p_{\rm{T}}$ -- high $|y|$ regions. Particles are identified using the energy loss signal in the \TPC. The selection is done within a 3$\sigma$ band around the expected d$E$/d$x$ value for each particle type.

In addition, for the case of $\rm \Lambda$ and $\rm \Omega$, we have excluded candidates falling into $\pm 10$~MeV/$c^2$ the mass window of the $\rm {K}_{s}^{0}$ (in case of $\rm \Lambda$) or $\rm \Xi$ (in case of $\rm \Omega$) nominal mass. The result is an improvement of the $S/B$ ratio by a factor of $\approx$1.5. 

\begin{table*}[tb]\footnotesize
\caption{Selection criteria for the $\rm \Lambda$, charged $\rm \Xi$ and $\rm \Omega$ candidates.}
\centering
\begin{tabular}{ p{4.2cm} c c c c c c c c c c c}
\toprule
\addlinespace
& \multicolumn{3}{c}{$\sqrt{s} = 0.9$~TeV} & & \multicolumn{3}{c}{$\sqrt{s} = 2.76$~TeV} & &\multicolumn{3}{c}{$\sqrt{s} = 7$~TeV} \\
\cmidrule{2-4} \cmidrule{6-8} \cmidrule{10-12}
& $\rm \Lambda$ & \multicolumn{2}{c}{$\rm \Xi$} & & $\rm \Lambda$ & $\rm \Xi$ & $\rm \Omega$ & & $\rm \Lambda$ & $\rm \Xi$ & $\rm \Omega$\\
\addlinespace
\midrule
DCA of V0 daughter track to primary vertex (cm) & $>$~0.05 &\multicolumn{2}{c}{$>$~0.01} & & $>$~0.05 & $>$~0.02 & $>$~0.02 & & $>$~0.05 & $>$~0.02 & $>$~0.02\\
\addlinespace
DCA between V0 daughter tracks (cm) & $<$~0.5 &\multicolumn{2}{c}{$<$~3.0} & & $<$~1.5 & $<$~2.0 & $<$~0.4 & & $<$~1.5 & $<$~2.0 & $<$~0.4\\
\addlinespace
Cosine of V0 pointing angle & $>$~0.9 &\multicolumn{2}{c}{$>$~0.97} & & $>$~0.95 & $>$~0.97 & $>$~0.97 & & $>$~0.98 & $>$~0.97 & $>$~0.97\\
\addlinespace
Minimum V0 transverse decay radius (cm) & $=$~0.2 &\multicolumn{2}{c}{$=$~0.2}& & $=$~0.2 & $=$~1.0 & $=$~1.0 & & $=$~0.2 & $=$~1.0 & $=$~1.0\\
\addlinespace
DCA of bachelor track to primary vertex (cm) & - &\multicolumn{2}{c}{$>$~0.01} & & - & $>$~0.03 & $>$~0.03 & & - & $>$~0.03 & $>$~0.03\\
\addlinespace
DCA of V0 in cascade to primary vertex (cm) & - & \multicolumn{2}{c}{$>$~0.001} & & - &$>$~0.05 & $>$~0.05 & & - & $>$~0.05 & $>$~0.05\\
\addlinespace
DCA between V0 and bachelor track (cm) & - &\multicolumn{2}{c}{$<$~3.0} & & - &$<$~2.0 & $<$~0.5 & & - &$<$~2.0 & $<$~0.5\\
\addlinespace
Cosine of cascade pointing angle & - &\multicolumn{2}{c}{$>$~0.85} & & - &$>$~0.97 & $>$~0.98 & & - &$>$~0.97 & $>$~0.98\\
\addlinespace
Minimum cascade transverse decay radius (cm) & - &\multicolumn{2}{c}{$=$~0.2} & & - & $=$~0.04 & $=$~0.04 & & - & $=$~0.04 & $=$~0.04\\
\bottomrule
\end{tabular}
\label{tab:Cutspp}
\end{table*}

Finally, the phase space used for each of the analysed baryons is summarised in Table~\ref{tab:PhaseSpace}. The lower limits are chosen based on the low signal to background ratio, while the upper values are driven by the limited statistics.
 
\begin{table*}[tb]\footnotesize
   \caption{Rapidity and \pt~ranges used for each baryon in this analysis.}
\centering
\begin{tabular}{ l c c c c}
\toprule
$\sqrt{s}$~(TeV)&\pbarp & \LbarL & \XbarX & \ObarO\\
\midrule
0.9 & \begin{tabular}[c]{@{}c@{}}$|y|<$0.5\\0.45~$<p_{\rm{T}}(\rm{GeV}/c)<$~1.05\end{tabular} & \begin{tabular}[c]{@{}c@{}}$|y|<$0.8\\0.5~$<p_{\rm{T}}(\rm{GeV}/c)<$~4.0\end{tabular} & \begin{tabular}[c]{@{}c@{}}$|y|<$0.8\\0.5~$<p_{\rm{T}}(\rm{GeV}/c)<$~3.5\end{tabular} & -\\
\addlinespace
2.76 & \begin{tabular}[c]{@{}c@{}}$|y|<$0.5\\0.45~$<p_{\rm{T}}(\rm{GeV}/c)<$~1.05\end{tabular} & \begin{tabular}[c]{@{}c@{}}$|y|<$0.8\\0.5~$<p_{\rm{T}}(\rm{GeV}/c)<$~4.5\end{tabular}  &
\begin{tabular}[c]{@{}c@{}}$|y|<$0.8\\0.5~$<p_{\rm{T}}(\rm{GeV}/c)<$~4.5\end{tabular} & \begin{tabular}[c]{@{}c@{}}$|y|<$0.8\\1.0~$<p_{\rm{T}}(\rm{GeV}/c)<$~4.5\end{tabular}\\ 
\addlinespace
7 & \begin{tabular}[c]{@{}c@{}}$|y|<$0.5\\0.45~$<p_{\rm{T}}(\rm{GeV}/c)<$~1.05\end{tabular} & \begin{tabular}[c]{@{}c@{}}$|y|<$0.8\\0.5~$<p_{\rm{T}}(\rm{GeV}/c)<$~10.5\end{tabular} & 
\begin{tabular}[c]{@{}c@{}}$|y|<$0.8\\0.5~$<p_{\rm{T}}(\rm{GeV}/c)<$~5.5\end{tabular} &
\begin{tabular}[c]{@{}c@{}}$|y|<$0.8\\1.0~$<p_{\rm{T}}(\rm{GeV}/c)<$~5.5\end{tabular} \\ 
\bottomrule
\end{tabular}
\label{tab:PhaseSpace}
\end{table*}

\begin{figure}
\begin{center}
\includegraphics[width=0.6\linewidth]{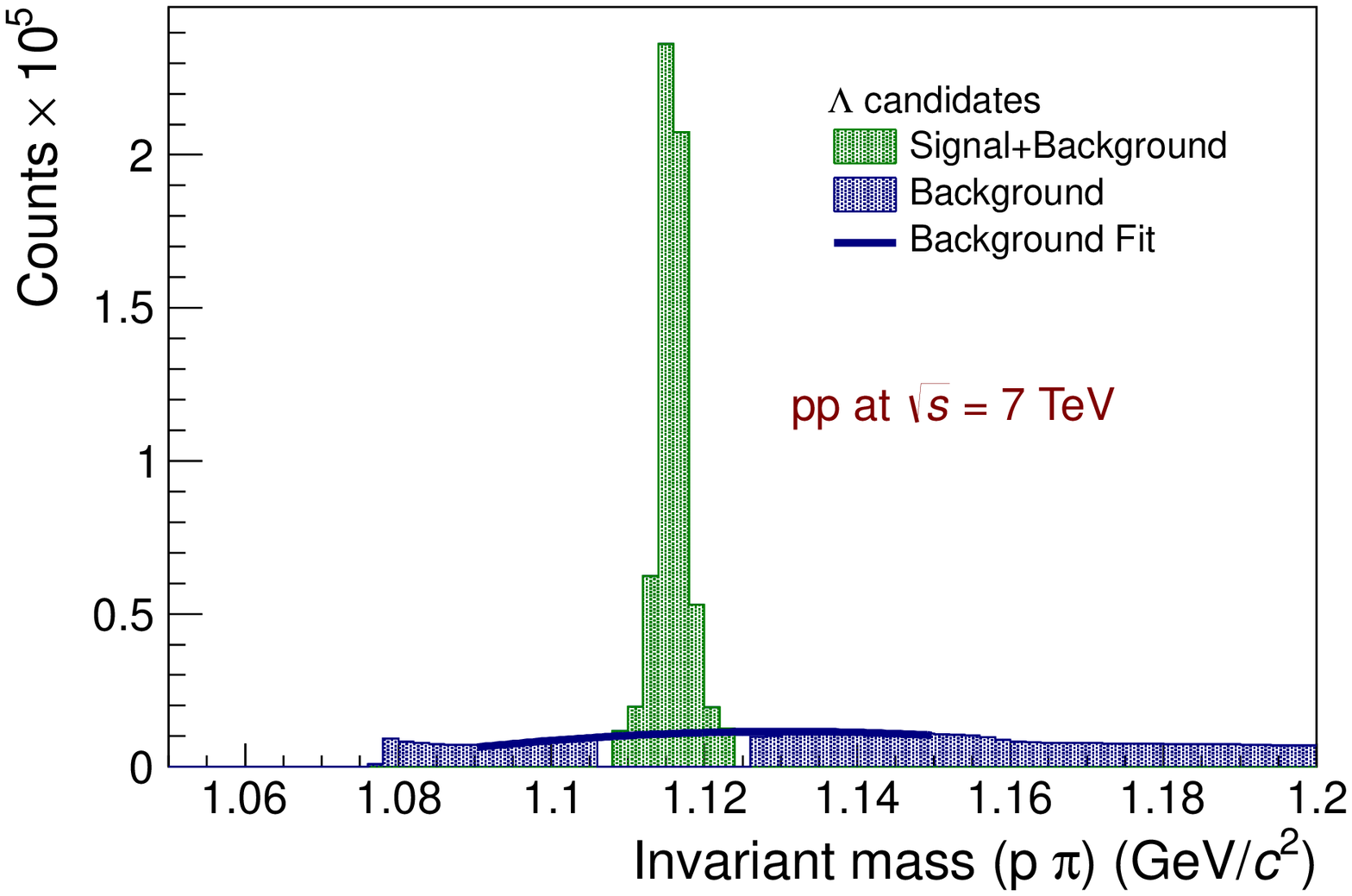}
\includegraphics[width=0.6\linewidth]{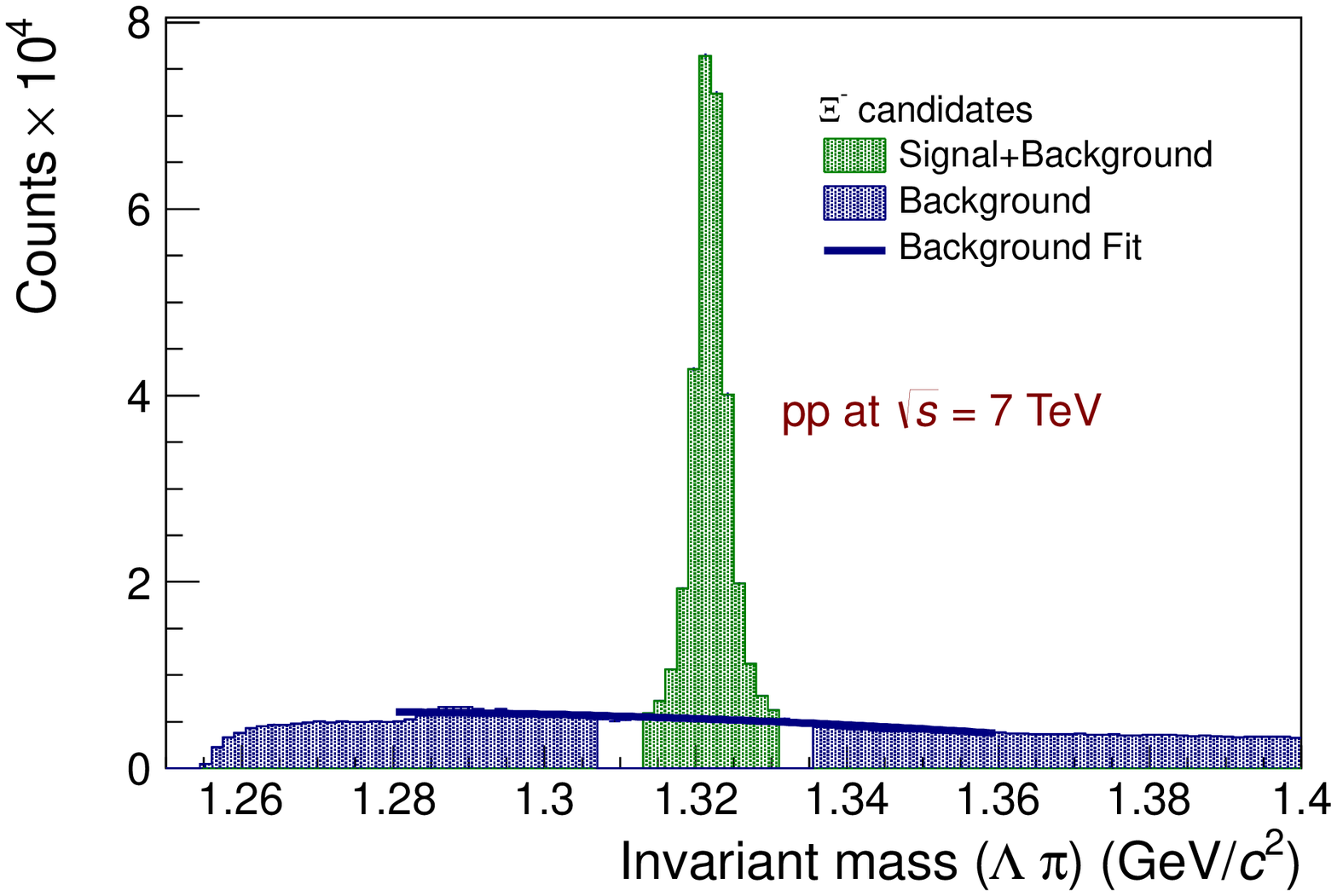}
\includegraphics[width=0.6\linewidth]{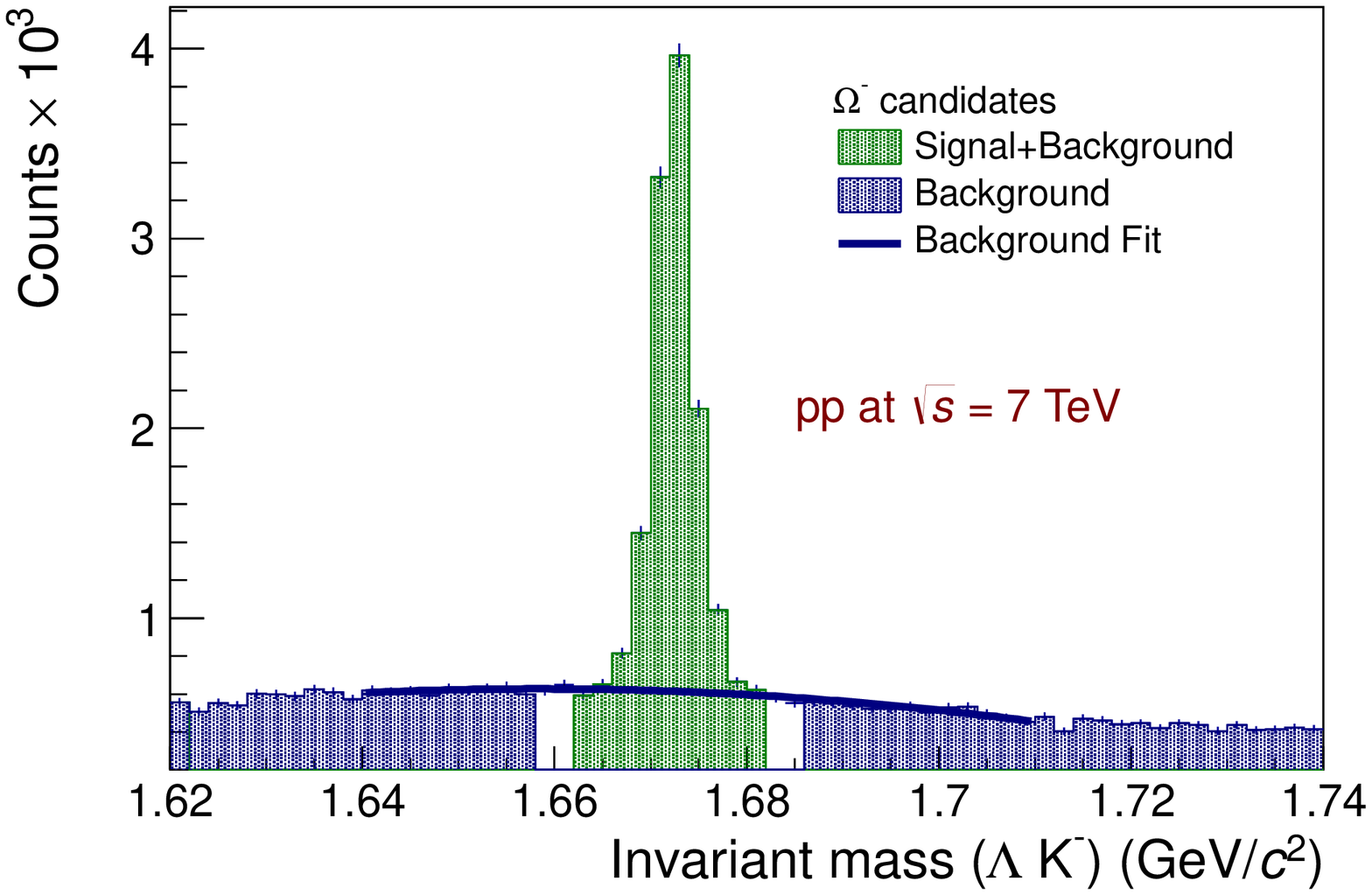}
\caption{(Colour online) The invariant mass distributions for $\rm \Lambda$ (top), $\rm \Xi$ (middle) and 
$\rm \Omega$ (bottom) in pp at $\sqrt{s} = 7$~TeV. Areas considered as signal and background (green) or pure 
background (blue) are shown. The lines corresponds to a polynomial fit to the background areas.}
\label{fig:invMassDistributions}
\end{center}
\end{figure}

The resulting invariant mass distributions for $\rm \Lambda$, $\rm \Xi$ and $\rm \Omega$ candidates in pp collisions 
at $\sqrt{s} = 7$~TeV are presented in Fig.~\ref{fig:invMassDistributions} in the top, middle and bottom plots, respectively.
The raw particle yields are extracted from these distributions divided in different \pt~bins by subtracting the contribution 
of the background (blue areas) from the peak regions (green areas), where both signal and background are located. Both areas are defined by first fitting the peak region with a Gaussian function and extracting the mean ($\mu$) and the width ($\sigma$). The 
sum of the signal and background ($S+B$) is sampled in the region defined by $\mu \pm 4\sigma$~\cite{Ref:ALICEstrangeness}, while the 
background is sampled on each side of the peak region using the areas that are more than $6\sigma$~\cite{Ref:ALICEstrangeness} away from the 
Gaussian mean.

The background is estimated by either simultaneously fitting both sides with a polynomial function or by simply counting 
the number of entries, the so-called ``bin-counting'' method. Both methods give similar results, however the 
first method is used as the default one. The $S/B$ is analysis and \pt~dependent and is summarised in Table \ref{tab:SignalBackground}.

\begin{table}[h]
\caption{Signal to background ratio for different hyperons and \pt~bins.}
\centering
\begin{tabular}{l c c c}
\toprule
\pt~(\rm{GeV}/c) & 1.0 -- 1.5& 3.0 -- 3.5& 5.0 -- 5.5\\ 
\midrule
$\rm \Lambda$, $\overline{\rm \Lambda}$ & 8 &15 &12\\
$\rm \Xi^{-}$, $\overline{\rm \Xi}^{+}$ & 4& 6& 6\\
$\rm \Omega^{-}$, $\overline{\rm \Omega}^{+}$ & 3& 2& 2\\
\bottomrule
\end{tabular}
\label{tab:SignalBackground}
\end{table}

%% file: Corrections.tex
\section{Corrections}
\label{Sec:Corrections}

The \TPC~\cite{Ref:ALICETPC} is symmetric around mid-rapidity and has full azimuthal coverage, hence many 
detector effects are the same for particles and anti-particles and thus cancel out in the ratio. However, there are 
mechanisms that affect the two particle types differently and need to be accounted for by applying the relevant 
corrections. These corrections are extracted from a detailed Monte Carlo simulation based on the GEANT3 transport code 
\cite{Ref:GEANT} and from data driven methods. The effects considered in this analysis are:

\begin{itemize}
\item the difference in the interactions of baryons and anti-baryons with the material of the detector, resulting in larger 
absorption of the latter particle type,
\item the inelastic cross section of $\overline{\rm p}$--A and $\rm K^{-}$--A interactions accounting for 
the wrong parametrisation that GEANT3 employs \cite{Ref:Alicepbarp},
\item the difference in the elastic cross section for p--A and $\overline{\rm p}$--A, resulting in differences in 
the cut efficiency,
\item the contamination from background particles (mainly $\rm p$ and $\rm \Lambda$) originating from the interaction 
of other particles with the material,
\item finally, the feed-down from secondary (anti-) baryons e.g. $\rm p$($\overline{\rm p}$) originating from the weak decay 
of a $\rm \Lambda$($\overline{\rm \Lambda}$).
\end{itemize}

\noindent Each one of these corrections is described separately in the following paragraphs.

\subsection{Absorption correction}
The inelastic p--A cross section is measured to be significantly different than for $\overline{\rm p}$--A \cite{Ref:XSection}. As a result, different fractions of p and $\overline{\rm p}$ are absorbed when interacting with 
the detectors' material. Similar assumption could be made for the hyperons, however, the corresponding 
cross section values have not been measured experimentally. The absorption correction factors rely on the proper 
description of the inelastic cross sections of p and $\overline{\rm p}$ used as input by the transport model (GEANT3) and on 
the accurate description of the material budget in the simulation.

In \cite{Ref:Alicepbarp} it was pointed out that GEANT3 uses an incorrect parametrisation of the inelastic cross section 
for $\overline{\rm p}$--A interactions. In particular, GEANT3 overestimates the experimentally measured cross sections \cite{Ref:XSection} by a factor of two for $p \approx 1$~GeV/$c$, a value that represents the mean $\overline{\rm p}$ momentum. 
This factor increases for lower momentum values. Also in \cite{Ref:Alicepbarp}, it was reported that FLUKA \cite{Ref:FLUKA} 
describes the data very well. In addition, it was found that a small difference between the input GEANT3 parametrisation 
and the experimentally measured values exists also for the case of $\rm K^-$--A interactions. The latter is important if 
one considers the decay mode of the $\rm \Omega$.

\begin{figure}[htbp]
\begin{center}
\includegraphics[width=0.5\linewidth]{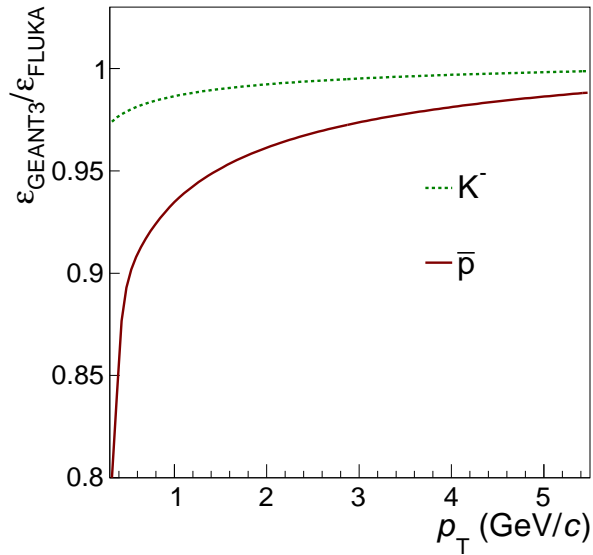}
\caption{(Colour online) The ratio of the detection efficiency for $\overline{\rm p}$ (solid line) and $\rm K^-$ (dashed line) 
calculated from GEANT3 to the one calculated from FLUKA as a function of the hadron's \pt.}
\label{fig:G3F}
\end{center}
\end{figure}

To account for these differences, a full detector Monte Carlo simulation with FLUKA as a transport code was 
used. This simulation was used to scale the absorption correction extracted from GEANT3 to match the correct (i.e. FLUKA) cross section 
parametrisation. The ratio of the detection efficiency calculated using GEANT3 as a transport code ($\rm \varepsilon_{GEANT3}$) to the one using FLUKA ($\rm \varepsilon_{FLUKA}$) as a function of the \pt~is presented in Fig.~\ref{fig:G3F}. The two curves represent the parametrisation of the ratio 

\begin{equation}
f(p_{\rm{T}}) = 1-A\times \exp{(B\times p_{\rm{T}})}+C+D\times \frac{\ln(p_{\rm{T}})}{p_{\rm{T}}^{n}},
\label{eq:G3Fantiproton}
\end{equation}

\noindent used to extrapolate the differences to higher values of \pt. The solid line 
corresponds to $\overline{\rm p}$ [$n = 0.2$ in Eq.~(\ref{eq:G3Fantiproton})] while $\rm K^-$ [$n = 0.15$ in 
Eq.~(\ref{eq:G3Fantiproton})] is represented by the dashed line. It is seen that for both hadrons the curves exhibit a significant \pt~dependence, which is more pronounced for the case of $\overline{\rm p}$. The resulting correction is of the order of $8 \%$ 
for the low-\pt~region (at $p_{\rm{T}} = 0.45$~GeV/$c$) for $\overline{\rm p}$, decreasing with increasing \pt. For the case of the $\rm K^-$, the corresponding correction is smaller ($\approx 2\%$ for $p_{\rm{T}} > 0.4$~GeV/$c$)).

The amount of material in the central part of ALICE is corresponding to about $10 \%$ of a 
radiation length on average between the vertex and the active volume of the \TPC. It has been studied with collision 
data and adjusted in the simulation based on the analysis of photon conversions. The current simulation reproduces 
the amount and spatial distribution of reconstructed conversion points in great detail, with a relative accuracy of a few 
percent.

\begin{figure}[htbp]
\begin{center}
  \includegraphics[width=0.49\linewidth]{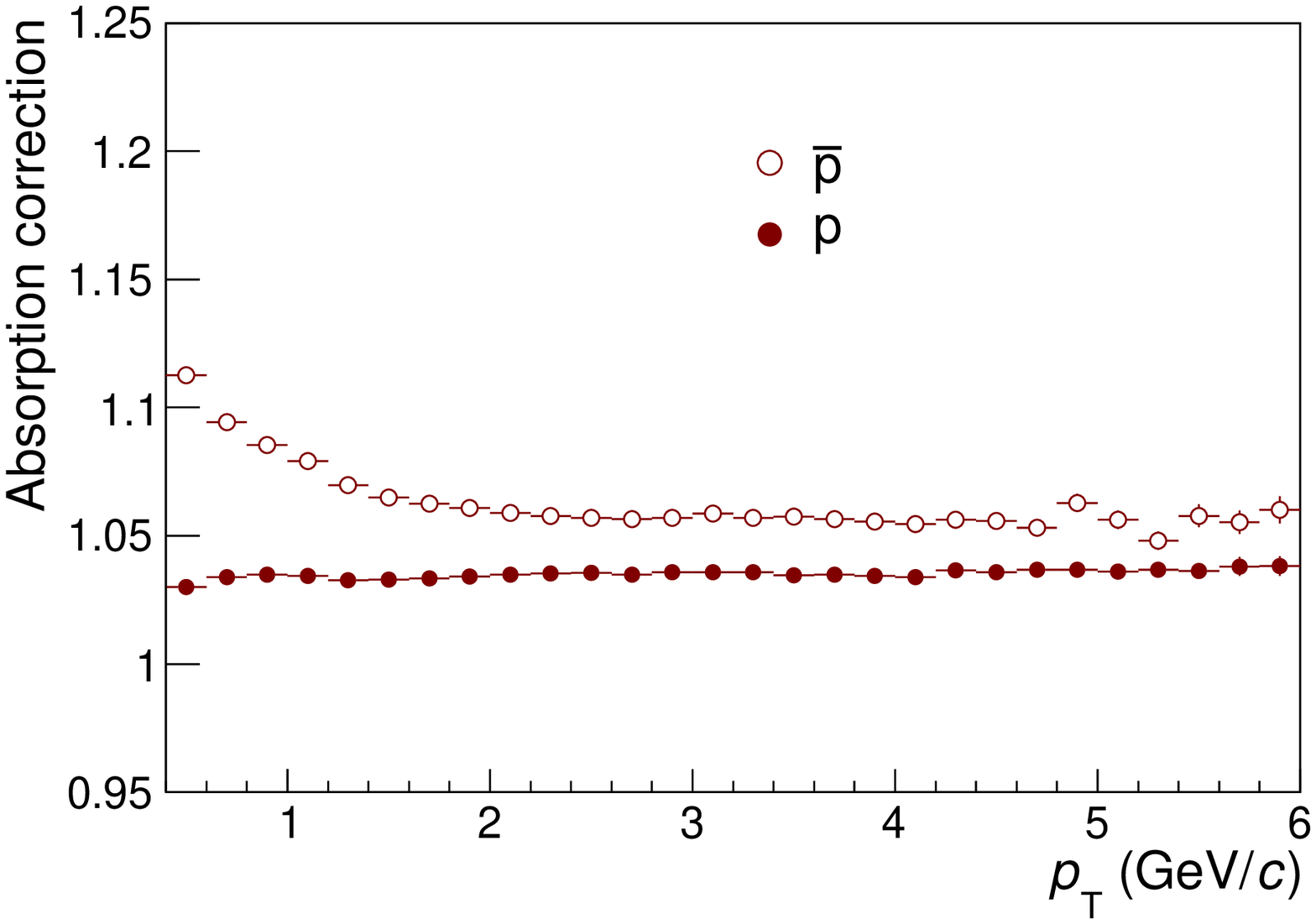}
  \includegraphics[width=0.49\linewidth]{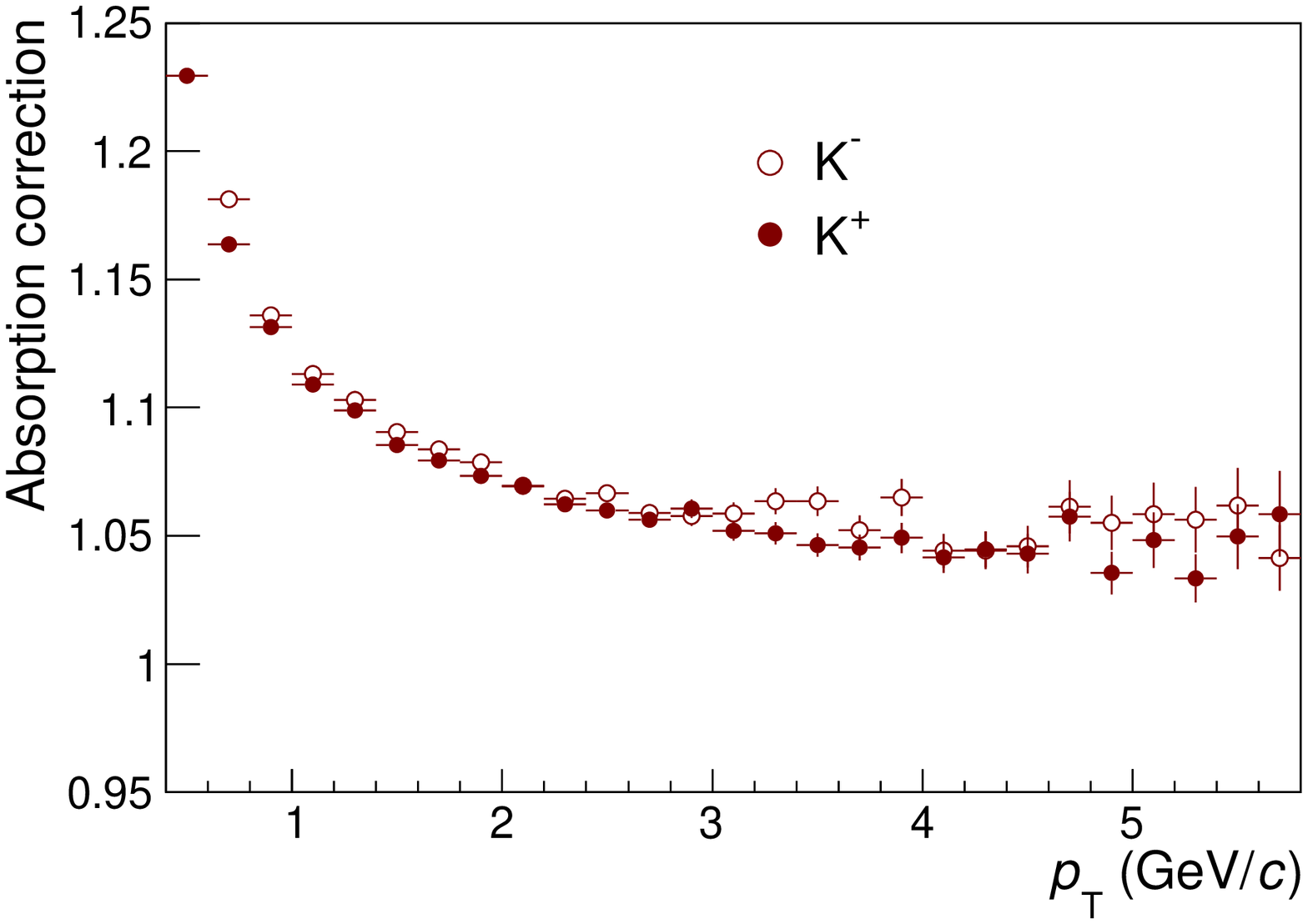}
  \caption{Absorption correction factors for protons, anti-protons and charged kaons\new{.}}
  \label{fig:AbsorptionDaughter}
\end{center}
\end{figure}

The \pt~dependence of the correction due to absorption for p, $\overline{\rm p}$, and charged kaons is presented in Fig.~\ref{fig:AbsorptionDaughter}. The aforementioned scaling for the corrections of $\overline{\rm p}$ and 
$\rm K^{-}$, is already applied. The resulting correction factors vary from $\approx 12\%$
 at low-\pt~to $\approx 6\%$ at high-\pt~for the case of $\overline{\rm p}$, while for p it is $\approx 3\%$, independently of \pt. The corresponding values for $\rm K^{\pm}$ also vary from $\approx 17\%$ to $\approx 5\%$, depending on \pt. The difference in the absorption of the positive and negative $\pi$ was found to be negligible.

Similar corrections were also applied to the hyperons. The correction factors are on the order of $\approx 1\%$. Due to the lack of experimental values of the corresponding inelastic cross sections we rely on the input parametrisation of GEANT3. 

Finally, we also considered the absorption of the daughter candidates for the hyperon decays, and in particular the (anti-)proton daughter, while for the case of the \ObarO~ratio the absorption of the kaon bachelor particle was also considered. This was done using the aforementioned correction factors.

\subsection{Correction for cut efficiency}

In addition to the previous effect on baryons and anti-baryons induced by absorption, it was reported in 
\cite{Ref:Alicepbarp}, that a relevant correction is needed to account for the differences in the cut efficiencies 
between p and $\overline{\rm p}$. The reason for the observed charge asymmetry is that particles undergoing 
elastic scattering in the inner detectors can still be reconstructed in the TPC but the corresponding ITS hits will in 
general not be associated to the track if the scattering angle is large. This in turns results from the corresponding 
differences in the elastic cross sections for p and $\overline{\rm p}$.

For the elastic cross section a limited set of experimentally measured values is available. It was found that 
GEANT3 cross sections are about $25\%$ above FLUKA, the latter being again closer to the measurements. 
Hence, we used the FLUKA results to account for the difference of $\rm{p}$ and $\overline{\rm{p}}$ cross sections. 
The resulting correction was estimated to be $\approx 3.5\%$ \cite{Ref:Alicepbarp}.

\subsection{Correction for secondary and background particles}

In order to distinguish between primary, secondary (i.e. products of the weak decay of particles) and background 
(i.e. particles emitted from the interaction of other particles with the material of the detectors) particles, we employ 
a data driven method based on distributions where these three categories of particles exhibit distinct differences.

Primary protons can be distinguished from secondary and background particles using the DCA distribution. The same distribution can be used for the case of 
$\overline{\rm p}$ for which the contribution from background particles (i.e. $\overline{\rm p}$ originating from the 
material) is negligible. Primary particles point to the primary vertex in contrast to the majority of the background, 
which can be removed by applying a DCA cut (described in the previous section). Secondary (anti-)protons point to the primary vertex with a DCA distribution that is wider than that of primaries. To account 
for the residual contamination from both sources, we determine the shape of the DCA distributions from Monte 
Carlo simulations, adjusting the amount to the data at large DCA values. The correction is calculated and 
applied differentially as a function of $y$ and \pt, and varies between $9\%$ for the lowest and less than 
$0.5\%$ for the highest \pt~bins for the background. For the feed-down corrections, the relevant 
values are $20\%$ and $17\%$ for the lowest and highest \pt~bins, respectively.

A similar procedure was applied for the case of $\rm \Lambda$ and $\overline{\rm \Lambda}$, using the information 
of the cosine of the pointing angle. These secondaries are mainly produced by primary $\rm K^{0}_{L}$ and 
charged kaons. The procedure resulted in a correction that varies between $8\%$ for the lowest and 
less than $0.5\%$ for the highest \pt~bins for the background.

The contamination of the $\rm \Xi^{\pm}$ sample from background particles was found to be negligible ($<$~0.5\%), based on Monte Carlo studies.

For the feed-down correction of $\rm \Lambda$ from $\rm \Xi$ decays, we rely on Monte Carlo simulations. The ratio $r_{\rm feed-down}$ of the reconstructed $\rm \Xi$ candidates to the number of reconstructed $\rm \Lambda$ candidates from $\rm \Xi$ decays is

\begin{equation}
r_{\rm feed-down} = \frac{{(N_{{\rm \Xi}^{-}})}_{\rm MC}}{{(N_{{\rm \Xi} \rightarrow  \rm \Lambda  })}_{\rm MC}}.
\end{equation}
Assuming that this ratio is the same in both Monte Carlo and data, the whole feed-down contribution to the spectra is estimated by dividing the number of reconstructed $\rm \Xi$ in data by the ratio extracted from Monte Carlo.

\begin{equation}
{(N_{{\rm \Xi}  \rightarrow \rm \Lambda  })}_{\rm data} = \frac{{(N_{{\rm \Xi}^{-}})}_{\rm data}}{r_{\rm feed-down}}.
\end{equation}

The overall fractions of $\rm \Lambda$ and $\overline{\rm \Lambda}$ coming from the $\rm \Xi$ decays for different $\sqrt{s}$ are summarized in Table~\ref{tab:FeedDown}. The uncertainty of the \LbarL~ratio resulting from the feed-down correction is based on our measurement of \XbarX~ratio and is described in Section~\ref{Sec:Systematics}.

\begin{table}[htbp]
\caption{Feed-down fraction of $\rm \Lambda$ from $\rm \Xi$ decays}
\centering
\begin{tabular}{l c c c}
\toprule
 &0.9~TeV &2.76~TeV &7~TeV\\ 
\midrule
$\rm \Lambda$ & 0.22 & 0.24 & 0.23\\
$\overline{\rm \Lambda}$ & 0.21 & 0.24  & 0.23\\
\bottomrule
\end{tabular}
\label{tab:FeedDown}
\end{table}

The contribution from $\rm \Omega$ decays was found to be negligible. It should be noted that since $\rm\Lambda$ ($\overline{\rm \Lambda}$) from electromagnetic $\rm \Sigma^{0}$ ($\overline{\rm \Sigma^{0}}$) decays cannot be distinguished from the primary ones, the identified $\rm\Lambda$ ($\overline{\rm \Lambda}$) also include these contributions.

The feed-down contamination of the $\rm \Xi$ sample from decays $\rm \Omega^{\pm} \rightarrow \rm \Xi^{\pm} + \pi^{0}$ considering the branching ratio for this decay and the $\rm \Omega/\Xi$ ratio reported in \cite{Ref:ALICEstrangeness} is $<$~1\% and thus negligible.

%% file: Systematics.tex
\section{Systematic uncertainties}
\label{Sec:Systematics}

Although the dominant sources of systematic uncertainties in this analysis are due to the corrections employed, uncertainties in the analysis procedure also contribute. Uncertainties arising from the correction procedures for elastic and inelastic cross-section parametrisation and for secondary particles produced in the beam pipe and detector material have been found to be very small. We have identified and estimated systematic uncertainties from the following sources:

\begin{itemize}
\item the amount of material of the central barrel;
\item the experimental values of the elastic and inelastic cross section implemented in the transport code (FLUKA);
\item the applied corrections for background and secondary particles;
\item the track and topological selections;
\item the hyperon signal extraction procedure. 
\end{itemize}

\noindent These are discussed in more detail in the following and the final uncertainty estimates are present in Tables \ref{tab:systematicsPr} and \ref{tab:systematicsHy} for protons and hyperons, respectively.

\subsection{Systematic uncertainties due to material budget, inelastic and elastic cross sections}
The amount of material in the central part of the detector is known, based on studies with $\gamma$ conversions, with a precision of 7\% \cite{Ref:Alicepbarp}. Dedicated simulations varying the material budget by this amount were used to determine the uncertainty from this source. The absorption corrections were recomputed using the output of these simulations and an uncertainty of 0.5 \% was found in the final ratios, calculated as half of the difference between the highest and the lowest values of each ratio. 

In addition, the experimental $\overline{\rm p}$--A inelastic cross sections are measured with an accuracy typically better than 5\% \cite{Ref:XSection}. We assign an uncertainty of 10\% to the absorption cross section calculated with FLUKA, resulting into an uncertainty of 0.8\% on the final measured ratio.

The inelastic hyperon--A cross sections have not been measured experimentally, so absorption corrections for pre-decay hyperons must rely on the cross section parametrisation implemented in GEANT3. Assuming that these have an uncertainty of 100\%, we find an error of 0.5\% on the \LbarL~ratio and 1\% on the \XbarX~and \ObarO.

By comparing GEANT3 and FLUKA with the experimentally measured elastic cross section, the corresponding uncertainty on \pbarp~ratio was estimated to be 0.8\%, which corresponds to the difference between the correction factors calculated with the two models.

\subsection{Systematic uncertainties due to corrections for secondary and background particles}
The uncertainty resulting from the subtraction of secondary protons and from the feed-down corrections was estimated to be 0.6\% by using various functional forms for the background subtraction and for the contributions of the weak decay products.
The uncertainty resulting from the subtraction of secondary $\rm \Lambda$ was estimated to be 0.4\% by using various methods for the background subtraction. The feed-down fractions of $\rm \Lambda$ and $\overline{\rm \Lambda}$ were estimated to be $\approx0.2$ (see Table \ref{tab:FeedDown}). The total uncertainties of the measured \XbarX~ratios were propagated into \LbarL~systematic uncertainty using this fraction, resulting in an uncertainty of 1\% at $\sqrt{s}=0.9$~TeV and 0.4\% for higher energies.

\subsection{Systematic uncertainties due to track and topological selections}
The systematic effects of the track quality criteria and the topological selections used in the hyperon reconstruction, the ``tightness'' of the PID cut, and ranges of additional cuts have been investigated. The selections were varied one-by-one using reasonably looser and tighter values for each parameter. The final systematic uncertainty was calculated as half of the difference between the highest and the lowest values. The final estimated systematic error presented in Tables \ref{tab:systematicsPr} and \ref{tab:systematicsHy} is the quadratic sum of the contributions from the variation of

\begin{itemize}
\item the width of N-$\sigma$ area used for the particle identification ($\pm 1 \sigma$); 
\item the minimum number of TPC clusters ($\pm 10$ clusters);
\item the topological selections used in the reconstruction of the V0 and cascade vertexes;
\item the width of the mass window around $\rm{{K}_{s}^{0}}$ or $\rm \Xi$ nominal mass in case of $\rm \Lambda$ and $\rm \Omega$ ($\pm 2$~MeV/$c$).
\end{itemize}

\subsection{Systematic uncertainties due to signal extraction} 

Two methods for signal extraction have been presented in Section~\ref{Sec:Analysis}. The final ratios differ by $\approx 0.4\%$ depending on the method used. This difference is due to the approximation of the background using different functions and is included here as a systematic uncertainty. The difference of the ratios due to the change of the fit range and width of the considered signal area by $\pm 1 \sigma$ was found to be negligible.

\begin{table*}
   \caption{Systematic uncertainty for the \pbarp~measurement quoted for each source separately}
\centering
\begin{tabular}{l l c c c}
\toprule
\multicolumn{2}{l}{Source} & \multicolumn{3}{c}{\pbarp}\\
\midrule
\multicolumn{2}{l}{Material budget} & \multicolumn{3}{c}{0.5\%}\\
\multicolumn{2}{l}{Inelastic cross section} & \multicolumn{3}{c}{0.8\%}\\
\multicolumn{2}{l}{Elastic cross section} & \multicolumn{3}{c}{0.8\%}\\
\multicolumn{2}{l}{Selections} & \multicolumn{3}{c}{0.4\%}\\
\multirow{1}{*}{Corrections} & Secondaries/Feed-down & \multicolumn{2}{c}{0.6\%}\\
\addlinespace
\midrule
\multicolumn{2}{l}{TOTAL} & \multicolumn{3}{c}{1.4\%}\\
\bottomrule
\end{tabular}
\label{tab:systematicsPr}
\end{table*}

\begin{table*}
   \caption{Systematic uncertainty for the \LbarL, \XbarX~and \ObarO~measurement quoted for each source separately. The uncertainties are shown for 0.9~TeV - 2.76~TeV - 7~TeV.}
\centering
\begin{tabular}{l l c c c}
\toprule
\multicolumn{2}{l}{Source} & \LbarL & \XbarX & \ObarO\\
\midrule
\multicolumn{2}{l}{Material budget} & 0.5\% & 0.5\% & 0.5\% \\
\multirow{2}{*}{Inelastic cross section} & $\overline{\rm p}$ & 0.8\% & 0.8\% & 0.8\%\\
& Hyperon & 0.5\% & 1.0\% & 1.0\%\\
\multicolumn{2}{l}{Selections} & 0.7\% - 0.1\% - 0.2\% & 4.2\% - 0.9\% - 0.7\% & 3.8\% - 1.7\%\\
\multicolumn{2}{l}{Signal extraction} & 0.3\% - 0.5\% - 0.3\% & 0.9\% - 0.4\% - 0.2\% & 1.7\% - 0.4\% \\
\multirow{2}{*}{Corrections} & Secondaries & 0.4\% & - & -\\
 & Feed-down & 1.0\% - 0.4\% - 0.4\%& - & -\\
\addlinespace
\midrule
\multicolumn{2}{l}{TOTAL} & 1.7\% - 1.3\% - 1.3\% & 4.5\% - 1.7\% - 1.6\% & 4.4\% - 2.2\% \\
\bottomrule
\end{tabular}
\label{tab:systematicsHy}
\end{table*}

%% file: Results.tex
\section{Results}
\label{Sec:Results}
\subsection{Rapidity and transverse momentum dependence}

\begin{figure*}[htb]
\includegraphics[width=0.5\linewidth]{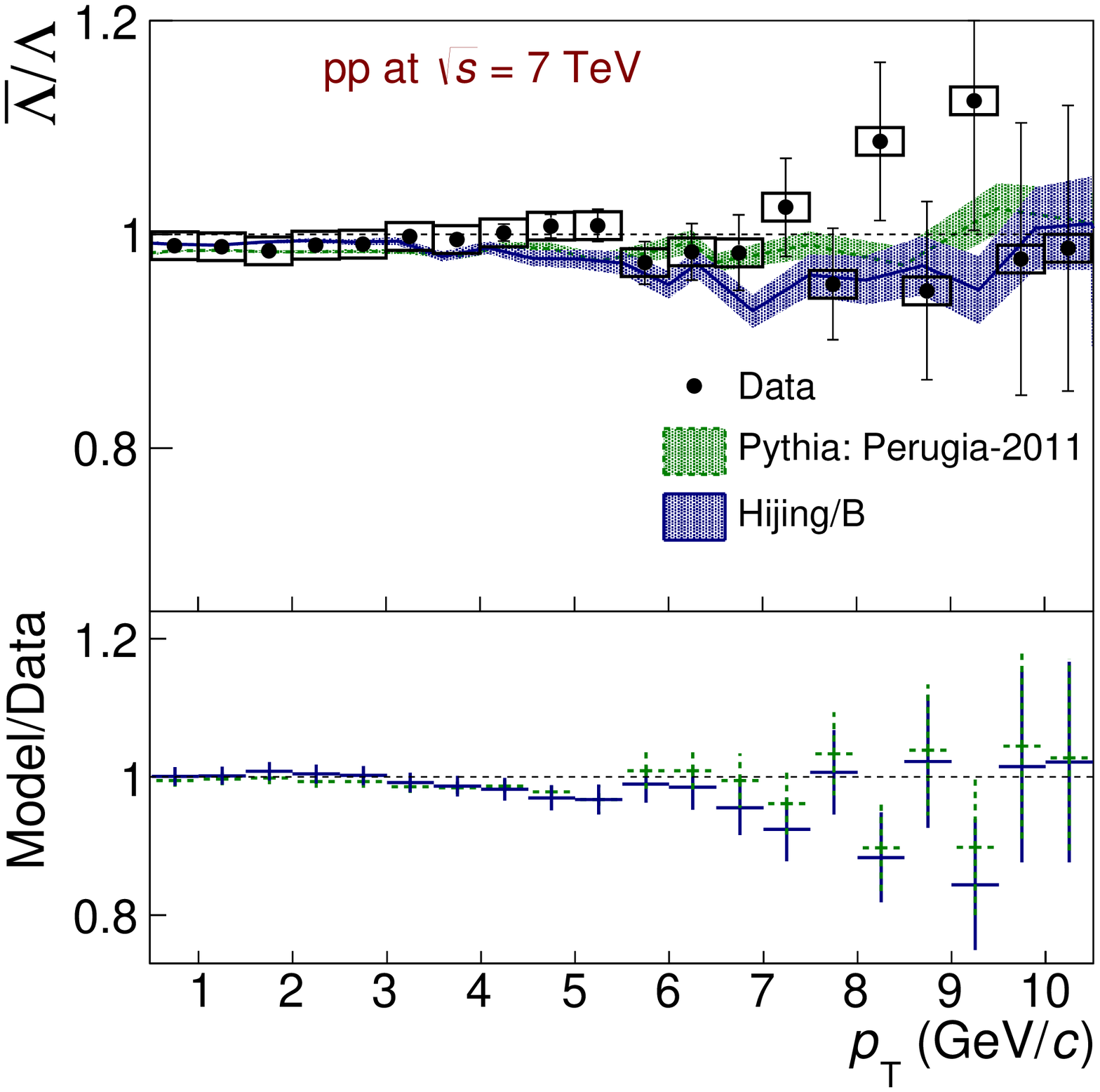}
\includegraphics[width=0.5\linewidth]{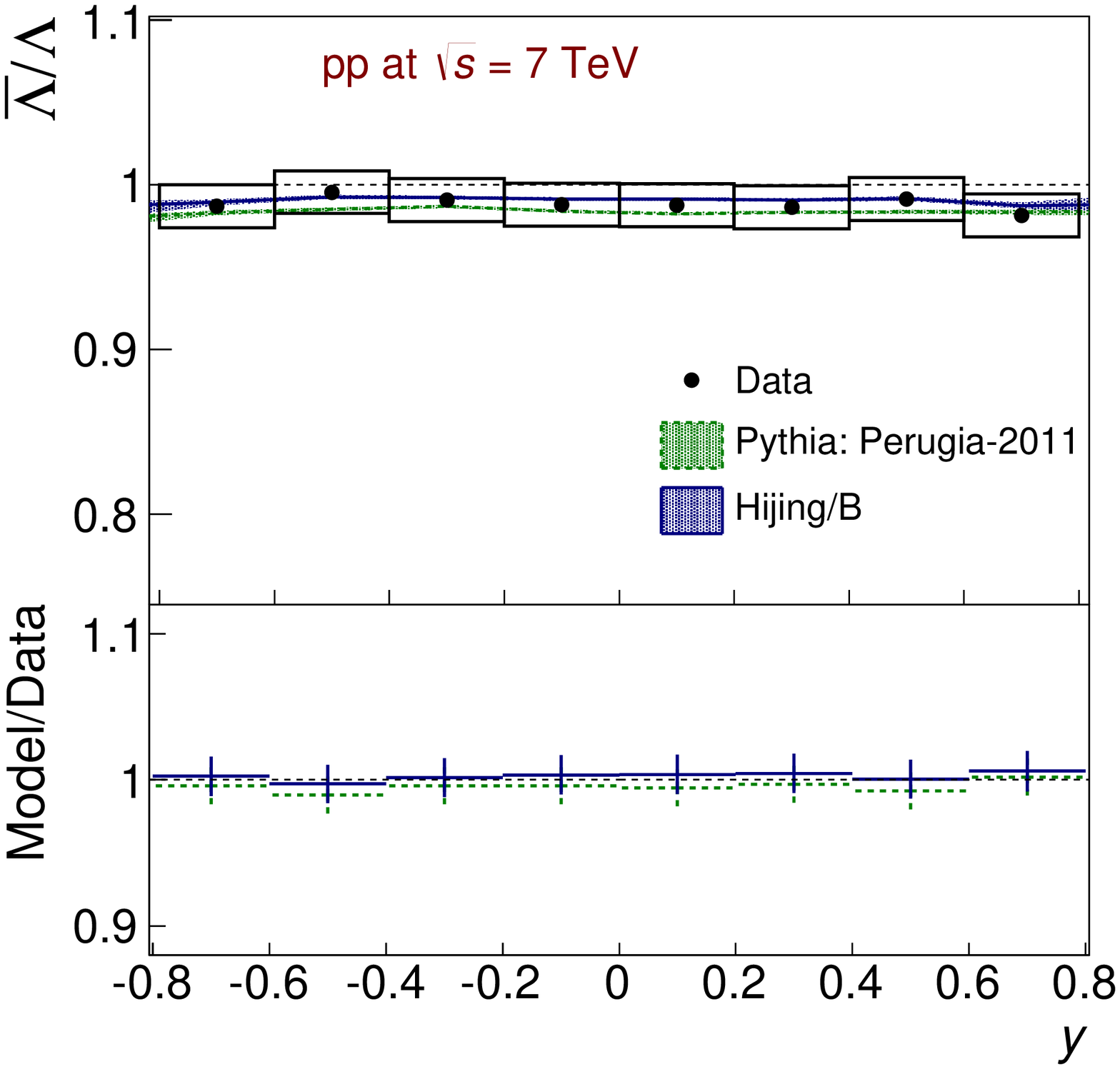}
\caption{(Colour online) The \LbarL~ratio at $\sqrt{s} = 7$~TeV as a function of \pt~(left) and rapidity (right). The data points are compared with different Monte Carlo generators. The vertical bars (boxes) represent the statistical (systematic) uncertainty, while the horizontal bars represent the width of the rapidity or \pt~bin. Ratio of model to data is shown below using uncertainties added in quadrature.}
\label{fig:lambdaRatio7}
\end{figure*}

Anti-baryon to baryon spectra ratios were measured as a function of rapidity and transverse momentum. We report results for the rapidity intervals $|y|<$~0.8 in the case of hyperons and $|y|<$~0.5 for $\overline{\rm p}/{\rm p}$. The available data were not statistically sufficient to determine the \ObarO~ratio at 0.9~TeV. For the same reason, the ratios were integrated over rapidity for \ObarO~at all remaining energies, for \XbarX~at 0.9 and 2.76~TeV and for \LbarL~at 7~TeV for \pt$>$~5.5 GeV/$c$ (i.e. rapidity dependence on Fig.~\ref{fig:lambdaRatio7}, bottom, is for \pt$<$~5.5 GeV/$c$).

\begin{figure*}[h]
\includegraphics[width=0.5\linewidth]{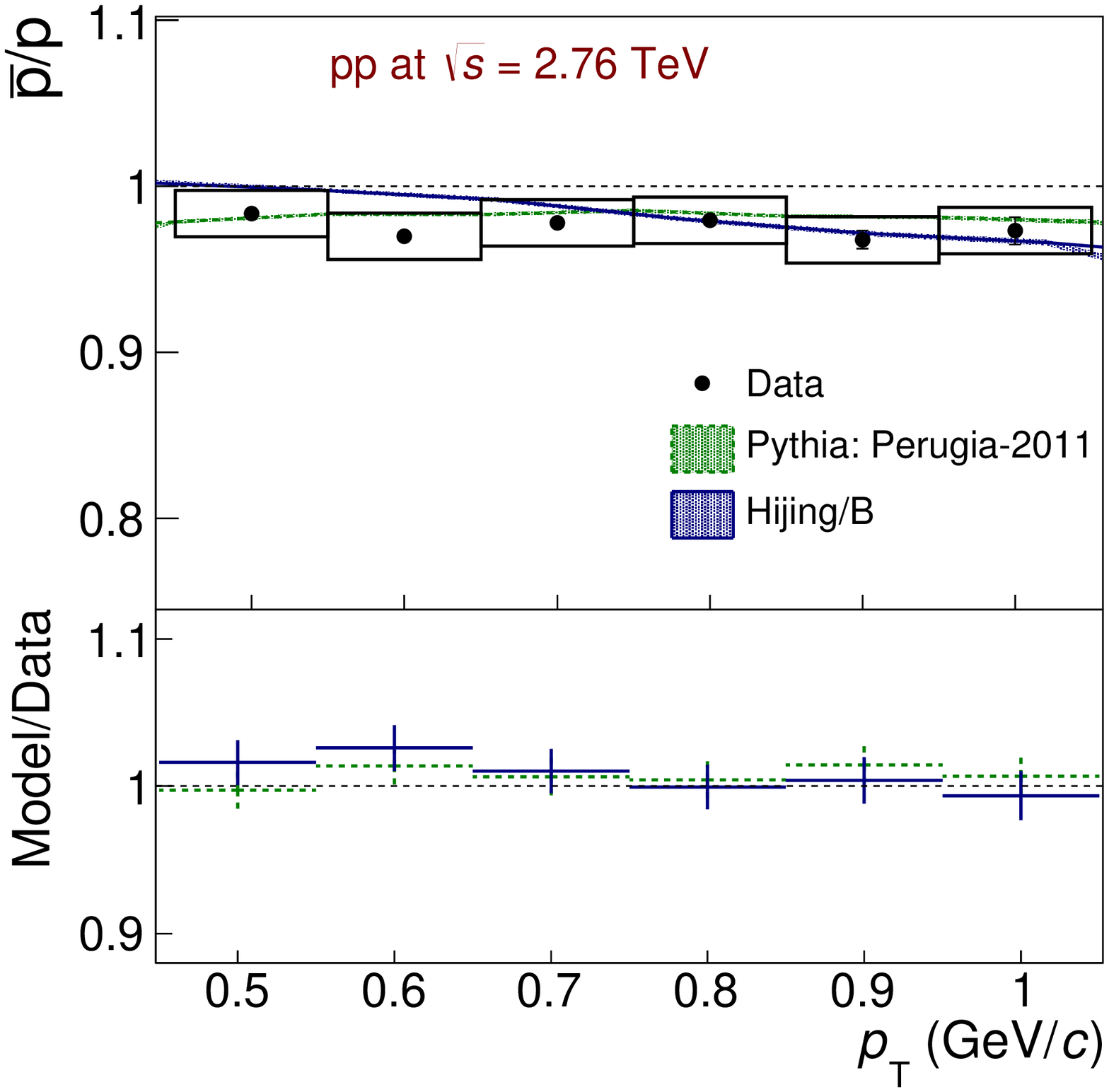}
\includegraphics[width=0.5\linewidth]{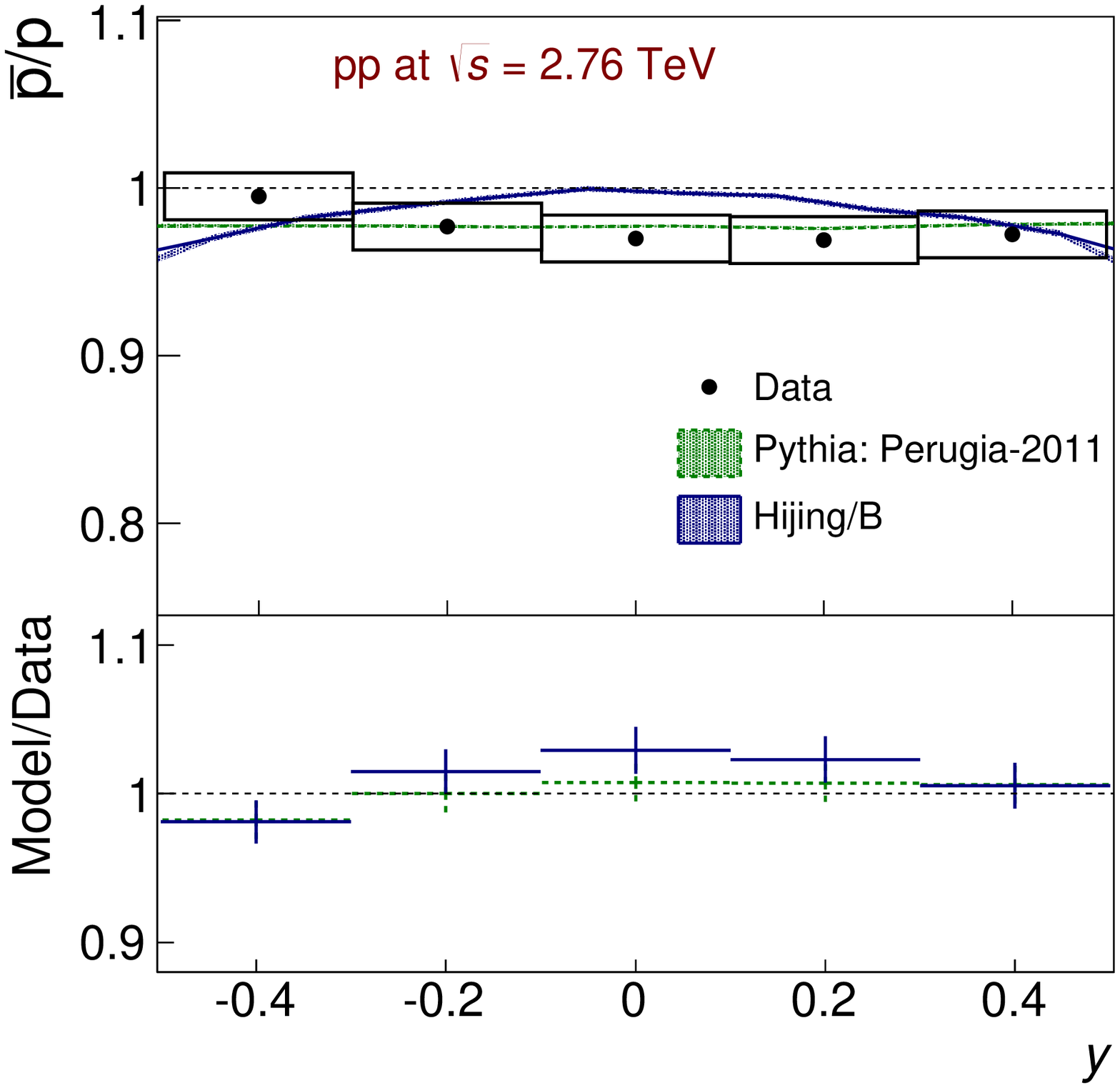}
\caption{(Colour online) The \pbarp~ratio at $\sqrt{s} = 2.76$~TeV as a function of \pt~(left) and rapidity (right). The data points are compared with different Monte Carlo generators. The vertical bars (boxes) represent the statistical (systematic) uncertainty, while the horizontal bars represent the width of the rapidity or \pt~bin. Ratio of model to data is shown below using uncertainties added in quadrature.}
\label{fig:protonRatio}
\end{figure*}

As can be seen in Fig.~\ref{fig:protonRatio}, there is no observed dependence on either rapidity or transverse momentum in the measured \pbarp~ratio at $\sqrt{s} = 2.76$~TeV which is consistent with previous ALICE measurements at $\sqrt{s} = 0.9$ and $7$~TeV \cite{Ref:Alicepbarp}. The data are described reasonably well by PYTHIA (Perugia2011, Tune 350) \cite{Ref:Pythia}. On the other hand, HIJING/B \cite{Ref:HijingB} is showing a decreasing ratio with increasing $p_{\rm T}$ (Fig.~\ref{fig:protonRatio}, left) and a slightly larger rapidity dependence than supported by the data (Fig.~\ref{fig:protonRatio}, right). Even though HIJING/B is showing different trends with \pt~and rapidity, compared to the data, the current uncertainties do not allow for any final conclusion yet.

\begin{figure*}[h]
\includegraphics[width=0.5\linewidth]{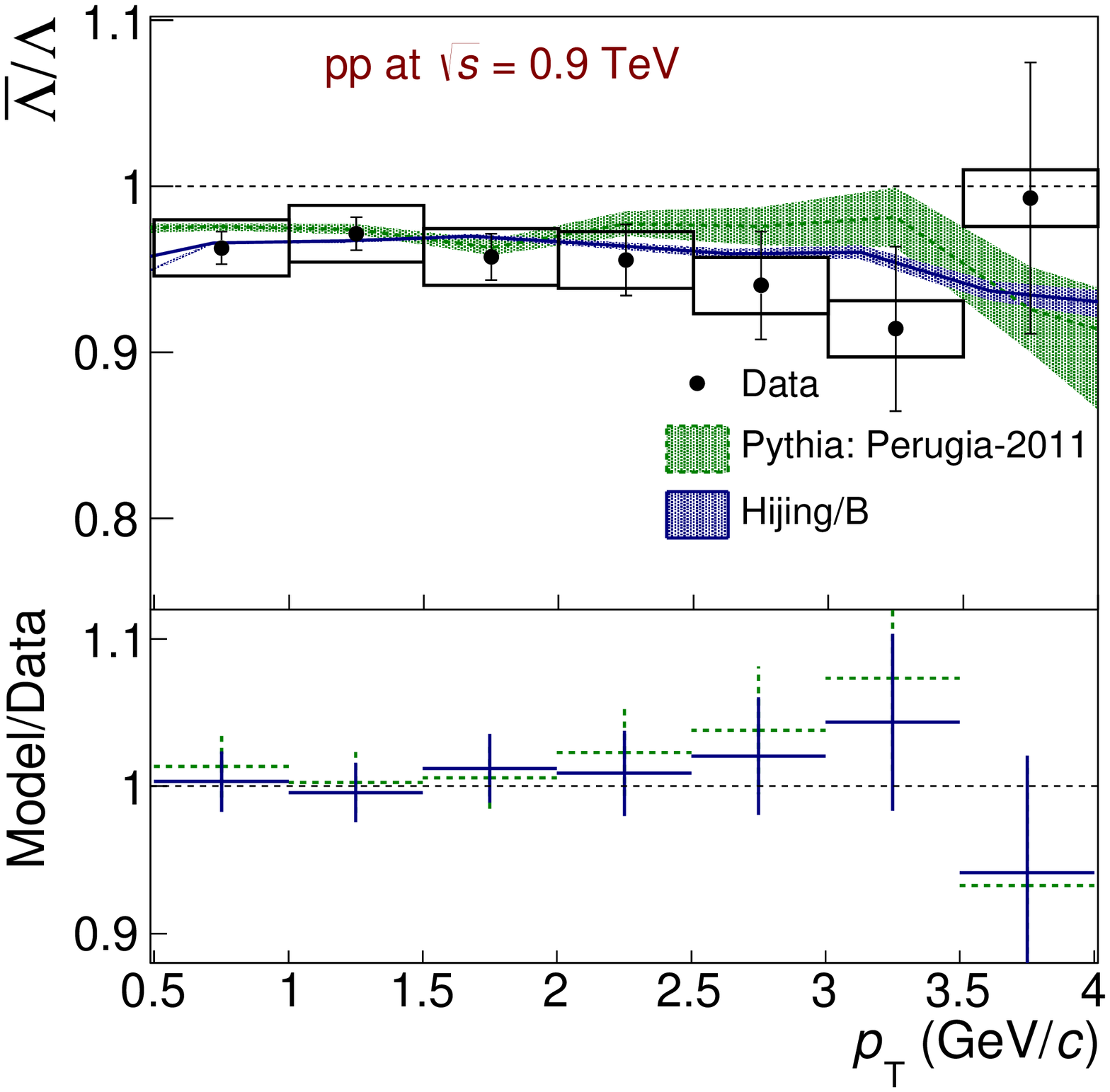}
\includegraphics[width=0.5\linewidth]{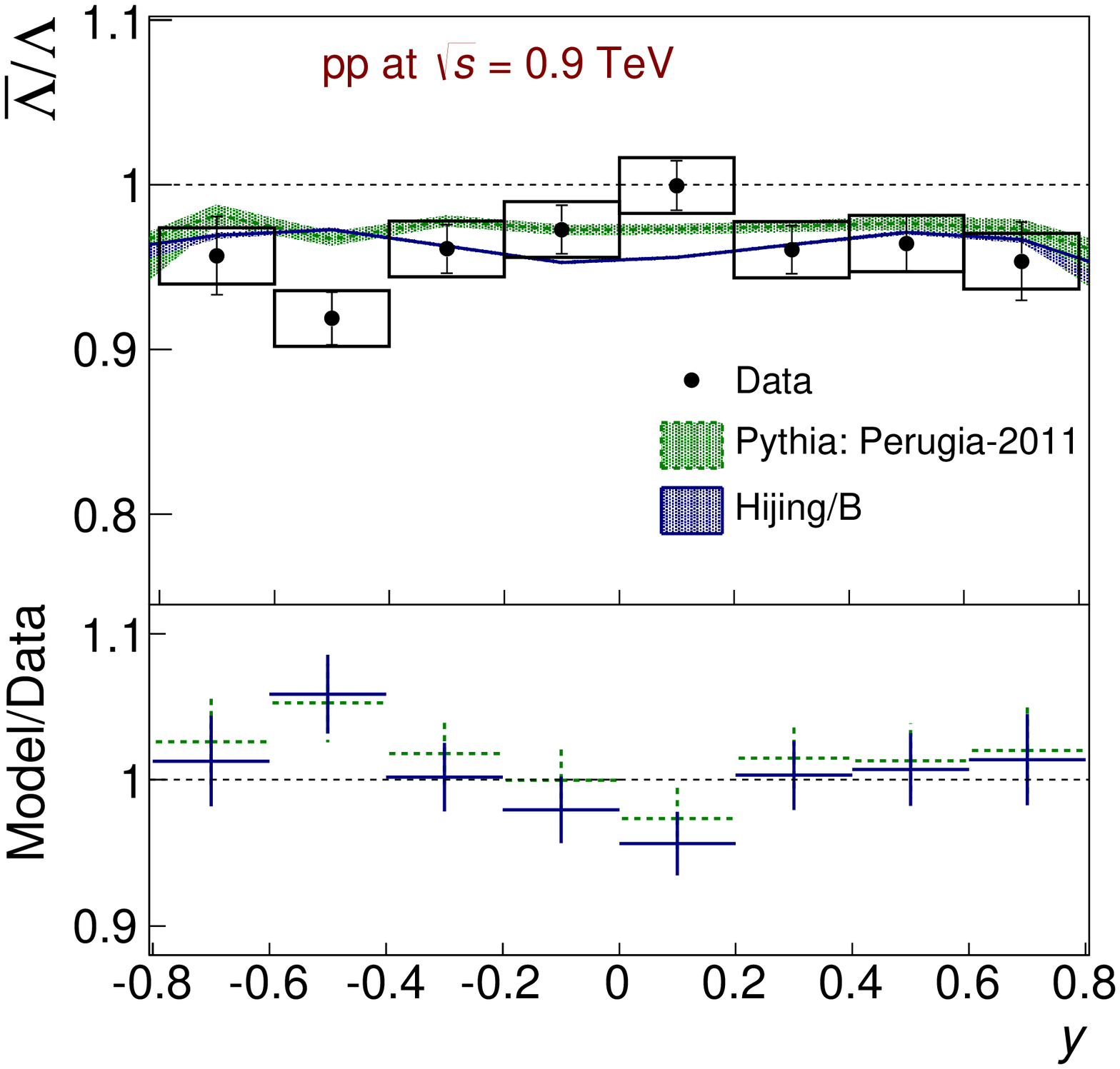}
\caption{(Colour online) The \LbarL~ratio at $\sqrt{s} = 0.9$~TeV as a function of \pt~(left) and rapidity (right). The data points are compared with different Monte Carlo generators. The vertical bars (boxes) represent the statistical (systematic) uncertainty, while the horizontal bars represent the width of the rapidity or \pt~bin. Ratio of model to data is shown below using uncertainties added in quadrature.}
\label{fig:lambdaRatio900}
\end{figure*}

\begin{figure*}[h]
\includegraphics[width=0.5\linewidth]{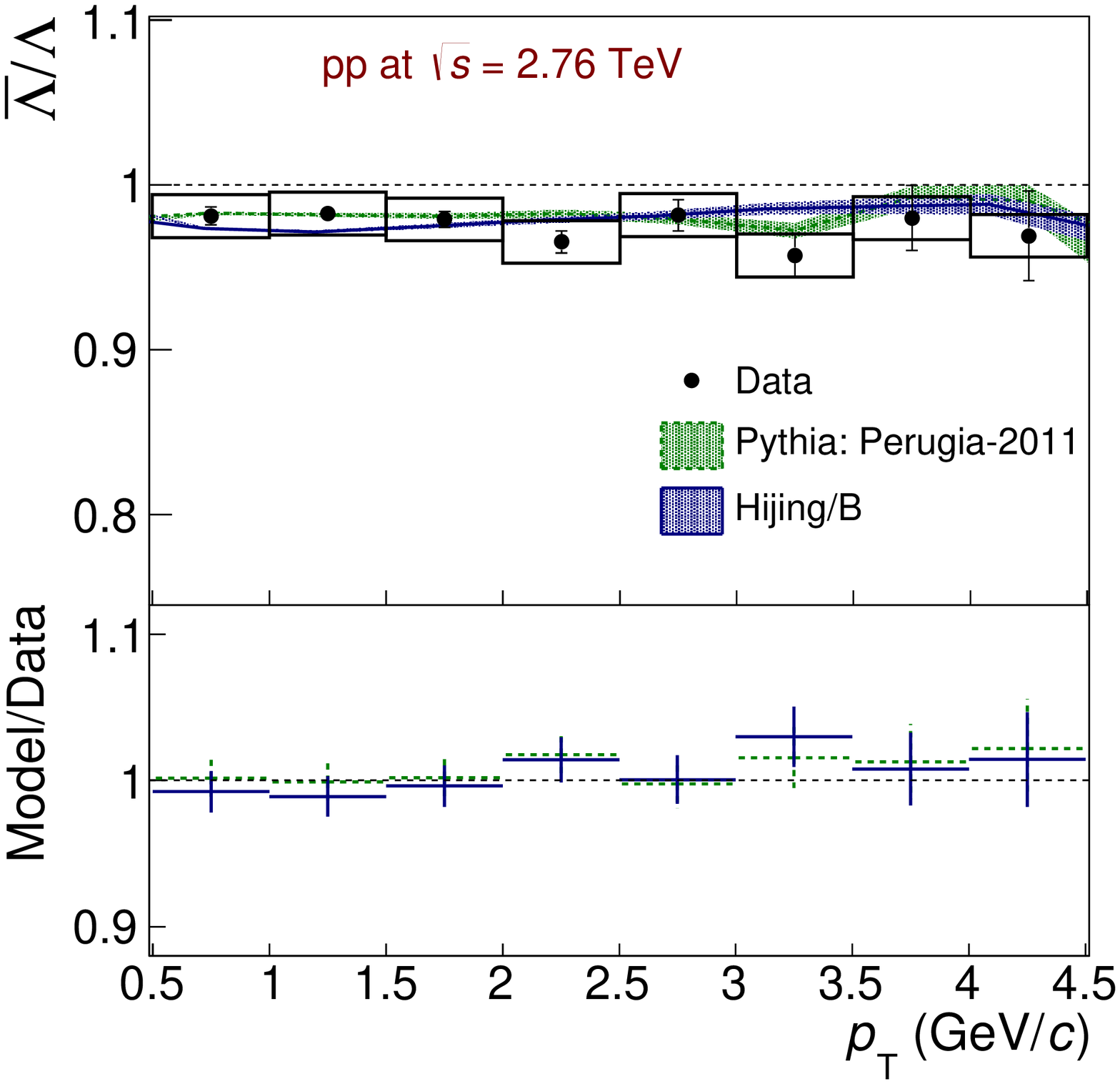}
\includegraphics[width=0.5\linewidth]{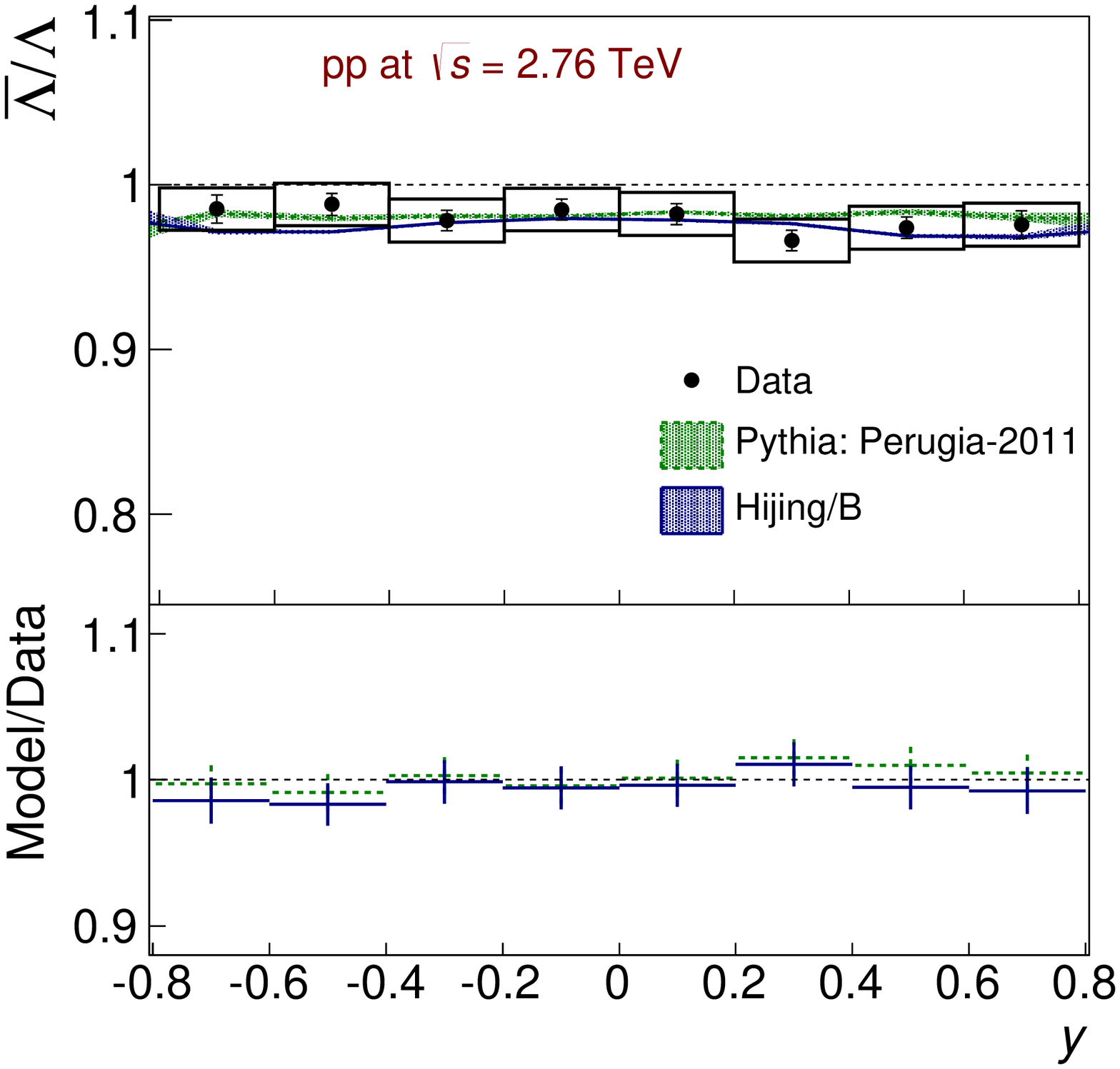}
\caption{(Colour online) The \LbarL~ratio at $\sqrt{s} = 2.76$~TeV as a function of \pt~(left) and rapidity (right). The data points are compared with different Monte Carlo generators. The vertical bars (boxes) represent the statistical (systematic) uncertainty, while the horizontal bars represent the width of the rapidity or \pt~bin. Ratio of model to data is shown below using uncertainties added in quadrature.}
\label{fig:lambdaRatio276}
\end{figure*}

\begin{figure}[htb]
\centering
\includegraphics[width=0.5\linewidth]{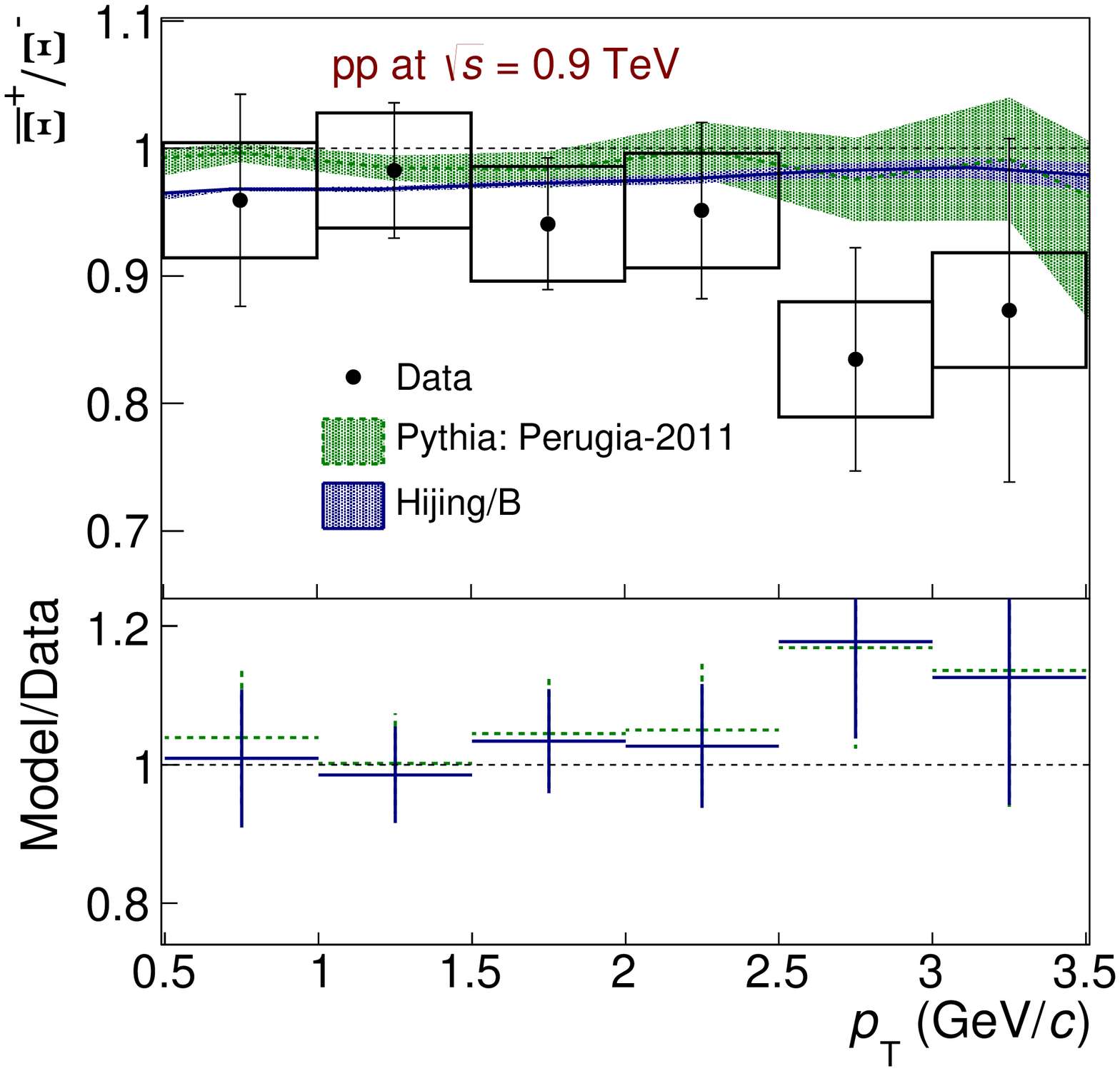}
\caption{(Colour online) The \XbarX~ratio at $\sqrt{s} = 0.9$~TeV integrated over $|y|<$ 0.8 as a function of \pt. The data points are compared with different Monte Carlo generators. The vertical bars (boxes) represent the statistical (systematic) uncertainty, while the horizontal bars represent the width of the \pt~bin. Ratio of model to data is shown below using uncertainties added in quadrature.}
\label{fig:xiRatio900}
\end{figure}

\begin{figure}[htb]
\centering
\includegraphics[width=0.5\linewidth]{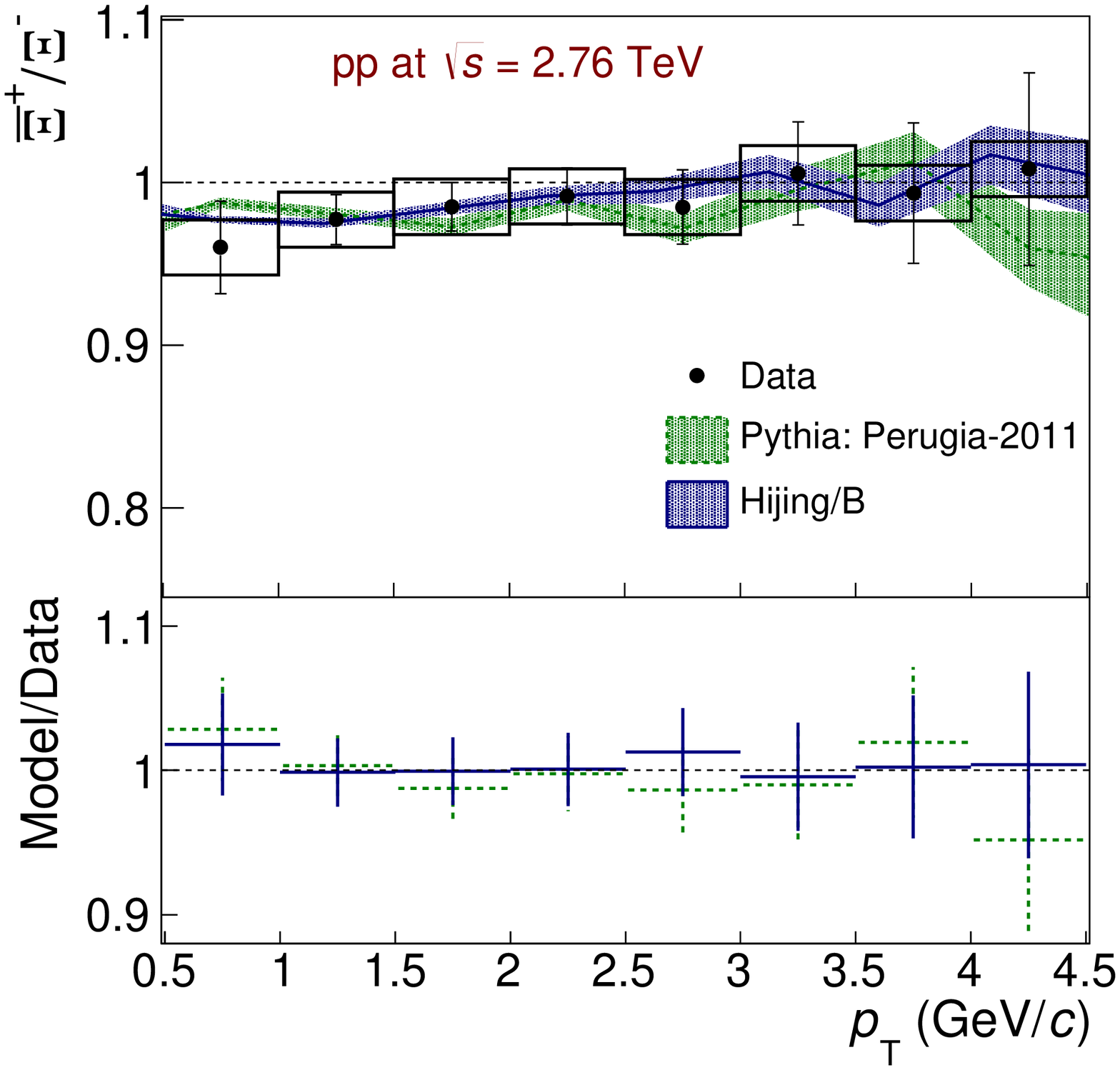}
\caption{(Colour online) The \XbarX~ratio at $\sqrt{s} = 2.76$~TeV integrated over $|y|<$ 0.8 as a function of \pt. The data points are compared with different Monte Carlo generators. The vertical bars (boxes) represent the statistical (systematic) uncertainty, while the horizontal bars represent the width of the \pt~bin. Ratio of model to data is shown below using uncertainties added in quadrature.}
\label{fig:xiRatio276}
\end{figure}

\begin{figure*}[htb]
\includegraphics[width=0.5\linewidth]{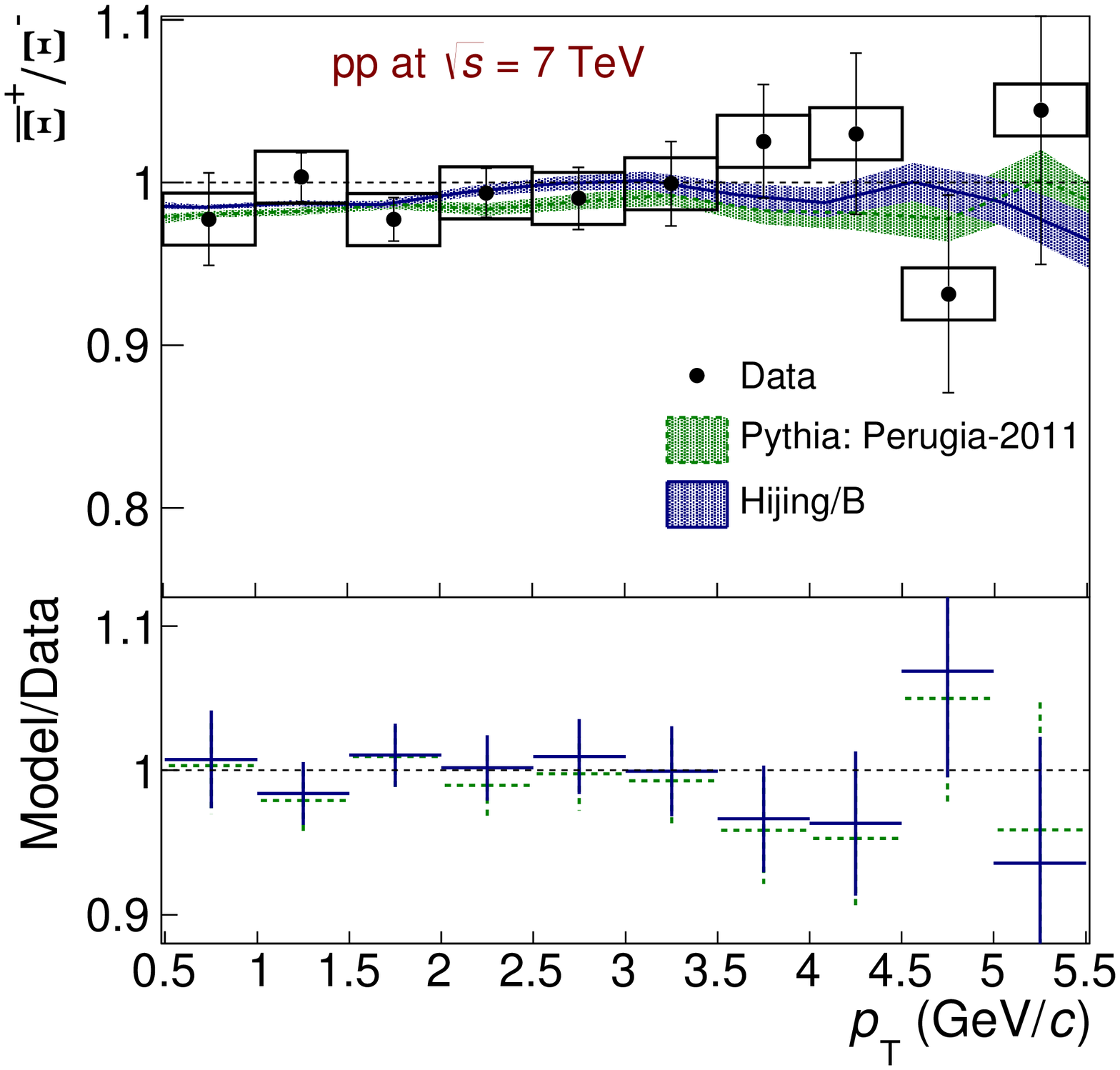}
\includegraphics[width=0.5\linewidth]{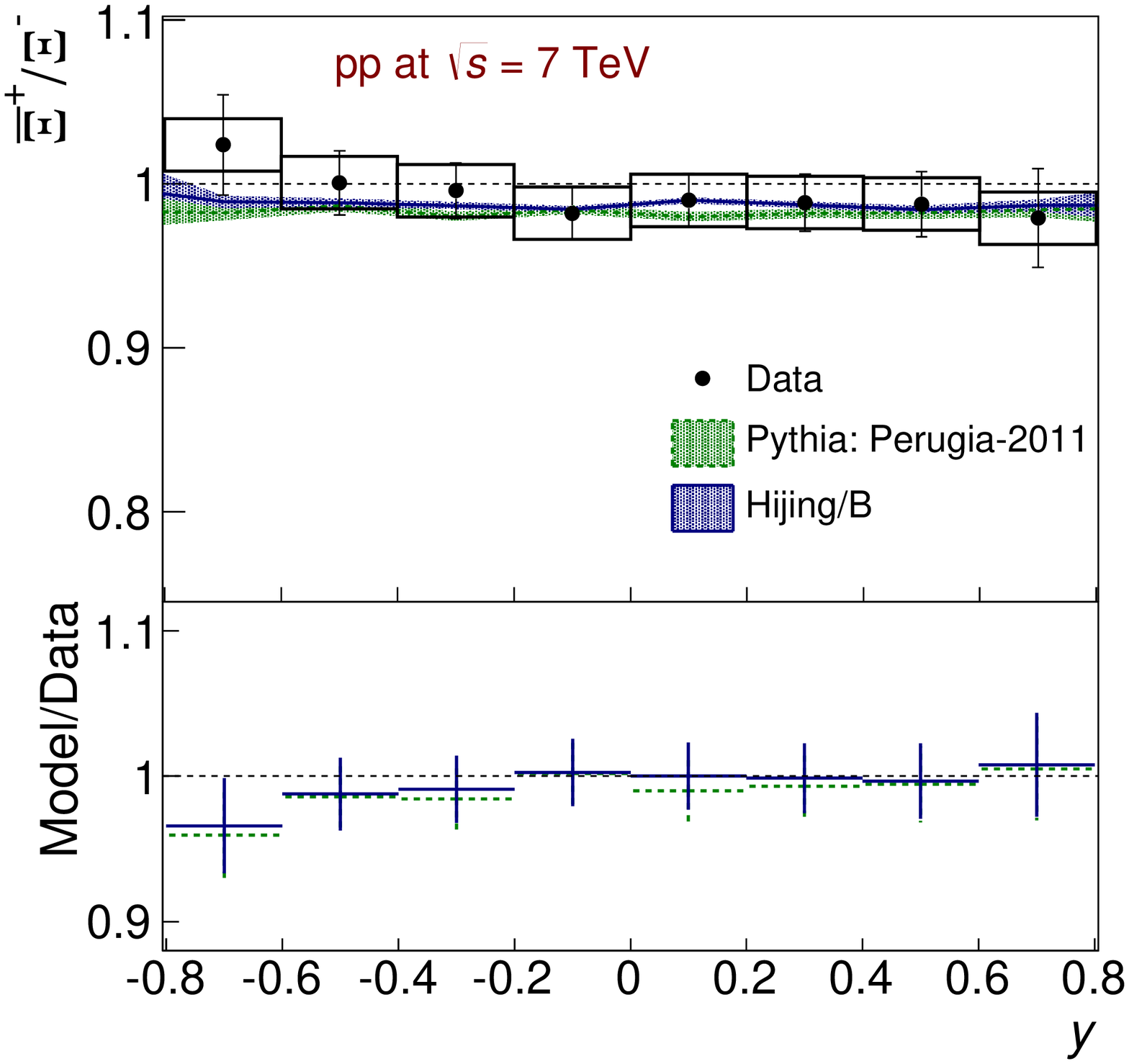}
\caption{(Colour online) The \XbarX~ratio at $\sqrt{s} = 7$~TeV as a function of \pt~(left) and rapidity (right). The data points are compared with different Monte Carlo generators. The vertical bars (boxes) represent the statistical (systematic) uncertainty, while the horizontal bars represent the width of the rapidity or \pt~bin. Ratio of model to data is shown below using uncertainties added in quadrature.}
\label{fig:xiRatio7}
\end{figure*}

\begin{figure}[htb]
\centering
\includegraphics[width=0.5\linewidth]{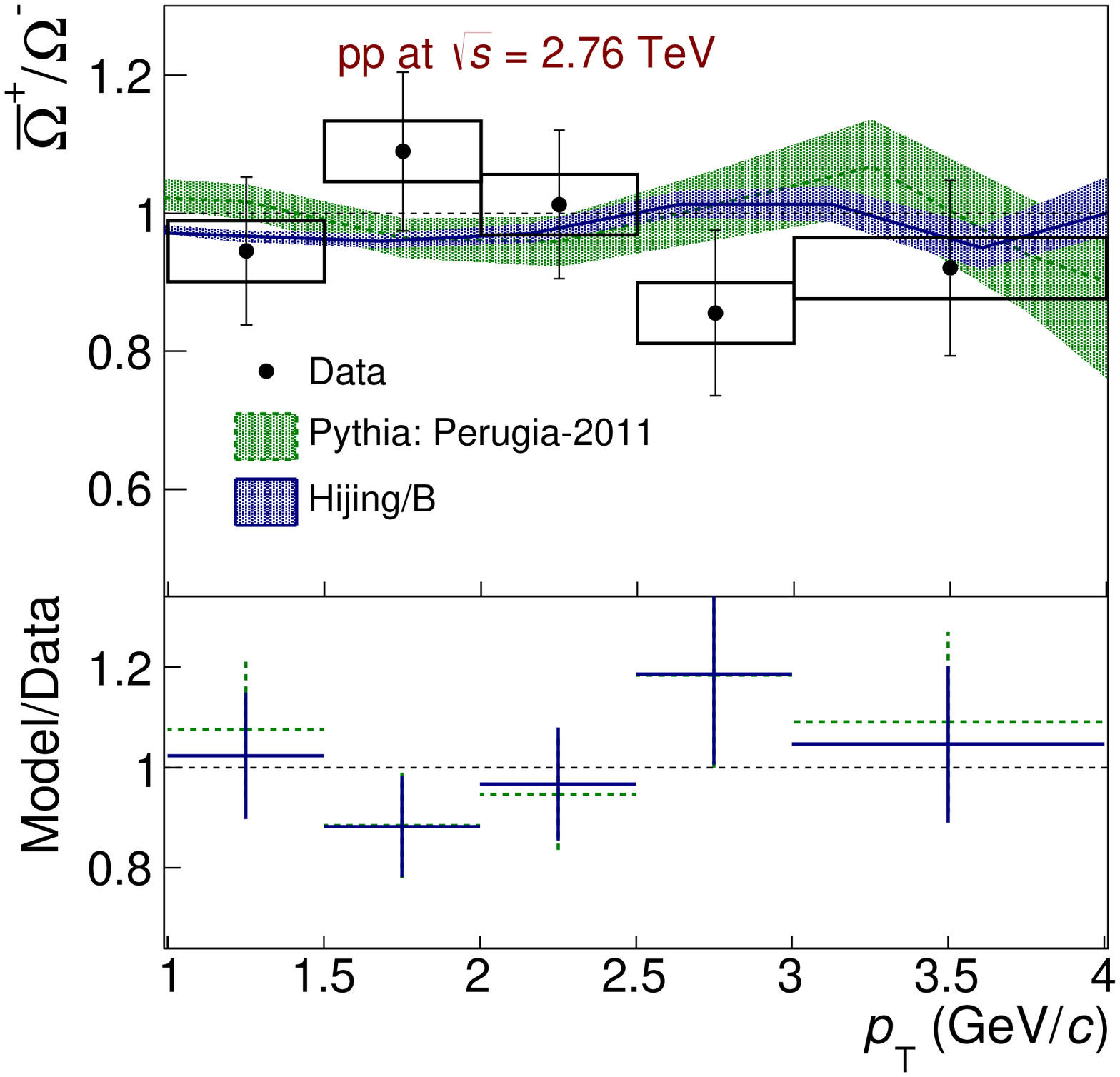}
\caption{(Colour online) The \ObarO~ratio at $\sqrt{s} = 2.76$~TeV integrated over $|y|<$ 0.8 as a function of \pt. The data points are compared with different Monte Carlo generators. The vertical bars (boxes) represent the statistical (systematic) uncertainty, while the horizontal bars represent the width of the \pt~bin. Ratio of model to data is shown below using uncertainties added in quadrature.}
\label{fig:omegaRatio276}
\end{figure}

\begin{figure}[htb]
\centering
\includegraphics[width=0.5\linewidth]{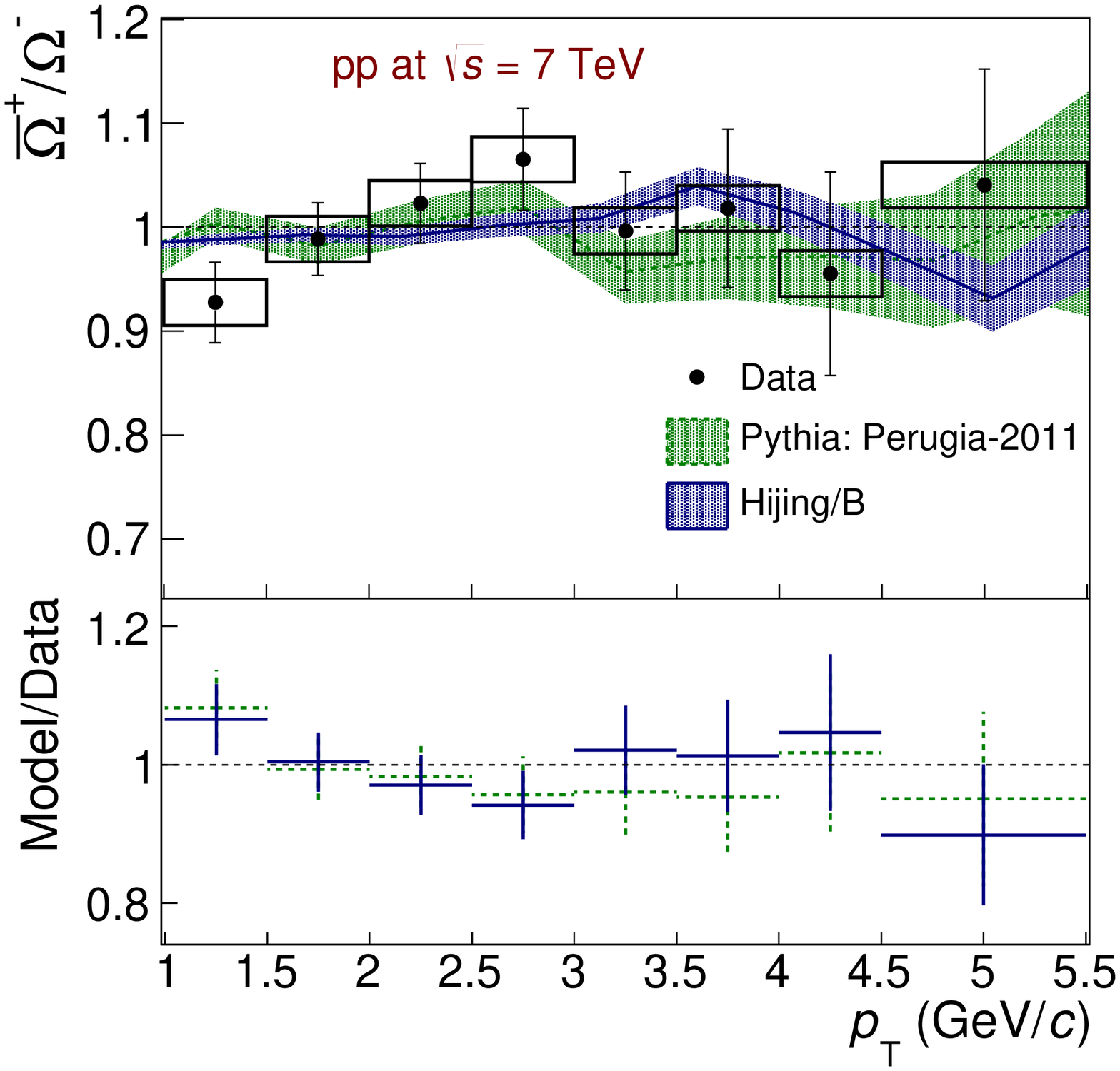}
\caption{(Colour online) The \ObarO~ratio at $\sqrt{s} = 7$~TeV integrated over $|y|<$ 0.8 as a function of \pt. The data points are compared with different Monte Carlo generators. The vertical bars (boxes) represent the statistical (systematic) uncertainty, while the horizontal bars represent the width of the \pt~bin. Ratio of model to data is shown below using uncertainties added in quadrature.}
\label{fig:omegaRatio7}
\end{figure}

Figures \ref{fig:lambdaRatio900}, \ref{fig:lambdaRatio276} and \ref{fig:lambdaRatio7} show the rapidity and \pt~independence of the \LbarL~ratios for all energies. The same measurements are shown for the \XbarX~ratios in Figures \ref{fig:xiRatio900}, \ref{fig:xiRatio276} and \ref{fig:xiRatio7} and for the \ObarO~ratios in Figures \ref{fig:omegaRatio276} and \ref{fig:omegaRatio7}. All hyperon measurements are described reasonably well by both PYTHIA (Perugia2011) and HIJING/B.

\subsection{Mid-rapidity ratios}

\begin{figure}
\includegraphics[width=\textwidth]{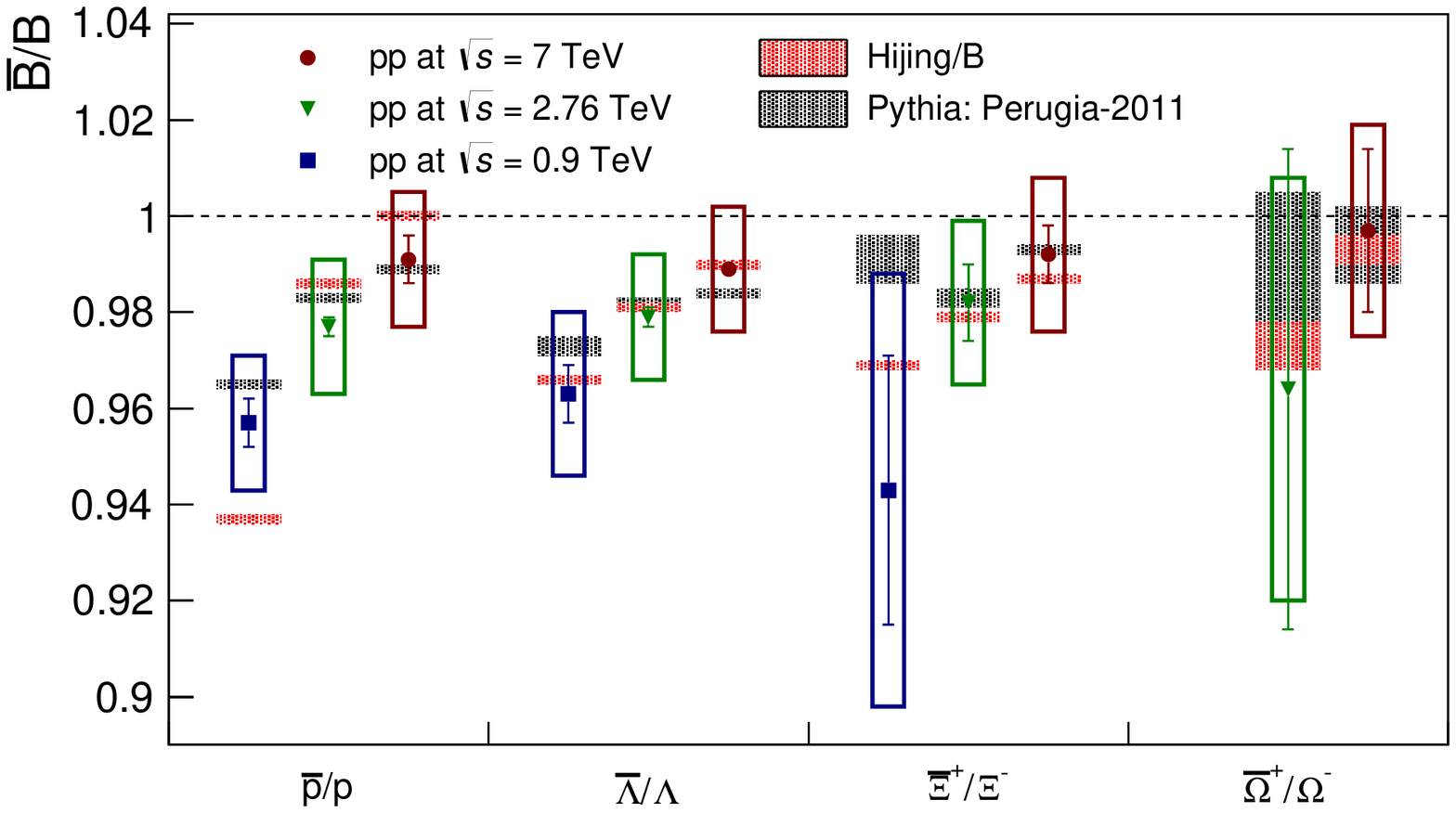}
\caption{The mid-rapidity yields ratio integrated over $|y| <0.5$ for \pbarp~and $|y| <0.8$ for \LbarL, \XbarX~and \ObarO. Squares, triangles and circles are for the data from pp at $\sqrt{s} = 0.9$, $2.76$ and $7$~TeV, respectively. The strangeness content increases along the abscissa.}
\label{fig:MidData}
\end{figure}

The corrected anti-baryon to baryon spectra ratios, integrated over the ALICE acceptance, in pp at $\sqrt{s} = 0.9$, $2.76$ and $7$~TeV are summarised in Table~\ref{tab:MidRatios}. Figure~\ref{fig:MidData} shows the measured \pbarp, \LbarL, \XbarX~and \ObarO~together with the same ratios extracted from PYTHIA (Perugia2011) and HIJING/B. HIJING/B models the baryon number stopping mechanism via string-junction transport; in contast, PYTHIA employs a pure multi-parton interaction model. The models reproduce the data reasonably well, although HIJING/B shows a steeper rise in the ratio as a function of beam energy for \pbarp~than the measured points. Within the uncertainties of our data, we cannot observe an increase of the ratio with the strangeness content, for the given energy. For all species (except the severely statistics limited \ObarO), the ratio increases with increasing beam energy, reaching values compatible with unity for $\sqrt{s} = 7$~TeV, which sets a stringent limit on the amount of baryon transport over 9 units in rapidity. The existence of a significant difference between the spectra of baryons and anti-baryons even at infinite energy \cite{Ref:Kopeliovich}, is therefore excluded. Various theory predictions using ${\alpha}_{\rm J}\approx 1$ are summarised in Table~\ref{tab:MidTheory}.

\begin{table*}
   \caption{Mid-rapidity anti-baryon to baryon yields ratios. The first uncertainty corresponds to the statistic, the second to the systematic one.}
\centering
\begin{tabular}{ l c c c c}
\toprule
$\sqrt{s}$ & \pbarp & \LbarL & \XbarX & \ObarO \\
\midrule
0.9~TeV & 0.957$\pm$0.006$\pm$0.014 & 0.963$\pm$0.006$\pm$0.017 & 0.938$\pm$0.028$\pm$0.045 & - \\
\addlinespace
2.76~TeV & 0.977$\pm$0.002$\pm$0.014 & 0.979$\pm$0.002$\pm$0.013 & 0.982$\pm$0.008$\pm$0.017 & 0.964$\pm$0.05$\pm$0.044\\
\addlinespace
7~TeV & 0.991$\pm$0.005$\pm$0.014 & 0.989$\pm$0.001$\pm$0.013 & 0.992$\pm$0.006$\pm$0.016 & 0.997$\pm$0.016$\pm$0.022 \\
\bottomrule
\end{tabular}
\label{tab:MidRatios}
\end{table*}

\begin{table}
\caption{Predictions for mid-rapidity anti-baryon to baryon yields ratios at $\sqrt{s}=7$~TeV}
\centering
\begin{tabular}{c c c c c}
\toprule
 &\pbarp & \LbarL & \XbarX & \ObarO \\
\midrule
\begin{tabular}[c]{@{}c@{}}Kopeliovich \cite{Ref:Kopeliovich}\\${\alpha}_{\rm J} = 1$\end{tabular} & \multicolumn{4}{c}{0.93} \\
\begin{tabular}[c]{@{}c@{}}QGSM \cite{Ref:QGSMMerino}\\${\alpha}_{\rm J} = 0.9$\end{tabular} & 0.946 & 0.945 & 0.958 & 0.958 \\
\bottomrule
\end{tabular}
\label{tab:MidTheory}
\end{table}

\subsection{Parametrisation of energy and rapidity dependence of the ratio}

\begin{figure*}
\centering
\includegraphics[width=0.49\textwidth]{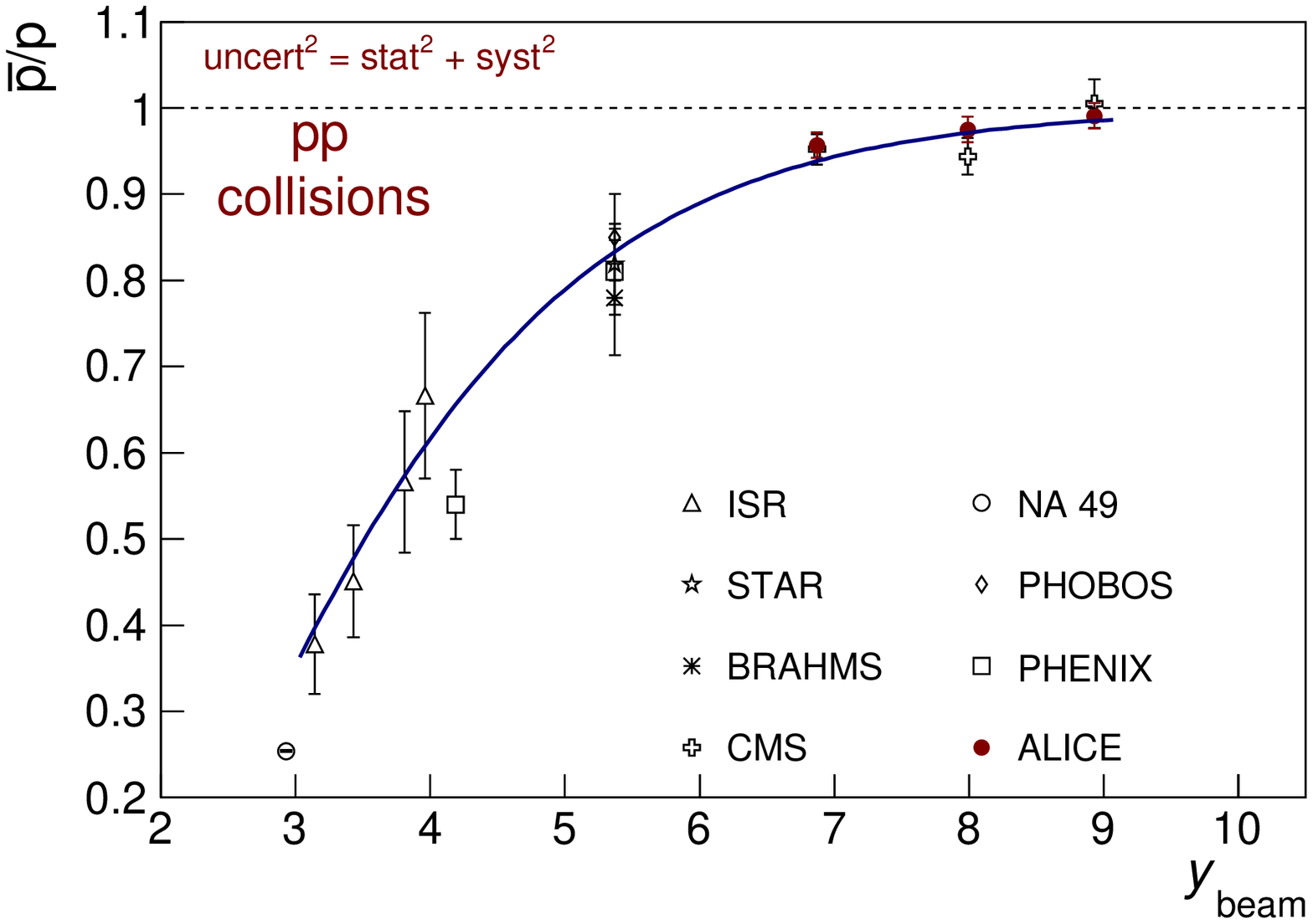}
\includegraphics[width=0.49\textwidth]{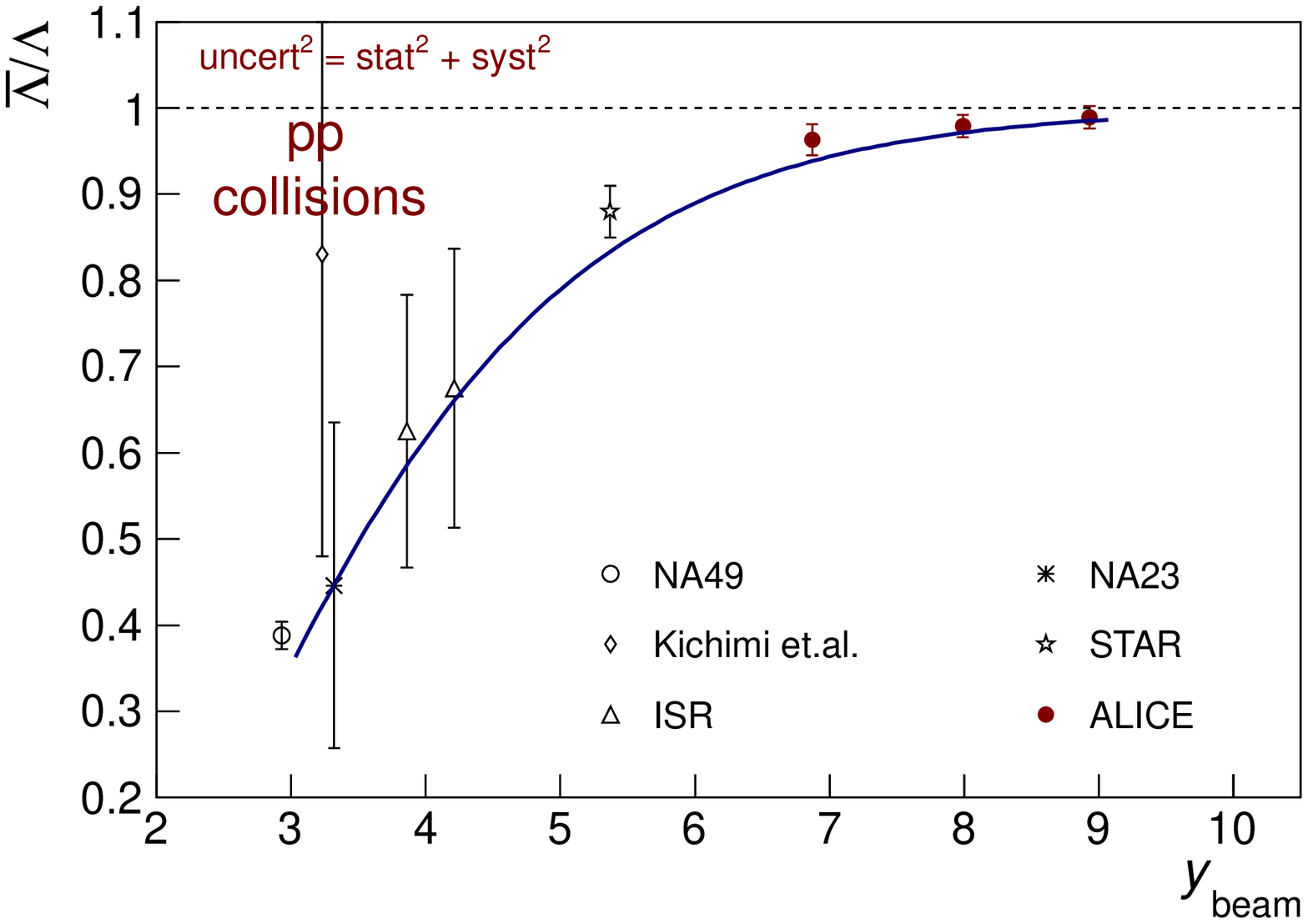}
\includegraphics[width=0.49\textwidth]{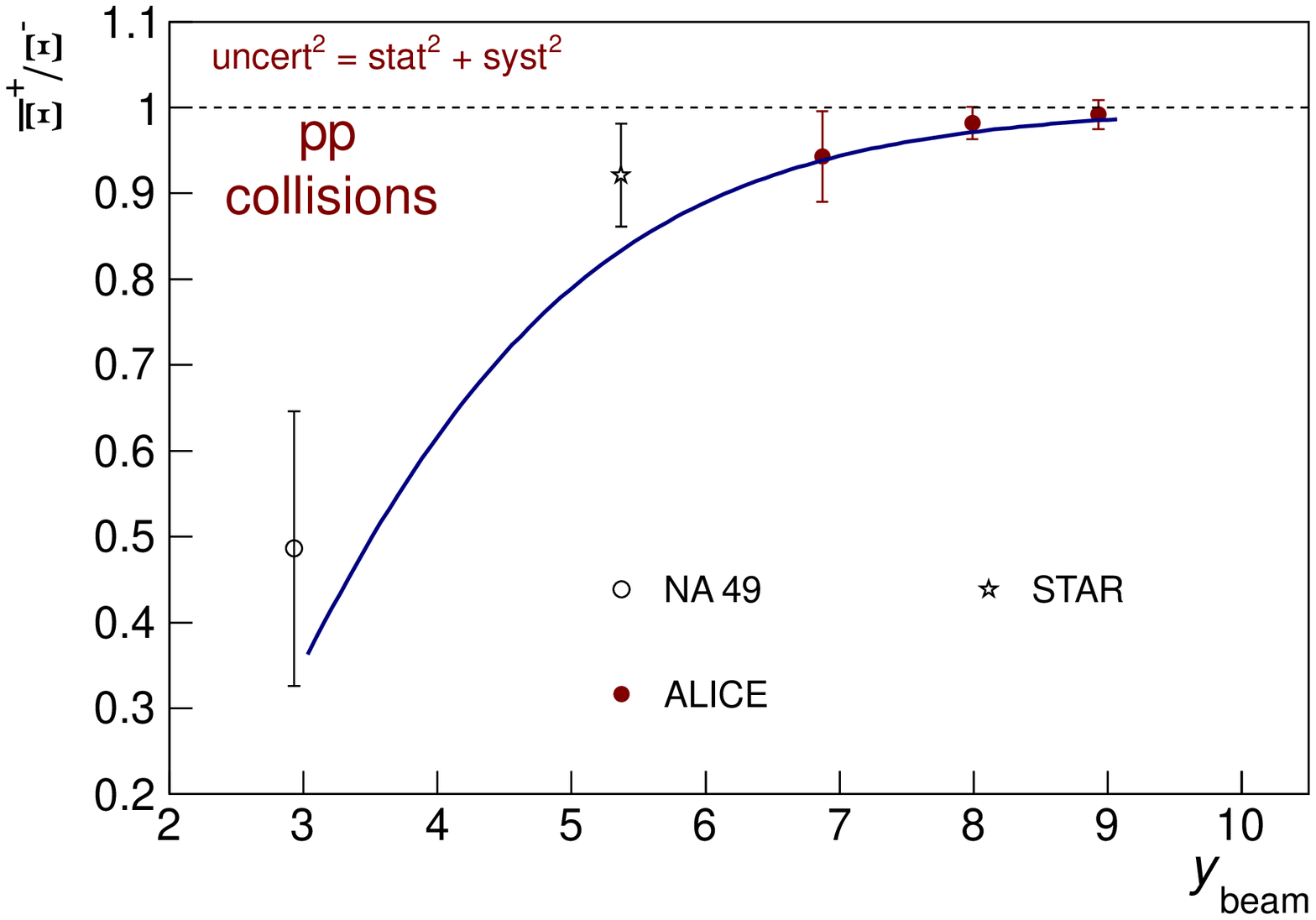}
\includegraphics[width=0.49\textwidth]{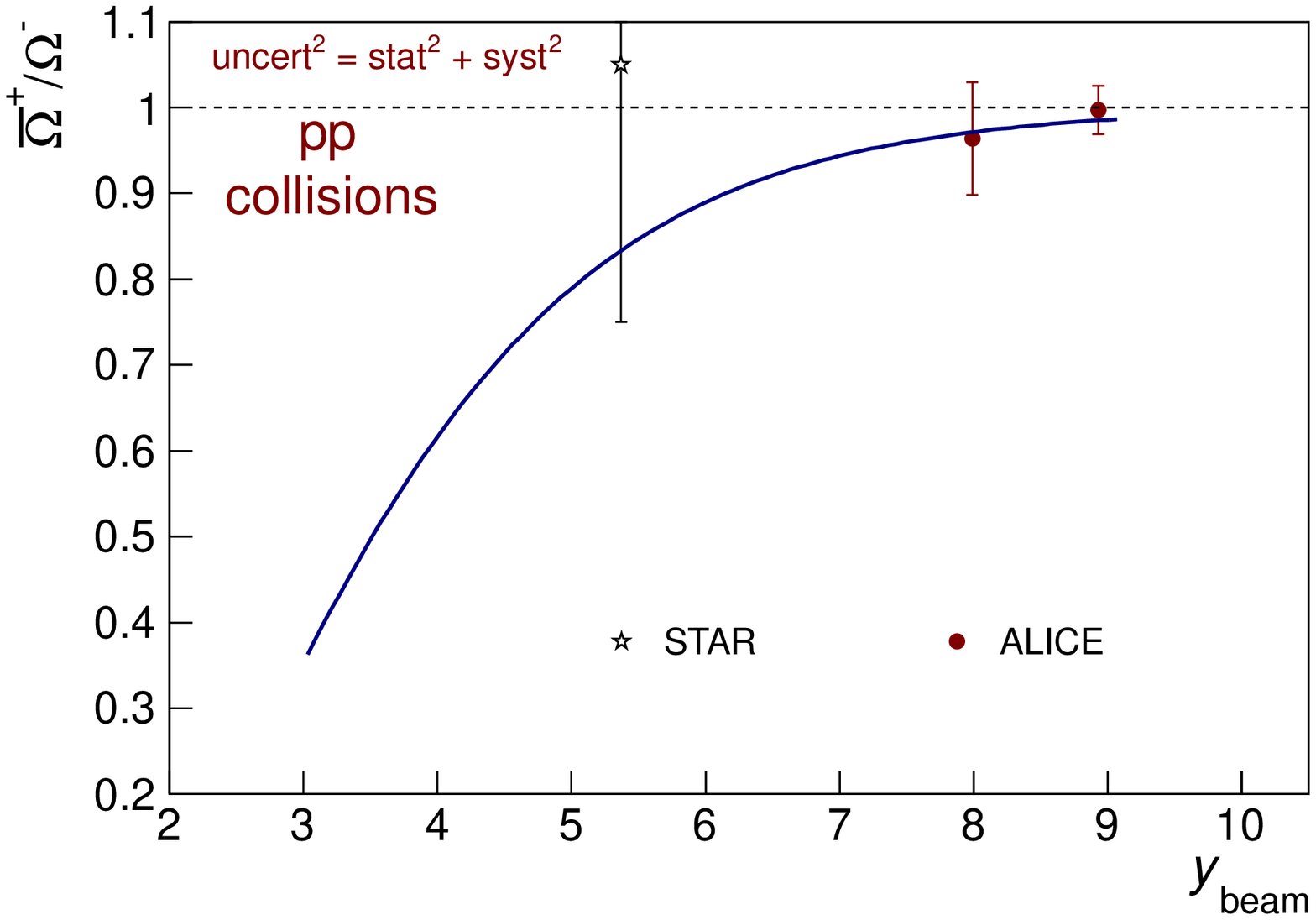}
\caption{(Colour online) Anti-baryon to baryon yields ratios as a function of beam rapidity for various baryons separately. The parametrisation with Eq.~(\ref{eq:Function1}) (blue line) is shown. The red points show the ALICE measurements.}
\label{fig:MidEnergy2}
\end{figure*}

Figure~\ref{fig:MidEnergy2} summarises the available data measured at mid-rapidity \cite{Ref:ExpCMS,Ref:ExpSPS,Ref:ExpISR,Ref:ExpRHIC} for \pbarp~(top left), \LbarL~(top right), \XbarX~(bottom left), and \ObarO~(bottom right) as a function of $y_{\rm beam}$.

As discussed in Section~\ref{Sec:Intro}, the behaviour of the $\overline{\mathrm{B}}/\mathrm{B}$ ratio as a function of $y_{\rm beam}$ and $y$ provides information on the mechanism responsible for baryon transport.

In pp collisions, baryons can be produced either from vacuum by baryon--anti-baryon pair production, or they can contain a quark, a di-quark or the string junction (or a combination of the latter three) of one of incoming protons. The probability of producing a baryon containing a valence quark or di-quark decreases exponentially with decreasing $|y|$. The anti-baryons are, in contrast, produced from the vacuum by baryon--anti-baryon pair production mechanisms. If the constituents of the incoming proton do not contribute at large rapidity intervals from the beam, one would expect, at asymptotic energies, the same yield of baryons and anti-baryons at mid-rapidity. The data favour this scenario. This fact is also complementary and/or is in agreement with lower energy experiments, where a similar $x$ dependence was observed for protons, neutrons and $\rm \Lambda$ and for anti-protons and $\overline{\rm \Lambda}$ at low and intermediate $x$-values ($x <$ 0.5).

We note that the \pt~cut-off used in this measurement for identifying baryons is higher than the mean \pt~of produced baryons. If a produced baryon contains a constituent from incoming protons, soft processes dominate its production and it likely has a \pt~lower than the mean \pt. Such \pt's are not in the \pt~range of our measurement.

An approximation of the $y_{\rm beam}$ and $y$ dependencies of the ratio can be derived in the Regge model. In this phenomenological approach, baryon-pair production at very high energy is governed by Pomeron exchange. The asymmetry between baryons and anti-baryons can be expressed by the string-junction transport and by an exchange with negative C-parity (e.g. $\rm\omega$ exchange). Following Refs.~\cite{Ref:Kharzeev} and ~\cite{Ref:Alicepbarp}, we parametrise the ratio, $R$, as a function of $y$ as follows:

\begin{equation}
R = \frac{1 + C_{1}\times\exp{({\alpha}_{\rm J} - {\alpha}_{\rm P}) y_{\rm beam}}\times\cosh{({\alpha}_{\rm J} - {\alpha}_{\rm P}) y}}{1 + C_{2}\times\exp{({\alpha}_{\rm J} - {\alpha}_{\rm P}) y_{\rm beam}}\times\cosh{({\alpha}_{\rm J} - {\alpha}_{\rm P}) y}},
\label{eq:Function1}
\end{equation}

\noindent where ${\alpha}_{\rm P}=1.2$ \cite{Ref:alpha1.2} is the Pomeron intercept and ${\alpha}_{\rm J}$ is the string-junction intercept, assumed to be 0.5 \cite{Ref:RossiVeneziano} and equal to the intercept of secondary Reggeons. If $C_{1}$ =0, the Eq.~\ref{eq:Function1} counts only the contribution of string junction and/or for the case when in the anti-proton spectrum the secondary Reggeons with positive C-parity (e.g. f exchange) have the same contribution as the secondary Reggeons with negative C-parity.

A fit to the data \pbarp~ratio at mid-rapidity gives $C_{2}=-C_{1}=$3.9$\pm$0.3. For the fit, we are using all the measurements with ${\rm \Delta} y>3$ i.e. the NA49 points are omitted, since in this region, contribution of other diagrams cannot be neglected \cite{Ref:Kharzeev}. The fit is shown as a solid lines in Fig.~\ref{fig:MidEnergy2} and gives a reasonable description of the data for all baryon species. This means a Reggeon with negative C-parity and ${\alpha}_{\rm J}$ = 0.5 is sufficient for describing the difference between baryons and anti-baryons at mid-rapidity. 

\begin{figure*}
\includegraphics[width=0.49\linewidth]{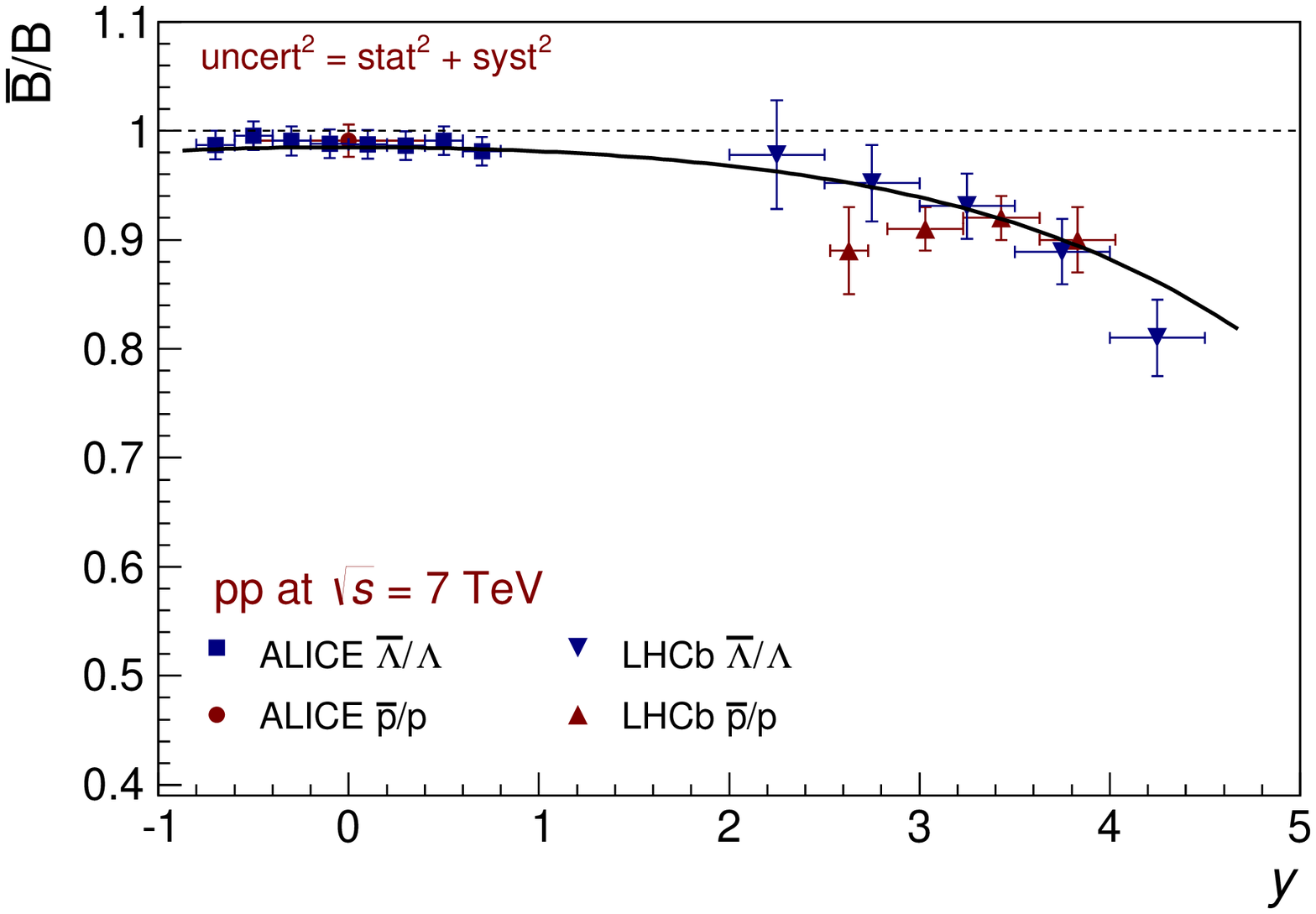}
\includegraphics[width=0.49\linewidth]{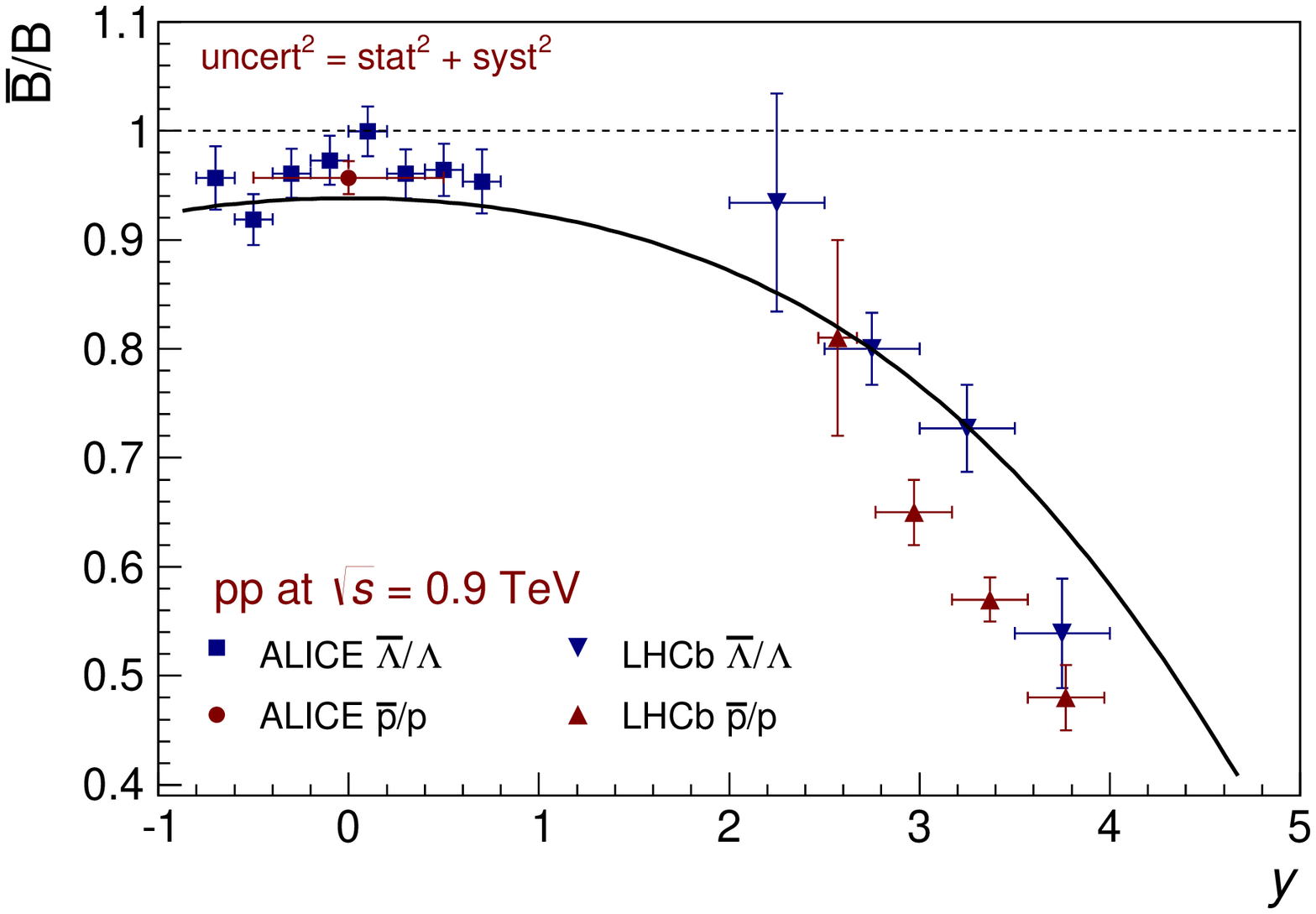}
\caption{(Colour online) \pbarp~and \LbarL~ratios as a function of rapidity at $\sqrt{s} = 0.9$ and $7$~TeV. The parametrisation with Eq.~(\ref{eq:Function1}) (black line) is shown.}
\label{fig:MidEnergy1}
\end{figure*}

In Fig.~\ref{fig:MidEnergy1} we show ALICE and LHCb \cite{Ref:ExpLHCb} data on \pbarp~and \LbarL~ratios as a function of rapidity at $\sqrt{s} = 0.9$ and $7$~TeV. The superimposed curve is obtained from Eq.~\ref{eq:Function1} using parameters $C_{2}=-C_{1}=$3.9 obtained from the fit of \pbarp~ratio at mid-rapidity as shown in Fig.~\ref{fig:MidEnergy2}. Again, a Reggeon with negative C-parity and ${\alpha}_{\rm J}$ = 0.5 is sufficient for describing the data, except for large values of rapidity where contribution of other diagrams cannot be neglected.

We can conclude that any significant contribution to anti-baryon to baryon ratio at mid-rapidity due to an exchange which is not suppressed with increasing rapidity interval is disfavoured. This picture is also supported by both PYTHIA (Perugia2011) and HIJING/B.

\subsection{Multiplicity dependence}
We have also investigated the dependence of the anti-baryon to baryon yields ratios on the charged-particle multiplicity density, ${\rm d}N_{\rm ch}/{\rm d}\eta$. The multiplicity measurement was based on the number of global tracks (which combine the information from the ITS and the TPC), and the number of tracklets (vectors connecting pairs of clusters each from one of the two SPD layers and pointing to the vertex but not part of a reconstructed global track) in $|\eta| < 0.5$. Using simulated events, it was verified, that this estimate is proportional to ${\rm d}N_{\rm ch}/{\rm d}\eta$. We present the anti-baryon to baryon ratios as a function of the relative charged-particle pseudorapidity density $({\rm d}N_{\rm ch}/{\rm d}\eta)/\langle {\rm d}N_{\rm ch}/{\rm d}\eta \rangle$, where $\langle {\rm d}N_{\rm ch}/{\rm d}\eta \rangle$ is a value measured for inelastic pp collisions with at least one charged particle in $|\eta| < 1$ (INEL $> 0_{|\eta|<1}$) \cite{Ref:ALICENchargedpapers} (see Table~\ref{tab:Inel}). The value at $\sqrt{s} = 2.76$~TeV was not measured: it is an interpolation of points at $\sqrt{s}$ = 0.9, 2.36 and 7~TeV using a power law function. The use of relative quantities was chosen in order to facilitate the comparison to other experiments, as well as to minimise systematic uncertainties. 

\begin{table}[htbp]
\caption{Charged-particle pseudorapidity densities}
\centering
\begin{tabular}{c c}
\toprule
$\sqrt{s}$~(TeV) & $\langle {\rm d}N_{\rm ch}/{\rm d}\eta \rangle$ (INEL $> 0_{|\eta|<1}$)\\
\midrule
0.9 & 3.81$\pm$~$0.01^{+0.07}_{-0.07}$\\
2.36 & 4.70$\pm$~$0.01^{+0.11}_{-0.08}$\\
2.76 & 4.88$\pm$~$0.01^{+0.13}_{-0.09}$\\
7 & 6.01$\pm$~$0.01^{+0.20}_{-0.12}$\\
\bottomrule
\end{tabular}
\label{tab:Inel}
\end{table}

\begin{figure}[h]
\centering
  \includegraphics[width=0.6\linewidth]{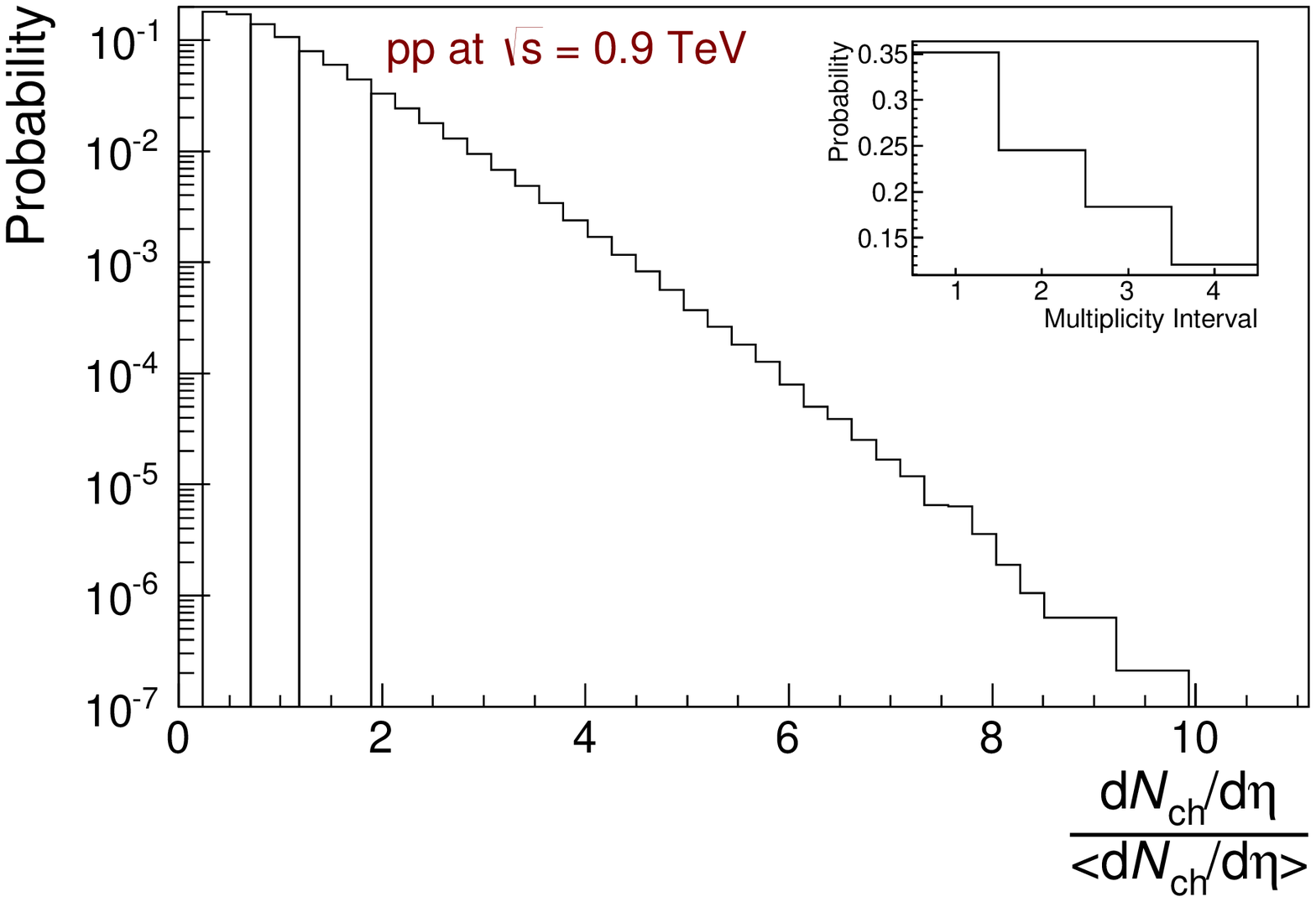}
  \includegraphics[width=0.6\linewidth]{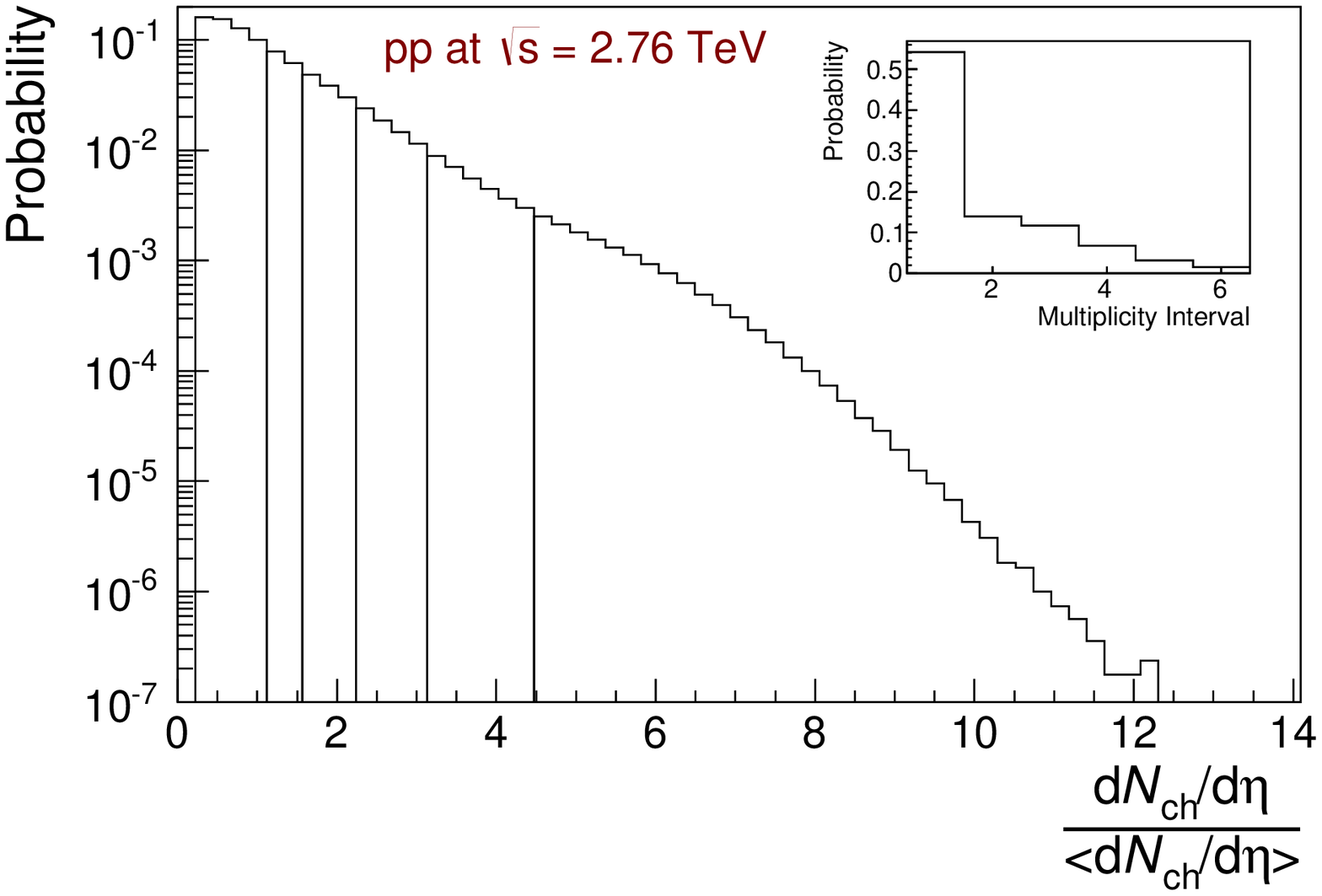}
  \includegraphics[width=0.6\linewidth]{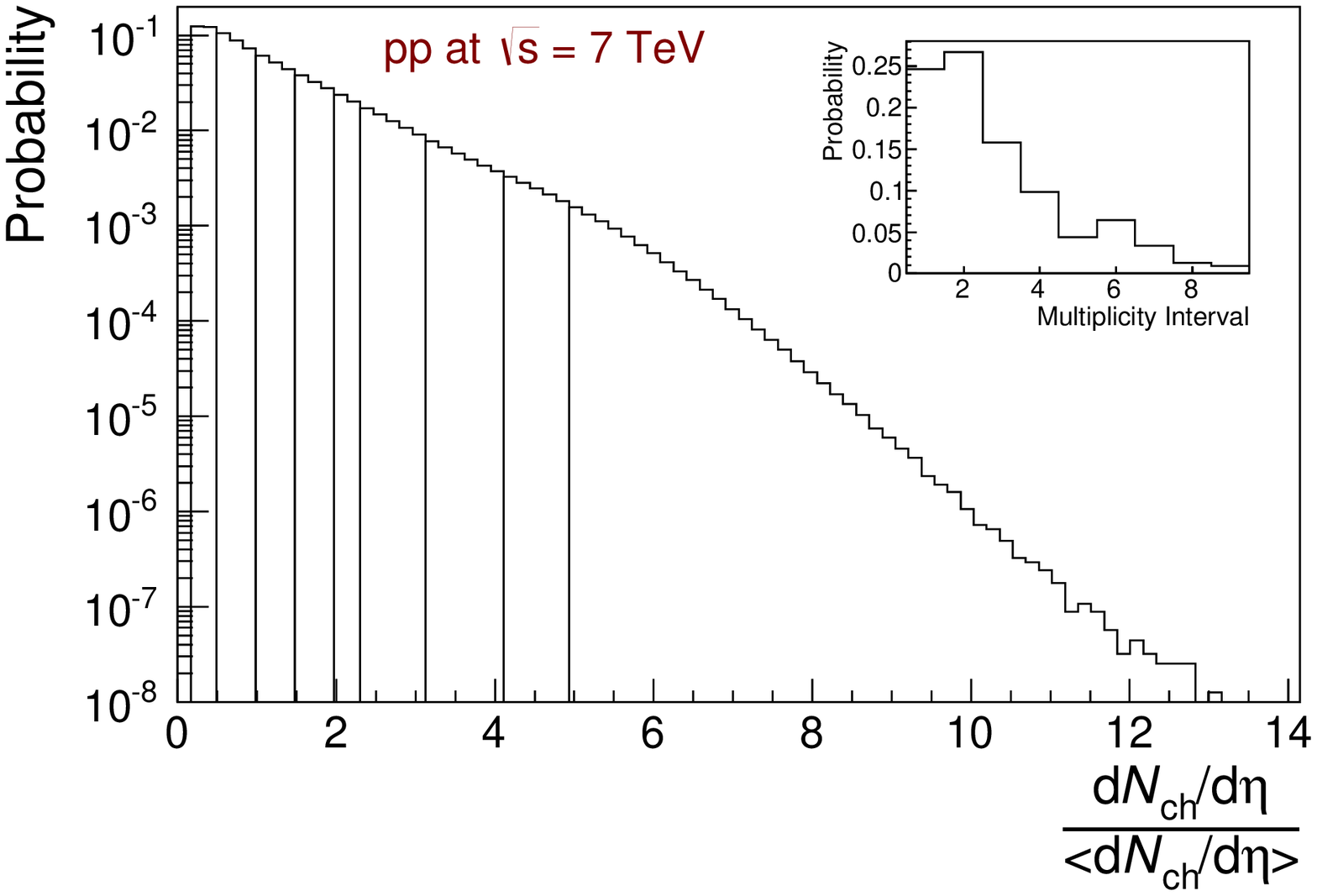}
  \caption{(Colour online) Charged particle multiplicity distributions. The event samples are divided according to multiplicity bins used in \pbarp~ratio analysis. The insets show the probability for different bins.}
  \label{fig:multiQM}
\end{figure}

\begin{figure}[h]
  \centering
  \includegraphics[width=0.6\linewidth]{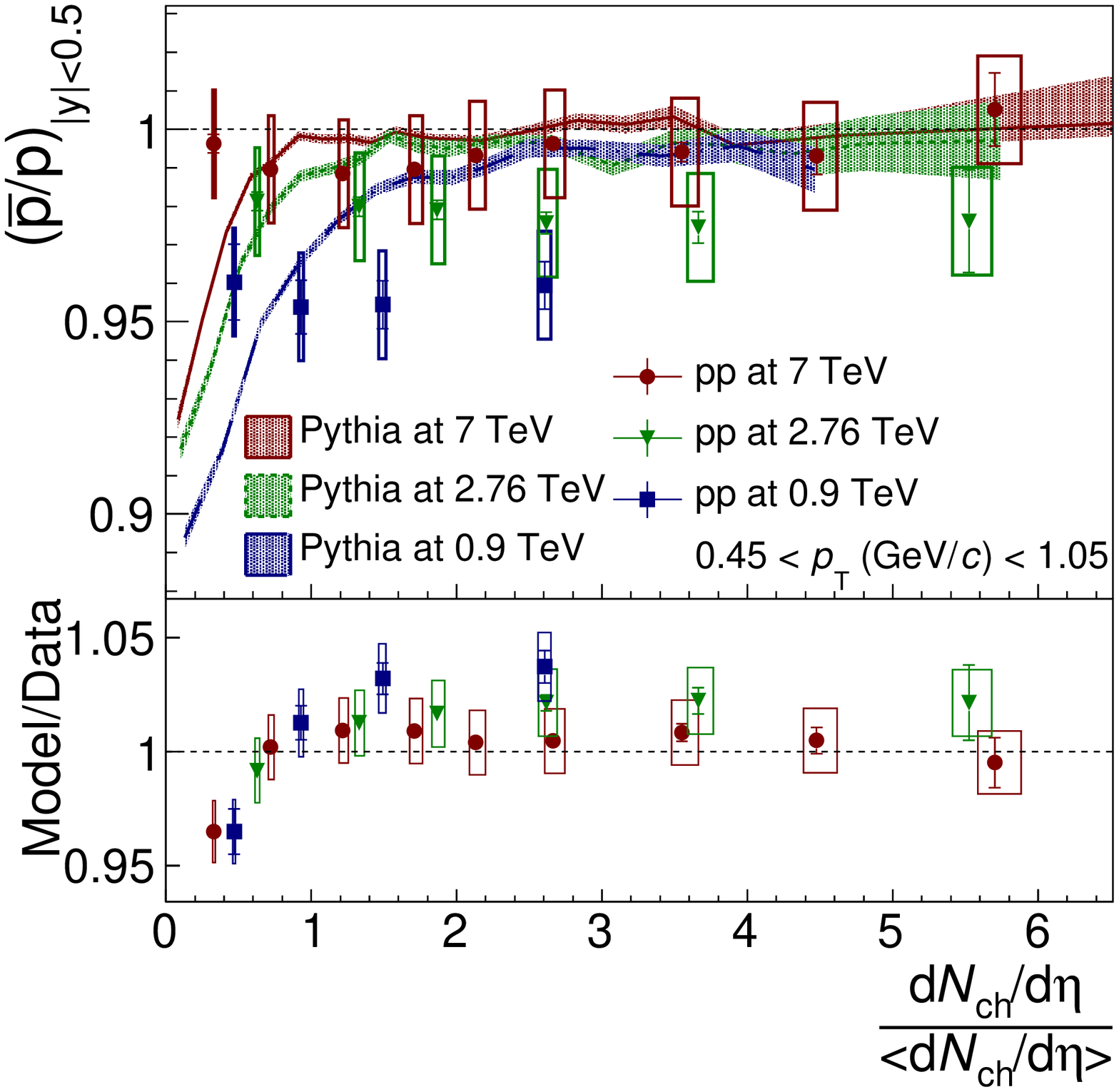}
  \caption{(Colour online) The \pbarp~ratio in pp collisions at $\sqrt{s} = 0.9$, $2.76$ and $7$~TeV as a function of the relative charged-particle pseudorapidity density. The data points are compared with prediction of PYTHIA (Perugia2011). The vertical bars (boxes) represent the statistical (systematic) uncertainty. Ratio of model to data is shown below using uncertainties added in quadrature.}
    \label{fig:MultiProton}
\end{figure}

\begin{figure}[h]
  \centering
  \includegraphics[width=0.6\linewidth]{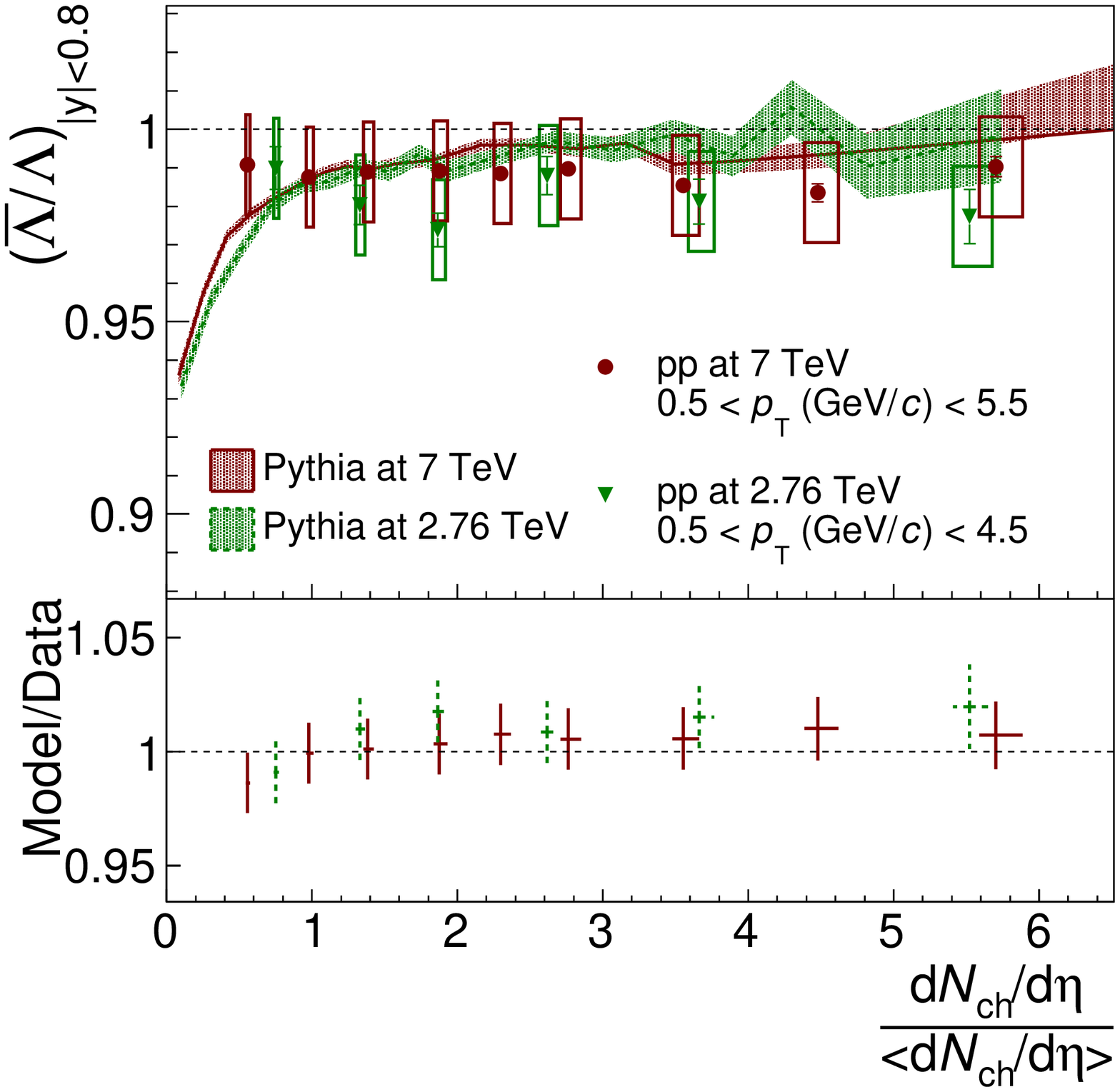}
  \caption{(Colour online) The \LbarL~ratio in pp collisions $\sqrt{s} = 2.76$ and $7$~TeV as a function of the relative charged-particle pseudorapidity density. The data points are compared with prediction of PYTHIA (Perugia2011). The vertical bars (boxes) represent the statistical (systematic) uncertainty. Ratio of model to data is shown below using uncertainties added in quadrature.}
    \label{fig:MultiLambda}
\end{figure}

\begin{figure}[h]
  \centering
  \includegraphics[width=0.6\linewidth]{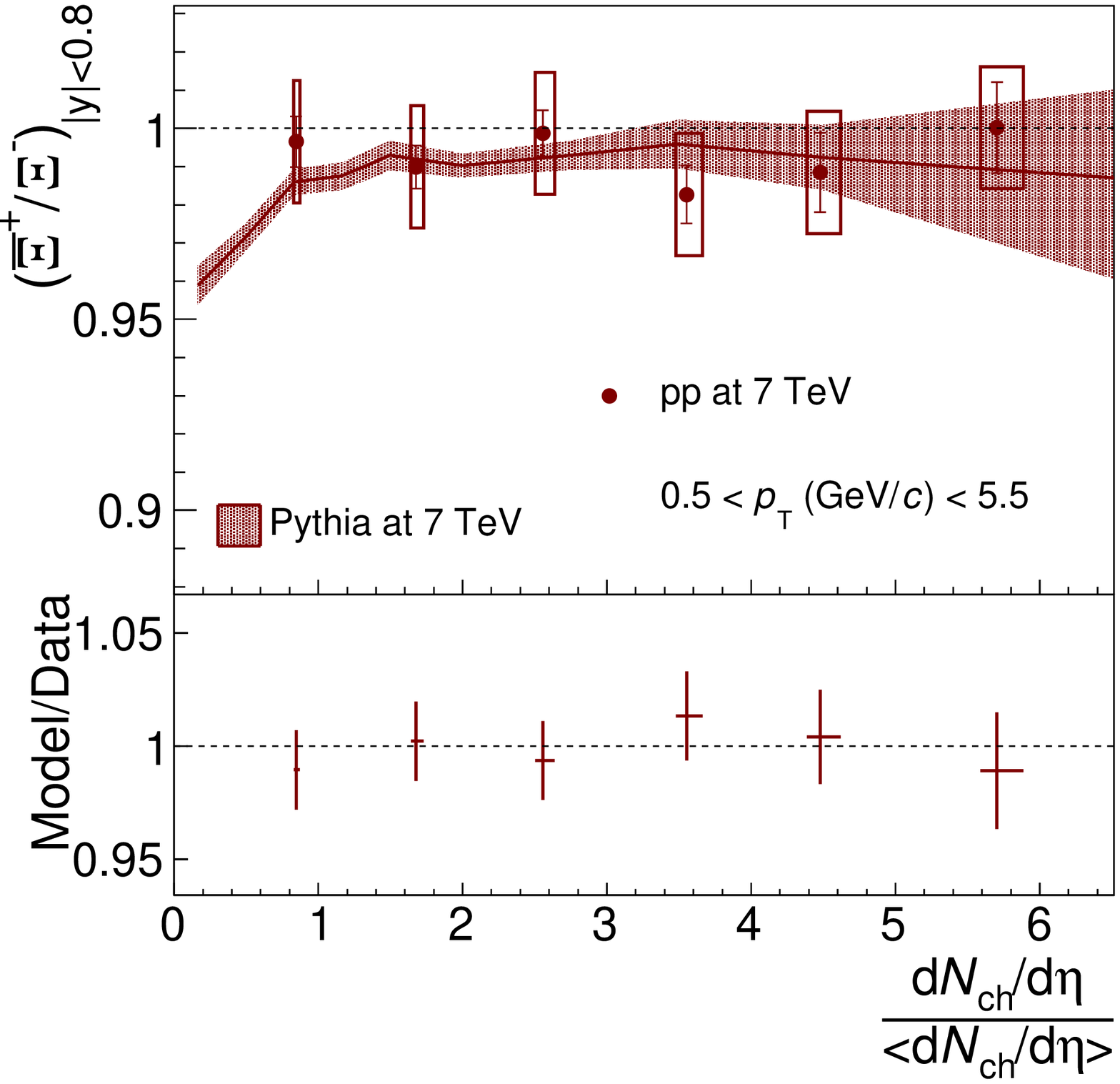}
  \caption{(Colour online) The \XbarX~ratio in pp collisions $\sqrt{s} = 7$~TeV as a function of the relative charged-particle pseudorapidity density. The data points are compared with prediction of PYTHIA (Perugia2011). The vertical bars (boxes) represent the statistical (systematic) uncertainty. Ratio of model to data is shown below using uncertainties added in quadrature.}
    \label{fig:MultiXi}
\end{figure}

The relative multiplicity densities are shown in Fig.~\ref{fig:multiQM}. The sizes of bins were chosen so that they all have sufficient event population. The ratios \pbarp, \LbarL, and \XbarX~are presented in Figs.~\ref{fig:MultiProton}, \ref{fig:MultiLambda} and \ref{fig:MultiXi}. The \ObarO~ratio had to be omitted due to insufficient statistics for this analysis. The weighted mean of the multiplicity distribution in the bin range was set as centre of the bin. The uncertainty on this quantity is due to the uncertainty on the measured ${\rm d}N_{\rm ch}/{\rm d}\eta$.

As can be seen, the ratios for \pbarp, \LbarL~and \XbarX~exhibit no dependence on $({\rm d}N_{\rm ch}/{\rm d}\eta)/\langle {\rm d}N_{\rm ch}/{\rm d}\eta \rangle$. 
On the other hand, PYTHIA (Perugia2011) is showing a steep rise of the ratio for low multiplicities, followed by a saturation, which is not present in our data. The most significant disagreement can be seen in case of \pbarp~at $\sqrt{s} = 0.9$~TeV.
Possible explanation of the discrepancy between our data and PYTHIA (Perugia2011) can be the following: the baryon--anti-baryon pair production is increasing as a function of multiplicity and since we do not see any multiplicity dependence of the ratio in the data, the baryon number transfer has to increase as well in the same way. PYTHIA (Perugia2011) is not in favour with this picture, predicting a constant or a slower increase of the baryon number transfer with multiplicity than the baryon--anti-baryon pair production, resulting into a (steep) rise of the ratio followed by a saturation at unity.


%% file: Summary.tex
\section{Summary}
\label{Sec:Summary}

Within the ALICE acceptance the \pbarp, \LbarL, \XbarX~and \ObarO~ratios in pp collisions at $\sqrt{s}$ = 0.9, 2.76 and 7~TeV are found to be independent of rapidity, transverse momentum, and charged particle multiplicity.

At $\sqrt{s} = 0.9$~TeV we see a small excess of baryons over anti-baryons for the \pbarp, \LbarL~and \XbarX~ratios. The ratios increase with increasing beam energy, reaching values compatible with unity for $\sqrt{s} = 7$~TeV. Within the uncertainties of our measurement, we do not observe an increase of the ratio with the strangeness content, for the given energy.

These results are consistent with model predictions describing the asymmetry between baryons and anti-baryons by the string-junction transport and/or by an exchange with negative C-parity (e.g. $\rm \omega$ exchange) using intercept of ${\alpha}_{\rm J} \approx$~0.5. These data are not consistent with models predicting a significant difference between the spectra of baryons and anti-baryons at large ${\rm \Delta} y$ (${\rm \Delta y} > 8$) in pp collisions.

\FloatBarrier

%% file: acknowledgements_march2013.tex
The ALICE collaboration would like to thank all its engineers and technicians for their invaluable contributions to the construction of the experiment and the CERN accelerator teams for the outstanding performance of the LHC complex.
\\
The ALICE collaboration acknowledges the following funding agencies for their support in building and
running the ALICE detector:
 \\
State Committee of Science,  World Federation of Scientists (WFS)
and Swiss Fonds Kidagan, Armenia,
 \\
Conselho Nacional de Desenvolvimento Cient\'{\i}fico e Tecnol\'{o}gico (CNPq), Financiadora de Estudos e Projetos (FINEP),
Funda\c{c}\~{a}o de Amparo \`{a} Pesquisa do Estado de S\~{a}o Paulo (FAPESP);
 \\
National Natural Science Foundation of China (NSFC), the Chinese Ministry of Education (CMOE)
and the Ministry of Science and Technology of China (MSTC);
 \\
Ministry of Education and Youth of the Czech Republic;
 \\
Danish Natural Science Research Council, the Carlsberg Foundation and the Danish National Research Foundation;
 \\
The European Research Council under the European Community's Seventh Framework Programme;
 \\
Helsinki Institute of Physics and the Academy of Finland;
 \\
French CNRS-IN2P3, the `Region Pays de Loire', `Region Alsace', `Region Auvergne' and CEA, France;
 \\
German BMBF and the Helmholtz Association;
\\
General Secretariat for Research and Technology, Ministry of
Development, Greece;
\\
Hungarian OTKA and National Office for Research and Technology (NKTH);
 \\
Department of Atomic Energy and Department of Science and Technology of the Government of India;
 \\
Istituto Nazionale di Fisica Nucleare (INFN) and Centro Fermi -
Museo Storico della Fisica e Centro Studi e Ricerche "Enrico
Fermi", Italy;
 \\
MEXT Grant-in-Aid for Specially Promoted Research, Ja\-pan;
 \\
Joint Institute for Nuclear Research, Dubna;
 \\
National Research Foundation of Korea (NRF);
 \\
CONACYT, DGAPA, M\'{e}xico, ALFA-EC and the EPLANET Program
(European Particle Physics Latin American Network)
 \\
Stichting voor Fundamenteel Onderzoek der Materie (FOM) and the Nederlandse Organisatie voor Wetenschappelijk Onderzoek (NWO), Netherlands;
 \\
Research Council of Norway (NFR);
 \\
Polish Ministry of Science and Higher Education;
 \\
National Authority for Scientific Research - NASR (Autoritatea Na\c{t}ional\u{a} pentru Cercetare \c{S}tiin\c{t}ific\u{a} - ANCS);
 \\
Ministry of Education and Science of Russian Federation, Russian
Academy of Sciences, Russian Federal Agency of Atomic Energy,
Russian Federal Agency for Science and Innovations and The Russian
Foundation for Basic Research;
 \\
Ministry of Education of Slovakia;
 \\
Department of Science and Technology, South Africa;
 \\
CIEMAT, EELA, Ministerio de Econom\'{i}a y Competitividad (MINECO) of Spain, Xunta de Galicia (Conseller\'{\i}a de Educaci\'{o}n),
CEA\-DEN, Cubaenerg\'{\i}a, Cuba, and IAEA (International Atomic Energy Agency);
 \\
Swedish Research Council (VR) and Knut $\&$ Alice Wallenberg
Foundation (KAW);
 \\
Ukraine Ministry of Education and Science;
 \\
United Kingdom Science and Technology Facilities Council (STFC);
 \\
The United States Department of Energy, the United States National
Science Foundation, the State of Texas, and the State of Ohio.

%% file: authorlist-2013-03-29-cernpreprint.tex
\begingroup
\small
\begin{flushleft}
E.~Abbas\Irefn{org36632}\And
B.~Abelev\Irefn{org1234}\And
J.~Adam\Irefn{org1274}\And
D.~Adamov\'{a}\Irefn{org1283}\And
A.M.~Adare\Irefn{org1260}\And
M.M.~Aggarwal\Irefn{org1157}\And
G.~Aglieri~Rinella\Irefn{org1192}\And
M.~Agnello\Irefn{org1313}\textsuperscript{,}\Irefn{org1017688}\And
A.G.~Agocs\Irefn{org1143}\And
A.~Agostinelli\Irefn{org1132}
% ...
\renewcommand\labelenumi{\textsuperscript{\theenumi}~}
\section*{Affiliation notes}
\renewcommand\theenumi{\roman{enumi}}
\begin{Authlist}
\item \Adef{0}Deceased
\item \Adef{M.V.Lomonosov Moscow State University, D.V.Skobeltsyn Institute of Nuclear Physics, Moscow, Russia}Also at: M.V.Lomonosov Moscow State University, D.V.Skobeltsyn Institute of Nuclear Physics, Moscow, Russia
\item \Adef{University of Belgrade, Faculty of Physics and "Vinvca" Institute of Nuclear Sciences, Belgrade, Serbia}Also at: University of Belgrade, Faculty of Physics and "Vinvca" Institute of Nuclear Sciences, Belgrade, Serbia
\item \Adef{Deceased}Also at: Deceased
\item \Adef{Institute of Theoretical Physics, University of Wroclaw, Wroclaw, Poland}Also at: Institute of Theoretical Physics, University of Wroclaw, Wroclaw, Poland
\end{Authlist}
\section*{Collaboration Institutes}
\renewcommand\theenumi{\arabic{enumi}~}
\begin{Authlist}
\item \Idef{org36632}Academy of Scientific Research and Technology (ASRT), Cairo, Egypt
\item \Idef{org1332}A. I. Alikhanyan National Science Laboratory (Yerevan Physics Institute) Foundation, Yerevan, Armenia
\item \Idef{org1279}Benem\'{e}rita Universidad Aut\'{o}noma de Puebla, Puebla, Mexico
\item \Idef{org1220}Bogolyubov Institute for Theoretical Physics, Kiev, Ukraine
\item \Idef{org20959}Bose Institute, Department of Physics and Centre for Astroparticle Physics and Space Science (CAPSS), Kolkata, India
\item \Idef{org1262}Budker Institute for Nuclear Physics, Novosibirsk, Russia
\item \Idef{org1292}California Polytechnic State University, San Luis Obispo, California, United States
\item \Idef{org1329}Central China Normal University, Wuhan, China
\item \Idef{org14939}Centre de Calcul de l'IN2P3, Villeurbanne, France
\item \Idef{org1197}Centro de Aplicaciones Tecnol\'{o}gicas y Desarrollo Nuclear (CEADEN), Havana, Cuba
\item \Idef{org1242}Centro de Investigaciones Energ\'{e}ticas Medioambientales y Tecnol\'{o}gicas (CIEMAT), Madrid, Spain
\item \Idef{org1244}Centro de Investigaci\'{o}n y de Estudios Avanzados (CINVESTAV), Mexico City and M\'{e}rida, Mexico
\item \Idef{org1335}Centro Fermi - Museo Storico della Fisica e Centro Studi e Ricerche ``Enrico Fermi'', Rome, Italy
\item \Idef{org17347}Chicago State University, Chicago, United States
\item \Idef{org1288}Commissariat \`{a} l'Energie Atomique, IRFU, Saclay, France
\item \Idef{org15782}COMSATS Institute of Information Technology (CIIT), Islamabad, Pakistan
\item \Idef{org1294}Departamento de F\'{\i}sica de Part\'{\i}culas and IGFAE, Universidad de Santiago de Compostela, Santiago de Compostela, Spain
\item \Idef{org1106}Department of Physics Aligarh Muslim University, Aligarh, India
\item \Idef{org1121}Department of Physics and Technology, University of Bergen, Bergen, Norway
\item \Idef{org1162}Department of Physics, Ohio State University, Columbus, Ohio, United States
\item \Idef{org1300}Department of Physics, Sejong University, Seoul, South Korea
\item \Idef{org1268}Department of Physics, University of Oslo, Oslo, Norway
\item \Idef{org1315}Dipartimento di Fisica dell'Universit\`{a} and Sezione INFN, Trieste, Italy
\item \Idef{org1145}Dipartimento di Fisica dell'Universit\`{a} and Sezione INFN, Cagliari, Italy
\item \Idef{org1312}Dipartimento di Fisica dell'Universit\`{a} and Sezione INFN, Turin, Italy
\item \Idef{org1285}Dipartimento di Fisica dell'Universit\`{a} `La Sapienza' and Sezione INFN, Rome, Italy
\item \Idef{org1154}Dipartimento di Fisica e Astronomia dell'Universit\`{a} and Sezione INFN, Catania, Italy
\item \Idef{org1132}Dipartimento di Fisica e Astronomia dell'Universit\`{a} and Sezione INFN, Bologna, Italy
\item \Idef{org1270}Dipartimento di Fisica e Astronomia dell'Universit\`{a} and Sezione INFN, Padova, Italy
\item \Idef{org1290}Dipartimento di Fisica `E.R.~Caianiello' dell'Universit\`{a} and Gruppo Collegato INFN, Salerno, Italy
\item \Idef{org1103}Dipartimento di Scienze e Innovazione Tecnologica dell'Universit\`{a} del Piemonte Orientale and Gruppo Collegato INFN, Alessandria, Italy
\item \Idef{org1114}Dipartimento Interateneo di Fisica `M.~Merlin' and Sezione INFN, Bari, Italy
\item \Idef{org1237}Division of Experimental High Energy Physics, University of Lund, Lund, Sweden
\item \Idef{org1192}European Organization for Nuclear Research (CERN), Geneva, Switzerland
\item \Idef{org1227}Fachhochschule K\"{o}ln, K\"{o}ln, Germany
\item \Idef{org1122}Faculty of Engineering, Bergen University College, Bergen, Norway
\item \Idef{org1136}Faculty of Mathematics, Physics and Informatics, Comenius University, Bratislava, Slovakia
\item \Idef{org1274}Faculty of Nuclear Sciences and Physical Engineering, Czech Technical University in Prague, Prague, Czech Republic
\item \Idef{org1229}Faculty of Science, P.J.~\v{S}af\'{a}rik University, Ko\v{s}ice, Slovakia
\item \Idef{org1184}Frankfurt Institute for Advanced Studies, Johann Wolfgang Goethe-Universit\"{a}t Frankfurt, Frankfurt, Germany
\item \Idef{org1215}Gangneung-Wonju National University, Gangneung, South Korea
\item \Idef{org20958}Gauhati University, Department of Physics, Guwahati, India
\item \Idef{org1212}Helsinki Institute of Physics (HIP) and University of Jyv\"{a}skyl\"{a}, Jyv\"{a}skyl\"{a}, Finland
\item \Idef{org1203}Hiroshima University, Hiroshima, Japan
\item \Idef{org1254}Indian Institute of Technology Bombay (IIT), Mumbai, India
\item \Idef{org36378}Indian Institute of Technology Indore, Indore, India (IITI)
\item \Idef{org1266}Institut de Physique Nucl\'{e}aire d'Orsay (IPNO), Universit\'{e} Paris-Sud, CNRS-IN2P3, Orsay, France
\item \Idef{org1277}Institute for High Energy Physics, Protvino, Russia
\item \Idef{org1249}Institute for Nuclear Research, Academy of Sciences, Moscow, Russia
\item \Idef{org1320}Nikhef, National Institute for Subatomic Physics and Institute for Subatomic Physics of Utrecht University, Utrecht, Netherlands
\item \Idef{org1250}Institute for Theoretical and Experimental Physics, Moscow, Russia
\item \Idef{org1230}Institute of Experimental Physics, Slovak Academy of Sciences, Ko\v{s}ice, Slovakia
\item \Idef{org1127}Institute of Physics, Bhubaneswar, India
\item \Idef{org1275}Institute of Physics, Academy of Sciences of the Czech Republic, Prague, Czech Republic
\item \Idef{org1139}Institute of Space Sciences (ISS), Bucharest, Romania
\item \Idef{org27399}Institut f\"{u}r Informatik, Johann Wolfgang Goethe-Universit\"{a}t Frankfurt, Frankfurt, Germany
\item \Idef{org1185}Institut f\"{u}r Kernphysik, Johann Wolfgang Goethe-Universit\"{a}t Frankfurt, Frankfurt, Germany
\item \Idef{org1177}Institut f\"{u}r Kernphysik, Technische Universit\"{a}t Darmstadt, Darmstadt, Germany
\item \Idef{org1256}Institut f\"{u}r Kernphysik, Westf\"{a}lische Wilhelms-Universit\"{a}t M\"{u}nster, M\"{u}nster, Germany
\item \Idef{org1246}Instituto de Ciencias Nucleares, Universidad Nacional Aut\'{o}noma de M\'{e}xico, Mexico City, Mexico
\item \Idef{org1247}Instituto de F\'{\i}sica, Universidad Nacional Aut\'{o}noma de M\'{e}xico, Mexico City, Mexico
\item \Idef{org1308}Institut Pluridisciplinaire Hubert Curien (IPHC), Universit\'{e} de Strasbourg, CNRS-IN2P3, Strasbourg, France
\item \Idef{org1182}Joint Institute for Nuclear Research (JINR), Dubna, Russia
\item \Idef{org1199}Kirchhoff-Institut f\"{u}r Physik, Ruprecht-Karls-Universit\"{a}t Heidelberg, Heidelberg, Germany
\item \Idef{org20954}Korea Institute of Science and Technology Information, Daejeon, South Korea
\item \Idef{org1017642}KTO Karatay University, Konya, Turkey
\item \Idef{org1160}Laboratoire de Physique Corpusculaire (LPC), Clermont Universit\'{e}, Universit\'{e} Blaise Pascal, CNRS--IN2P3, Clermont-Ferrand, France
\item \Idef{org1194}Laboratoire de Physique Subatomique et de Cosmologie (LPSC), Universit\'{e} Joseph Fourier, CNRS-IN2P3, Institut Polytechnique de Grenoble, Grenoble, France
\item \Idef{org1187}Laboratori Nazionali di Frascati, INFN, Frascati, Italy
\item \Idef{org1232}Laboratori Nazionali di Legnaro, INFN, Legnaro, Italy
\item \Idef{org1125}Lawrence Berkeley National Laboratory, Berkeley, California, United States
\item \Idef{org1234}Lawrence Livermore National Laboratory, Livermore, California, United States
\item \Idef{org1251}Moscow Engineering Physics Institute, Moscow, Russia
\item \Idef{org1322}National Centre for Nuclear Studies, Warsaw, Poland
\item \Idef{org1140}National Institute for Physics and Nuclear Engineering, Bucharest, Romania
\item \Idef{org1017626}National Institute of Science Education and Research, Bhubaneswar, India
\item \Idef{org1165}Niels Bohr Institute, University of Copenhagen, Copenhagen, Denmark
\item \Idef{org1109}Nikhef, National Institute for Subatomic Physics, Amsterdam, Netherlands
\item \Idef{org1283}Nuclear Physics Institute, Academy of Sciences of the Czech Republic, \v{R}e\v{z} u Prahy, Czech Republic
\item \Idef{org1264}Oak Ridge National Laboratory, Oak Ridge, Tennessee, United States
\item \Idef{org1189}Petersburg Nuclear Physics Institute, Gatchina, Russia
\item \Idef{org1170}Physics Department, Creighton University, Omaha, Nebraska, United States
\item \Idef{org1157}Physics Department, Panjab University, Chandigarh, India
\item \Idef{org1112}Physics Department, University of Athens, Athens, Greece
\item \Idef{org1152}Physics Department, University of Cape Town and  iThemba LABS, National Research Foundation, Somerset West, South Africa
\item \Idef{org1209}Physics Department, University of Jammu, Jammu, India
\item \Idef{org1207}Physics Department, University of Rajasthan, Jaipur, India
\item \Idef{org1200}Physikalisches Institut, Ruprecht-Karls-Universit\"{a}t Heidelberg, Heidelberg, Germany
\item \Idef{org1017688}Politecnico di Torino, Turin, Italy
\item \Idef{org1325}Purdue University, West Lafayette, Indiana, United States
\item \Idef{org1281}Pusan National University, Pusan, South Korea
\item \Idef{org1176}Research Division and ExtreMe Matter Institute EMMI, GSI Helmholtzzentrum f\"ur Schwerionenforschung, Darmstadt, Germany
\item \Idef{org1334}Rudjer Bo\v{s}kovi\'{c} Institute, Zagreb, Croatia
\item \Idef{org1298}Russian Federal Nuclear Center (VNIIEF), Sarov, Russia
\item \Idef{org1252}Russian Research Centre Kurchatov Institute, Moscow, Russia
\item \Idef{org1224}Saha Institute of Nuclear Physics, Kolkata, India
\item \Idef{org1130}School of Physics and Astronomy, University of Birmingham, Birmingham, United Kingdom
\item \Idef{org1338}Secci\'{o}n F\'{\i}sica, Departamento de Ciencias, Pontificia Universidad Cat\'{o}lica del Per\'{u}, Lima, Peru
\item \Idef{org1155}Sezione INFN, Catania, Italy
\item \Idef{org1313}Sezione INFN, Turin, Italy
\item \Idef{org1271}Sezione INFN, Padova, Italy
\item \Idef{org1133}Sezione INFN, Bologna, Italy
\item \Idef{org1146}Sezione INFN, Cagliari, Italy
\item \Idef{org1316}Sezione INFN, Trieste, Italy
\item \Idef{org1115}Sezione INFN, Bari, Italy
\item \Idef{org1286}Sezione INFN, Rome, Italy
\item \Idef{org36377}Nuclear Physics Group, STFC Daresbury Laboratory, Daresbury, United Kingdom
\item \Idef{org1258}SUBATECH, Ecole des Mines de Nantes, Universit\'{e} de Nantes, CNRS-IN2P3, Nantes, France
\item \Idef{org35706}Suranaree University of Technology, Nakhon Ratchasima, Thailand
\item \Idef{org1304}Technical University of Split FESB, Split, Croatia
\item \Idef{org1017659}Technische Universit\"{a}t M\"{u}nchen, Munich, Germany
\item \Idef{org1168}The Henryk Niewodniczanski Institute of Nuclear Physics, Polish Academy of Sciences, Cracow, Poland
\item \Idef{org17361}The University of Texas at Austin, Physics Department, Austin, TX, United States
\item \Idef{org1173}Universidad Aut\'{o}noma de Sinaloa, Culiac\'{a}n, Mexico
\item \Idef{org1296}Universidade de S\~{a}o Paulo (USP), S\~{a}o Paulo, Brazil
\item \Idef{org1149}Universidade Estadual de Campinas (UNICAMP), Campinas, Brazil
\item \Idef{org1239}Universit\'{e} de Lyon, Universit\'{e} Lyon 1, CNRS/IN2P3, IPN-Lyon, Villeurbanne, France
\item \Idef{org1205}University of Houston, Houston, Texas, United States
\item \Idef{org20371}University of Technology and Austrian Academy of Sciences, Vienna, Austria
\item \Idef{org1222}University of Tennessee, Knoxville, Tennessee, United States
\item \Idef{org1310}University of Tokyo, Tokyo, Japan
\item \Idef{org1318}University of Tsukuba, Tsukuba, Japan
\item \Idef{org21360}Eberhard Karls Universit\"{a}t T\"{u}bingen, T\"{u}bingen, Germany
\item \Idef{org1225}Variable Energy Cyclotron Centre, Kolkata, India
\item \Idef{org1017687}Vestfold University College, Tonsberg, Norway
\item \Idef{org1306}V.~Fock Institute for Physics, St. Petersburg State University, St. Petersburg, Russia
\item \Idef{org1323}Warsaw University of Technology, Warsaw, Poland
\item \Idef{org1179}Wayne State University, Detroit, Michigan, United States
\item \Idef{org1143}Wigner Research Centre for Physics, Hungarian Academy of Sciences, Budapest, Hungary
\item \Idef{org1260}Yale University, New Haven, Connecticut, United States
\item \Idef{org15649}Yildiz Technical University, Istanbul, Turkey
\item \Idef{org1301}Yonsei University, Seoul, South Korea
\item \Idef{org1327}Zentrum f\"{u}r Technologietransfer und Telekommunikation (ZTT), Fachhochschule Worms, Worms, Germany
\end{Authlist}
\endgroup